\documentclass[11pt]{article}
\usepackage{amsmath,amsthm,amssymb}
\usepackage{graphicx}
\usepackage{tikz}
\usepackage{float}
\usepackage{hyperref}
\usepackage{dsfont}
\hypersetup{colorlinks,citecolor=blue,urlcolor=blue,linkcolor=blue}
\usepackage[citestyle=authoryear, backend=bibtex,natbib=true,firstinits=true,doi=false,isbn=false,url=false,eprint=false,maxbibnames=20]{biblatex}
\usepackage{bbm}
\usepackage{algorithm}
\usepackage{algorithmic}
\usepackage{caption}
\usepackage{subcaption}
\usepackage{hhline}
\usepackage{booktabs}
\usepackage{multirow}
\usepackage{array}
\usepackage{placeins}
\usepackage{makecell}
\usepackage{arydshln}

\bibliography{ref.bib}

\newcounter{supplementalalgorithm}
\renewcommand{\thesupplementalalgorithm}{A\arabic{supplementalalgorithm}}

\newcommand{\bX}{\mathbf{X}}
\newcommand{\bY}{\mathbf{Y}}
\newcommand{\bx}{\mathbf{x}}
\newcommand{\by}{\mathbf{y}}

\newcommand{\bSigma}{\mathbf{\Sigma}}
\newcommand{\bOmega}{\mathbf{\Omega}}
\newcommand{\bmu}{\boldsymbol{\mu}}
\newcommand{\bbeta}{\boldsymbol{\beta}}
\newcommand{\bA}{\mathbf{A}}
\newcommand{\ba}{\mathbf{a}}
\newcommand{\bB}{\mathbf{B}}

\newcommand{\bC}{\mathbf{C}}

\newcommand{\bD}{\mathbf{D}}

\newcommand{\bF}{\mathbf{F}}
\newcommand{\bG}{\mathbf{G}}
\newcommand{\bH}{\mathbf{H}}
\newcommand{\bI}{\mathbf{I}}
\newcommand{\bL}{\mathbf{L}}

\newcommand{\bP}{\mathbf{P}}
\newcommand{\bV}{\mathbf{V}}
\newcommand{\be}{\mathbf{e}}
\newcommand{\br}{\mathbf{r}}
\newcommand{\bu}{\mathbf{u}}
\newcommand{\bv}{\mathbf{v}}
\newcommand{\bw}{\mathbf{w}}
\newcommand{\bS}{\mathbf{S}}
\newcommand{\bT}{\mathbf{T}}
\newcommand{\bU}{\mathbf{U}}
\newcommand{\bz}{\mathbf{z}}
\newcommand{\bZ}{\mathbf{Z}}
\newcommand{\bzero}{\mathbf{0}}

\def\amp{\mathop{\:\:\,}\nolimits}
\def\tr{\mathop{\rm tr}\nolimits}

\def\log{\mathop{\rm log}\nolimits}
\def\det{\mathop{\rm det}\nolimits}
\newcommand{\bbR}{\mathbb{R}}
\newcommand{\ind}{\perp \!\!\! \perp}

\newcommand{\st}{\vert}

\newtheorem{definition}{Definition}
\newtheorem{thm}{Theorem}

\makeatletter
\def\@fnsymbol#1{\ensuremath{\ifcase#1\or *\or 1 \or 2 \or 3 \or 4 \or 5 \else\@ctrerr\fi}}
\makeatother

\title{Second-order group knockoffs with applications to GWAS}
\author{Benjamin B. Chu$^1$, Jiaqi Gu$^2$, Zhaomeng Chen$^3$, Tim Morrison$^3$,\\ Emmanuel Cand\`es$^{3,4,*}$, Zihuai He$^{2,5,*}$, Chiara Sabatti$^{1,3,}\footnote{Joint corresponding authors}$}
\date{
    $^1$Department of Biomedical Data Science, Stanford University\\
    $^2$Department of Neurology and Neurological Sciences, Stanford University\\
    $^3$Department of Statistics, Stanford University\\
    $^4$Department of Mathematics, Stanford University\\
    $^5$Quantitative Sciences Unit, Department of Medicine, Stanford University\\
}
\begin{document}
\maketitle

\begin{abstract}






Conditional testing via the knockoff framework allows one to identify---among large number of possible explanatory  variables---those that carry unique information about an outcome of interest, and also provides a false discovery rate guarantee on the selection. This approach is particularly well suited to the analysis of genome wide association studies (GWAS), which have the goal of identifying genetic variants which influence traits of medical relevance. 

While conditional testing can be both more powerful and precise than traditional GWAS analysis methods, its vanilla implementation encounters a difficulty common to all multivariate analysis methods: it is challenging to distinguish among multiple, highly correlated regressors. This impasse can be overcome by shifting the object of inference from single variables to groups of correlated variables. To achieve this, it is necessary to construct ``group knockoffs." While successful examples are already documented in the literature, this paper substantially expands the set of algorithms and software for group knockoffs. We focus in particular on  second-order knockoffs, for which we describe correlation matrix approximations that are appropriate for GWAS data and that result in considerable computational savings. We illustrate the effectiveness of the proposed methods with simulations and with the analysis of albuminuria data from the UK Biobank.

The described  algorithms  are implemented in an open-source Julia package \texttt{Knockoffs.jl}, for which both R and Python wrappers are available.

\end{abstract}

\section{Introduction}\label{sec:introduction}

A common problem, in our data rich era, is the selection of important features  among a large number of recorded variables, with reproducibility guarantees. For example, in genome-wide association studies (GWAS) one observes data on million of features,  corresponding to polymorphic loci in the human genome, and wants to identify those that carry information about a phenotype of medical relevance.

Much research in recent years has focused on testing conditional independent hypotheses with False Discovery Rate (FDR)  \citep{benjamini1995controlling} control, which provides findings that are informative, precise, and reproducible. An effective way of doing so is via the knockoff framework \citep{barber2015controlling,candes2018panning,sesia2019gene,he2021genome}. 
Given $p$ features $(X_1,...,X_p)$ and a response $Y$, the conditional independence hypothesis  $H_j$ is false only if variable $X_j$ carries information on $Y$ that is not provided by the other variables under consideration. Formally, \begin{align}
    H_j: X_j \ind Y | X_{-j},\label{ci}
\end{align}
where $X_{-j}$ indicates all variables except for the $j$th one. The knockoff framework tests these hypotheses by comparing the signal in $X_j $ with that of  artificial negative controls $\tilde{X}_j$. 
The variables $\tilde{X} = (\tilde{X}_1,...,\tilde{X}_p)$ are constructed without looking at $Y$, such that the joint distribution of $(X, \tilde{X})$ satisfies a pairwise exchangeability condition
\begin{eqnarray*}
    (X, \tilde{X})_{\operatorname{swap}(\mathcal{A})} \overset{d}{=} (X, \tilde{X})
\end{eqnarray*}
for any subset $\mathcal{A} \subseteq \{1,...,p\}$. Above, $(X, \tilde{X})_{\operatorname{swap}(\mathcal{A})}$ is obtained from $(X, \tilde{X})$ by swapping the entries $X_j$ and $\tilde{X}_j$ for each $j \in \mathcal{A}$. 
See \cite{candes2018panning} for an overall description of the approach, as well as detailed discussion of motivation and assumptions.

Variables discovered by rejecting conditional hypotheses \eqref{ci} are not ``guilty by association," but carry unique information among the features in the data. This precision, however, comes with a price: if a non null variable is strongly correlated with a null one, the power to reject the associated conditional hypothesis can be diminished in finite samples. 
To remedy this, one can shift the object of inference from individual variables to groups of (highly correlated) variables. The knockoff framework can be adapted to this task, by constructing artificial control variables that conserve the joint distribution $(X_j,X_k)$ only when $j$ and $k$ belong to different groups. 

This idea of group knockoffs
was introduced in \cite{dai2016knockoff} and developed by   \cite{katsevich2019multilayer}, who considered the model-X framework we focus on in this paper.   \cite{sesia2020multi} and \cite{sesia2021false} applied this extensively to the analysis of GWAS, showing how group knockoffs can be constructed when the distribution of $(X_1,...,X_p)$  can be described with a hidden Markov model. For Gaussian designs, \cite{spector2022powerful} carried out a proof-of-concept study utilizing more general group knockoff algorithms. 
However, their python package \texttt{knockpy} encounters a few computational challenges for application in large-scale datasets. All in all, there remains a paucity of efficient algorithms to generate group knockoffs. 

We address this need by focusing on second-order knockoffs \citep{candes2018panning}. These are constructed to ensure that the necessary exchangeability properties hold up to the first two moments of the joint distribution of $X$ and $\tilde{X}$, rather than for the entire distribution. This produces valid knockoffs when $X$ has a Gaussian distribution and is a good approximation when the distribution of  $X$ can be effectively approximated as normal \citep{barber2020robust}. Relying on second-order knockoffs has minimal impact on FDR control when the feature importance measures are based on second-order moments.  For example, the analysis of summary statistics coming from GWAS data \citep{he2022ghostknockoff} is particularly well suited to second-order knockoffs.

In this paper, we extend three optimization frameworks for the construction of knockoffs \citep{candes2018panning,gimenez2019improving,spector2022powerful} to develop new algorithms for generating group knockoffs. We then focus on the task of constructing group knockoffs for the analysis of GWAS summary statistics. Leveraging existing databases, we identify variance-covariance matrices representing the correlation between genetic variants. We  regularize these matrices, using a combination of existing methods and novel representations which lend themselves to efficient knockoff computations. 

We implemented the proposed algorithms in an open-sourced \texttt{Julia} \citep{bezanson2017julia} package \texttt{Knockoffs.jl} available at \url{https://github.com/biona001/Knockoffs.jl}. We also provide \texttt{R} and \texttt{Python} wrappers via the packages \texttt{knockoffsr} and \texttt{knockoffspy} available at \url{https://github.com/biona001/knockoffsr} and \url{https://github.com/biona001/knockoffspy}. To facilitate downloading and extraction of LD matrices featured in our real data analysis, we developed another software \texttt{EasyLD.jl} available at \url{https://github.com/biona001/EasyLD.jl}. Finally, scripts to reproduce our real and simulated experiments are available at \url{https://github.com/biona001/group-knockoff-reproducibility}.

The rest of the paper is organized as follows. In section \ref{sec:problem_statement}, we formally define group conditional independence hypotheses, group knockoffs, and second-order group knockoffs.  We then recall different objective functions that have been considered in the literature for selecting among all covariance matrices that satisfy the stated requirements for regular (non-grouped) knockoffs. In section \ref{sec:algs}, we describe the key steps of our algorithms to solve the associated optimization problems under the grouped setting. Section \ref{sec:condinp} illustrates how conditional independence among the features can be exploited to significantly reduce computational costs. Next, we turn to the analysis of covariance matrices associated with genetic data in section \ref{sec:summarystat_gwas}. Section \ref{sec:simulations} explores the performance of our proposals in terms of FDR, power, and computational costs with  simulations. We conclude in section \ref{sec:gwas} with a real data application from the UK Biobank that results in interesting new findings.

\section{Problem statement}\label{sec:problem_statement}



We are interested in discovering which among $p$ features $X = (X_1,...,X_p)$ carries information on a response  $Y$. We assume that the distribution of $X$ is known, with variance covariance  $\bSigma$, but make no assumptions on the distribution of $Y | X$. We consider a partition of 
 $[p] = \{1, 2,..., p\}$ into $g$ disjoint groups of variables. For $\gamma \in \{1, 2,..., g\}$, we denote by $ {\cal A}_{\gamma}$ and $X_{\gamma}$ respectively, the collection of indices corresponding to group $\gamma$ and the vector of random  variables in group $\gamma$, and by  $X_{-\gamma}$ the collection of all other features. Without loss of generality, we assume that the indices of all variables in the same group are contiguous.
 Given $n$ i.i.d.~samples, we use $\bX \in \bbR^{n \times p}$ to denote the stacked design matrix where each row is a sample. For other variables, we will use bolded uppercase text to denote matrices, unbolded uppercase for random variables, and bolded lowercase for vectors. 

Throughout this paper, the problem of interest is to test the \textit{group conditional independence hypotheses}
\begin{equation} \label{grouphypothesis} H_{\gamma}: X_{\gamma} \ind Y \st X_{-\gamma}
\end{equation} 
for $1 \leq \gamma \leq g$. The idea of exploring the joint influence of multiple correlated variables on an outcome has been proposed multiple times in the literature. When the data are not informative enough to distinguish between the contributions of single variable, moving the inference to the group level can yield an increase in power. For example, in the context of gene mapping, this approach is implemented in methods such as CAVIAR \citep{hormozdiari2014identifying} and SuSiE \citep{wang2020simple}.
The knockoffs procedure can be adapted to test hypotheses of the form (\ref{grouphypothesis}) by constructing artificial controls that are ``indistinguishable'' from the original variables at the group level. Specifically,  as in  \cite{katsevich2019multilayer},  {\bf group knockoffs }$\tilde{X} = (\tilde{X}_1,..., \tilde{X}_p)$ for the variables $X = (X_1,..., X_p)$ are defined by the following properties:
\begin{eqnarray}
  \text{Conditional independence:} &&  \tilde{X} \ind  Y \st X \label{GKi} \\ 
  \text{Group exchangeability:} && (X, \tilde{X})_{\text{swap}({\cal C})} \overset{d}{=}  (X, \tilde{X}) \;\;\;\; \forall\;\;  {\cal C}: \;\;{\cal C}=\cup_{\gamma\in {\cal S}} {\cal A}_{\gamma}
  \label{GKe}
  \end{eqnarray}
 where ${\cal S}$ is a set of groups $\gamma$, and  $(X, \tilde{X})_{\text{swap}({\cal C})}$ is formed by swapping $X_j$ with $\tilde{X}_j$ for every index $j$ in ${\cal C}$. 
Condition (\ref{GKe}) is what distinguishes group knockoffs from individual variable knockoffs. The latter require the distributional invariance to hold for all sets ${\cal C}$, not just those that can be described as a union of groups. Since fewer exchangeability restrictions are imposed on group knockoffs, they have added flexibility that can be leveraged for increased power.
For an illustration of this advantage, see supplemental section \ref{Wgroups}.

 As in  \cite{candes2018panning}, rather than requiring that $(X, \tilde{X})_{\text{swap}({\cal C})} $ and  $(X, \tilde{X})$ have the same distribution for any ${\cal C}$, we
can ask that they have the same first two moments, i.e., the same mean and covariance. This approximate construction is known as {\em second-order} knockoffs and is the focus of the present paper. If all variables have been standardized to have mean zero, the second-order version of condition (\ref{GKe}) translates to the following property of the  correlation matrix of $(X, \tilde{X})$: 
\begin{equation} \label{G_S} \text{Cov}(X, \tilde{X}) = \bG_{\bS} = \begin{bmatrix} \bSigma & \bSigma - \bS \\ \bSigma - \bS & \bSigma \end{bmatrix} \in \mathbb{R}^{2p \times 2p}, \end{equation} 
with $\bS$ block-diagonal, where the blocks correspond to the  groups $\gamma\in \{1,\ldots, g\}$. That is, $S_{ij} = 0$ whenever indices $i$ and $j$ are not in the same group of the partition. For ease of notation, we thus write $\bS = \text{diag}\{\bS_1,..., \bS_g\}$, where $\bS_i$ s the $p_i \times p_i$ symmetric matrix for the $i$th group.  Again, it is useful to contrast (\ref{G_S}) with the analogous requirement for second-order (single variables) knockoffs, where $\bS$ needs to be a diagonal matrix. When trying to make inferences at the group level, only correlations across groups need to be conserved, resulting in greater degrees of freedom in the construction of knockoffs. In addition, note that $\bS$ has to be chosen such that $\bG_{\bS}$ remains a valid $2p \times 2p$ covariance matrix, leading to the constraints that $\bS \succeq 0$\footnote{This means that $\bS$ is positive semidefinite.} and $2 \bSigma - \bS \succeq 0$, as in \cite{barber2015controlling, candes2018panning}. In particular all the blocks $\bS_i$ are positive semidefinite.

What choice of $\bS$ leads to strong performance? While any valid knockoff will guarantee FDR control, we aim for a construction that leads to high power. To develop intuition about this question, it is useful to recall how knockoffs are used for inference on the hypotheses (\ref{grouphypothesis}).

%
Once  valid group knockoffs $\tilde{\bX}$ have been constructed for each observation, the augmented data $(\bY, \bX, \tilde{\bX})$ is used to obtain a vector of feature importance statistics $(Z,\tilde{Z}) \in \bbR^{2g}$ where, for $\gamma \leq g$, $Z_{\gamma}$ and $\tilde{Z}_{\gamma}$ reflect the importance of group $\gamma$ and its corresponding set of knockoffs. The only requirement on $(Z,\tilde{Z})$ is that swapping all variables in group $\gamma$ with their corresponding knockoffs also swaps the values of $Z_{\gamma}$ and $\tilde{Z}_{\gamma }$. 
Next, define the knockoff score $W_{\gamma} = w(Z_{\gamma}, \tilde{Z}_{\gamma})$, where $w$ is any antisymmetric function (e.g., $w(x, y) = x - y$). Intuitively, $W_{\gamma}$ measures the relative importance of group $\gamma$ as compared to its group of knockoffs. $W_{\gamma}$ should therefore be large and positive for groups with true importance and is equally likely to be positive or negative for null groups. The procedure concludes by rejecting all groups $\gamma$ for which $W_{\gamma} \geq \tau$, where the data-dependent threshold $\tau$ is computed from the knockoff filter \citep{barber2015controlling}. 



 
 Given that the key to the selection of group $\gamma$ is the difference between the feature importance statistics $Z_{\gamma}$ and $Z_{\gamma + g}$, it is natural to aim for a construction that makes $X$ and $\tilde{X}$ as distinguishable as possible. Here, we describe from the literarure three popular criteria for this task and generalize them to the group setting. 

The first suggestion  implemented  in the literature is to {\bf minimize the  average correlation} between $X_j$ and $\tilde{X}_j$  \citep{barber2015controlling,candes2018panning}, 
 which \cite{dai2016knockoff} generalize in principle  to the group setting. We refer to this as the SDP criterion, since the underlying optimization problem requires solving a semi-definite programming problem. This can be formulated as 
\begin{equation} \label{SDPcriterion}\tag{SDP} \underset{0 \preceq \bS \preceq 2\bSigma}{\min} \sum_{\gamma =1}^g \frac{1}{|\mathcal{A}_{\gamma}|^2} \sum_{i, j \in {\cal A}_{\gamma}} |S_{ij} - \Sigma_{ij}|. \end{equation}
Here, the $|\mathcal{A}_{\gamma}|^2$ terms normalize each group by its size to prevent larger groups from dominating the criterion, since all groups ought to matter equally for power consideration. In fact, \cite{dai2016knockoff}  consider only a special case of this problem, where $\bS_{\gamma}=\tau \cdot \bSigma_{\gamma},$ with $\bSigma_{\gamma}$ the submatrix of $\bSigma$ corresponding to the variables in group $\gamma$. This simplification (which we refer to as equivariant SDP or eSDP)  amounts to choosing $\tau=\min \left\{1,2 \cdot \lambda_{\min }(\bB \bSigma \bB)\right\}$ where $ \bB=\operatorname{diag}\left\{\bSigma_{1}^{-1 / 2}, \ldots, \bSigma_{g}^{-1 / 2}\right\}$. 
Algorithms to solve the more general problem have not been described previously, nor their efficacy explored in depth.

Working with individual level knockoffs,  \cite{gimenez2019improving}  explored a different suggestion from \cite{candes2018panning}: minimizing the mutual information, or {\bf maximizing the entropy} (ME), between $X$ and $\tilde{X}$. In the case of Gaussian variables, the entropy of $(X, \tilde{X})$ is proportional to the log-determinant of its covariance matrix. In the single variable setting, the objective is 
\begin{equation}\underset{0 \preceq \bS \preceq 2\bSigma}{\min}  \log \det(\bG_{\bS}^{-1})\label{ME}\tag{ME}\end{equation}
with a diagonal matrix $\bS$. In the group setting, the diagonal constraint on $\bS$ translates to a group-block-diagonal constraint as before. In both cases, the entropy maximization program is  convex  in  $\bS$. 

Finally, \cite{spector2022powerful} introduce another criterion that, like ME, focuses on minimizing the reconstructability of $X_j$ from $\tilde{X}$.   Knockoffs that satisfy the {\bf minimum variance-based reconstructability} (MVR) criterion minimize
$$\sum_{j=1}^{p}(\mathbb{E}[\text{Var}(X_j \st X_{-j}, \tilde{X})])^{-1}$$
In the case of Gaussian variables, or under the second-order approximation, this is equivalent to 
\begin{equation} \underset{0 \preceq \bS \preceq 2\bSigma}{\min}  \text{Tr}(\bG_{\bS}^{-1}).\label{MVR}\tag{MVR}\end{equation}        

\cite{spector2022powerful}, working primarily with individual level knockoffs, show how ME and MVR can yield better power than SDP. 
To summarize, we consider three second-order group knockoff construction procedures, each minimizing a different  loss functions:
\begin{equation}\label{eq:ko_objs}
	L_{\text{SDP}}(\bS) = \sum_{\gamma= 1}^g \frac{1}{|\mathcal{A}_{\gamma}|^2} \sum_{i, j \in {\cal A}_{\gamma}} |S_{ij} - \Sigma_{ij}|, \quad L_{\text{MVR}}(\bS) = \text{Tr}(\bG_{\bS}^{-1}), \quad L_{\text{ME}}(\bS) = \log \det(\bG_{\bS}^{-1}),
\end{equation}
under the  constraints  that $0 \preceq \bS \preceq 2\bSigma$ and  $\bS = \text{diag}(\bS_1,..., \bS_{g})$. 

Before discussing our algorithms, we note that, in the interest of generality, we will work with 
the multiple knockoff filter of \cite{gimenez2019improving}. That is, instead of generating one knockoff copy of each observation, we will generate $m$, and suitably aggregate them for inferential purposes.  This approach was used extensively in \citep{he2021genome,he2022ghostknockoff} to improve the power and stability of model-X knockoffs. 

As in  \cite{gimenez2019improving}, in the case of second-order knockoffs, this simply means that we will need to use a variance-covariance matrix of the form \begin{eqnarray} \label{eq:G_Smult}
	\bG_{\bS} &=& 
	\begin{bmatrix}
		\bSigma & \bSigma - \bS & \cdots & \bSigma - \bS\\
		\bSigma - \bS & \bSigma & \cdots & \bSigma - \bS\\
		\vdots & \vdots & \ddots & \vdots\\
		\bSigma - \bS & \cdots & \cdots & \bSigma
	\end{bmatrix} \in \bbR^{p(m + 1) \times p(m + 1)}.
\end{eqnarray}
The optimization of the objectives in (\ref{eq:ko_objs}) will now be subject to 
 $0 \preceq \bS \preceq \frac{m+1}{m}\bSigma$ and $\bS = \text{diag}(\bS_1,...,\bS_g)$ (note that this recovers the original constraints when $m = 1$). 


\vspace*{.55cm}
\noindent {\huge \sc Methods}

\section{Algorithms for second-order group knockoffs}\label{sec:algs}

While  the modification of the objective from individual level to group knockoffs is formally straightforward, the fact that $\bS$ becomes block-diagonal poses computational challenges. The added degrees of freedom in the knockoff construction can translate to higher power, but also to a much larger number of variables to optimize over. Indeed, \cite{dai2016knockoff} who introduce the first of the group loss functions (\ref{eq:ko_objs}), immediately reduce the dimensions of the problem by imposing the structure $\bS_\gamma=\tau \cdot \bSigma_{\gamma}$ for all $\gamma$.  While this leads to a convenient one-dimensional problem, it counters the increased flexibility of group knockoffs, reducing power in a number of situations.

Our goal is to design computationally feasible algorithms that allow researchers to use any of the second-order group knockoff constructions in (\ref{eq:ko_objs}) while optimizing over all of $\bS$. To do so, we combine full coordinate descent with some coarser updates -- somewhat inspired by \cite{dai2016knockoff,askari2021fanok,spector2022powerful}. 

We start by describing a general coordinate descent strategy, which optimizes over every non-zero entry of $\bS$ by sequentially performing the updates
\begin{eqnarray}
    S_{ij}^{new} &=& S_{ij} + \delta_{ij} \label{cd}.
\end{eqnarray}
We impose symmetry on $\bS$, so we must also have $S_{ji}^{new} = S_{ij}^{new}$. {\cite{spector2022powerful} briefly explored a similar form of update in their \texttt{knockpy} software, although a detailed comparison is difficult since their algorithm is sparsely documented. Our approach here is derived independently.}

In our approach, there are three key steps in performing this update. The first is identifying the valid values of $\delta_{ij}$. If we write $\bD = \frac{m+1}{m}\bSigma - \bS$, then the positive definiteness constraints require that both $\bS^{new}$ and $\bD^{new}= \frac{m+1}{m}\bSigma - \bS^{new}$ are positive definite. If $\be_i, \be_j$ are the $i$th and $j$th basis vector, then  \ref{sec:feasible_region_diag} shows that this feasible region is
\begin{eqnarray}\label{eq:ccd_feasble_region}
    \begin{cases}
        \frac{-1}{\be_j^t\bS^{-1}\be_j} \le \delta_{ij} \le \frac{1}{\be_j^t\bD^{-1}\be_j} & \text{if } i = j\\
        \frac{-2}{\be_i^t\bS^{-1}\be_i + 2\be_i^t\bS^{-1}\be_j + \be_j^t\bS^{-1}\be_j} \le \delta_{ij} \le \frac{2}{\be_i^t\bD^{-1}\be_i + 2\be_i^t\bD^{-1}\be_j + \be_j^t\bD^{-1}\be_j} & \text{if } i \ne j
    \end{cases}
\end{eqnarray}
which contains $\delta_{ij} = 0$. To prevent $\delta_{ij}$ from reaching the boundary condition, which yields $\bS$ or $\bD$ numerically singular, in practice we make the feasible region slightly smaller by taking away $\epsilon=10^{-6}$ from both ends. 

The second step is to optimize for $\delta_{ij}$ in this region for different objective functions, as detailed in the supplemental section \ref{sec:supplementccd}. For example, in the case of the ME objective, we can solve for $\delta_{ij}$ by maximizing
\begin{eqnarray}\label{eq:ccd_me_update}
    \begin{split}
    g(\delta_{ij}) &=& \log\left((1 - \delta_{ij} \be_i^t\bD^{-1}\be_j)^2 - \delta^2_{ij} \be_i^t\bD^{-1}\be_i\be_j^t\bD^{-1}\be_j\right) + \\
    && m\log\left((1+\delta_{ij} \be_i^t\bS^{-1}\be_j)^2 - \delta^2_{ij} \be_j^t\bS^{-1}\be_j\be_i^t\bS^{-1}\be_i\right).
    \end{split}
\end{eqnarray}
If all constants such as $\be_i^t\bS^{-1}\be_i$ and $\be_i^t\bD^{-1}\be_i$ are stored, then $g(\delta_{ij})$ is a scalar-valued function with scalar inputs, and is therefore easy to optimize within an interval. 

The final step is to update the stored terms such as $\be_i^t\bS^{-1}\be_i$ and $\be_i^t\bD^{-1}\be_i$. We adopt the approach in \cite{askari2021fanok} and \cite{spector2022powerful} by precomputing and maintaining Cholesky factors of $\bD = \bL\bL^t$ and $\bS = \bC\bC^t$. Constants such as $\be_i^t\bD^{-1}\be_j$ can then be efficiently computed by noting that $\be_i^t\bD^{-1}\be_j = \be_i^t(\bL\bL^t)^{-1}\be_j = (\bL^{-1}\be_i)^t(\bL^{-1}\be_j) \equiv \bv^t\bu$, where $\bv$ and $\bu$ can be computed via forward-backward substitution on $\bL\bu = \be_j$ and $\bL\bv = \be_i$. After updating $S_{ij}$ and $S_{ji}$, we perform a rank-2 update to the Cholesky factors to maintain the equalities $\bL_{new}\bL_{new}^t = \frac{m+1}{m}\bSigma - \bS_{new}$ and $\bC_{new}\bC_{new}^t = \bS_{new}$. Full details are available in \ref{sec:efficiently_obtaining_const}.

The coordinate descent updates (\ref{cd}) allow optimization over all free parameters of $\bS$ rather than reducing the dimension of the problem. However, the local nature of these moves encounters two connected challenges. First, given a large number of optimization variables, a complete update of all the elements of $\bS$ can require substantial time (for example, note that if a  group comprises  $1000$ elements, then that single group would contribute $10^6$ optimization variables). Second, even if the objectives are convex, coordinate descent can become stuck at a local minimum due to the presence of constraints (positive definiteness of $\bS$ and $\bD$).

To alleviate these difficulties, we found it useful to augment our iterative procedure with `global' updates, which restrict the set of possible values for $\bS$, in a similar but less limiting way than what is described in  \cite{dai2016knockoff}. 
Specifically, we consider the following ``PCA updates":
\begin{eqnarray}
    \bS^{new} &=& \bS + \delta_i \bv_i\bv_i^t, \label{pca-cd}
\end{eqnarray}
where $i \in [p]$ and each $\bv_i$ are precomputed vectors such that the outer product $\bv_i\bv_i^t$ respects the block diagonal structure of $\bS$.  In practice, we choose $\bv_1,...,\bv_p$ to be the eigenvectors of the block-diagonal covariance matrix 
\begin{eqnarray}\label{eq:sigma_block}
    \bSigma_{blocked} &=&
    \begin{bmatrix} 
        \bSigma_1 & & \\ & \ddots & \\ & & \bSigma_g
    \end{bmatrix}
\end{eqnarray}
In the case of the ME criterion, we can explicitly compute
\begin{eqnarray}\label{eq:pca_me_update}
    \delta_i &=& \frac{m\bv_i^t\bS^{-1}\bv_i - \bv_i^t\bD^{-1}\bv_i}{(m+1)\bv_i^t\bS^{-1}\bv_i\bv_i^t\bD^{-1}\bv_i},
\end{eqnarray}
which conveniently lies in  the feasible region
\begin{eqnarray}\label{eq:pca_feasible_region}
    \frac{-1}{\bv_i^t\bS^{-1}\bv_i} \le \delta_i \le \frac{1}{\bv_i^t\bD^{-1}\bv_i}.
\end{eqnarray}
Each full iteration now requires the optimization of only $p$ variables, so the computational complexity is equivalent to that of coordinate descent model-X knockoffs \citep{askari2021fanok} in the non-grouped setting, scaling as $\mathcal{O}(p^3)$. Full details of such PCA updates are provided in Supplemental section \ref{sec:pca_ccd}.

In practice, we find that alternating between updates of types (\ref{cd}) and (\ref{pca-cd}) greatly reduces the possibility of the algorithm being trapped in a local minimum while maintaining a small overall computational cost. This immediately yields the question of how often one should alternate between general coordinate descent (\ref{cd}) and the faster but more restrictive PCA updates (\ref{pca-cd}). Our numerical experiments with a few alternating schemes produce no clear winner. Thus, by default, our software performs them back-to-back, and we leave it as an option for the user to adjust. The overall algorithm is summarized in Algorithm \ref{alg:CCD_concise}.

\begin{algorithm}
    \caption{Scheme of iterative solution of  (\ref{SDPcriterion}),(\ref{ME}),(\ref{MVR})}\label{alg:CCD_concise}
    \begin{algorithmic}[1]
	\STATE { \textbf{Input:} correlation matrix $\bSigma_{p \times p}$, group membership vector, and $m$ (number of knockoffs)}
        \STATE{ \textbf{Initialize:} $\bS = \frac{1}{2}\bS_{\text{eSDP}}$ with $\bS_{\text{eSDP}}$ from  \citep{dai2016knockoff}} and $\bD = \frac{m+1}{m}\bSigma - \bS$
        \STATE{ \textbf{Compute:} ($\bv_1,...,\bv_p$) based on eigendecomposition of $\bSigma_{block}$ in eq \eqref{eq:sigma_block}}
        \STATE{ \textbf{Compute:} Cholesky factors $\bL\bL^t = \text{Cholesky}(\bD)$ and $\bC\bC^t = \text{Cholesky}(\bS)$}
	\WHILE{Not converged}
            \FOR{$\bv_i \in \{\bv_1,...,\bv_p\}$} 
                \STATE Compute $\delta_i$ according to eq \eqref{eq:pca_me_update} for ME, \eqref{eq:pca_mvr_update} for MVR, or solving \eqref{eq:pca_sdp_update} for SDP
                \STATE Clamp $\delta_i$ to be within feasible region as in \eqref{eq:pca_feasible_region}.
                \STATE Perform rank-1 update: $\bS_{new} = \bS + \delta_i \bv_i\bv_i^t$
                \STATE Update Cholesky factors $\bL_{new}\bL_{new}^t = \frac{m+1}{m}\bSigma - \bS_{new}$ and $\bC_{new}\bC_{new}^t = \bS_{new}$
		\ENDFOR
            \FOR{$\gamma \in \{1,...,g\}$}
                \FOR{$(i,j)$ in group $\gamma$}
                    \STATE Compute $\delta_{ij}$ by solving eq \eqref{eq:ccd_me_update} for ME, \eqref{eq:ccd_mvr_update} for MVR, or \eqref{eq:ccd_sdp_update} for SDP
                    \STATE Clamp $\delta_{ij}$ to be within feasible region as in eq \eqref{eq:ccd_feasble_region}
                    \STATE Perform rank-2 update: $S^{new}_{ij} = S^{new}_{ji} = S_{ij} + \delta_{ij}$
                    \STATE Update Cholesky factors $\bL_{new}\bL_{new}^t = \frac{m+1}{m}\bSigma - \bS_{new}$ and $\bC_{new}\bC_{new}^t = \bS_{new}$
                \ENDFOR
            \ENDFOR
        \ENDWHILE
        \STATE \textbf{Output:} Group-block-diagonal matrix $\bS$ satisfying $\frac{m+1}{m}\bSigma - \bS \succeq 0$ and $\bS \succeq 0$.
	\end{algorithmic}
\end{algorithm}

\section{Exploiting conditional independence }\label{sec:condinp}

While the approaches just described allow us to identify $\bS$ that minimizes any of the three considered loss function,  $L_{\text{SDP}}$, $L_{\text{MVR}}$ or $L_{\text{ME}}$, 
depending on the size of the problem (defined both by $p$ and the size of the largest groups) the solution can be computationally intensive.  To avoid unnecessary costs, it is important to leverage  structure in the distribution of $X$, when this is possible. \cite{Metro} discuss the 
general computational advantages in knockoff construction associated with conditional independence and \cite{sesia2019gene} leverages these in the context of hidden Markov models (theorem 1), while \cite{sesia2020multi} extend the same construction to group knockoffs (see Proposition 2 in the supplement). 

We consider here a particular form of conditional independence, linked to the group structure.
\begin{definition} \label{gkci} Let $X$ be $p$ random variables, partitioned in groups $\gamma=1,\ldots g$. We say that their distribution $P_X$ has a group-key conditional independence property with respect to the partition $\{\mathcal{A}_\gamma\}_{\gamma=1}^g$ if 
for each $\gamma$  there exists  two disjoint subsets $\mathcal{A}^\dagger_\gamma$ and $\mathcal{A}^\star_\gamma$ such that
\begin{eqnarray*}
\mathcal{A}_\gamma& = &\mathcal{A}^\dagger_\gamma \cup \mathcal{A}^\star_\gamma\\
 	X^\dagger_\gamma & \ind &  X_{-\gamma}|X^\star_{\gamma},\quad \gamma\in [g],
\end{eqnarray*}
where $ X^\dagger_\gamma $ ($X^\star _\gamma$) collects  the variables  $X_i$ with $i \in \mathcal{A}^\dagger_\gamma$ ($\mathcal{A}^\star_\gamma$).
\end{definition}
Definition \ref{gkci} describes a setting where,  within each group $\gamma$, there is a subset of key variables that ``drive'' the dependence across groups, and conditioning on these all other variables are independent across groups. 
 Algorithm \ref{alg:Groupknockoff} capitalizes on this conditional independence for the construction of knockoffs. 
{\begin{algorithm}
		\caption{Group knockoffs under  group key conditional independence (Definition \ref{gkci}).}\label{alg:Groupknockoff}
		\begin{algorithmic}[1]
			\STATE { \textbf{Input:} Realizations of the variables $X$, and their joint distribution.}
			\STATE { Generate valid group knockoffs $\widetilde{X}^{\star}$ of $X^\star=(X^\star_1,\ldots,X^\star_g)$.}
			\STATE Compute the conditional distribution $F_{\gamma}$ of $X^\dagger_\gamma$ given $X^\star$ ($\gamma=1,\ldots,g$). 
			\STATE Generate knockoffs $\widetilde{X}_\gamma^{\dagger}\sim F_{\gamma}(\cdot|\widetilde{X}^{\star})$ ($\gamma=1,\ldots,g$).
			\STATE { \textbf{Output:} group knockoffs $(\widetilde{X}^{\star}_1,\widetilde{X}^{\dagger}_1),\ldots,(\widetilde{X}^{\star}_g,\widetilde{X}^{\dagger}_g)$.}
		\end{algorithmic}
\end{algorithm}}
While it resembles that adopted for HMM in 
\citet{sesia2019gene}, the conditional independence structure  it rests upon is different: for each $\gamma\in [g]$, the size of  both $\mathcal{A}^\dagger_\gamma$ and $\mathcal{A}^\star_\gamma$ is arbitrary and within each group the variables $X^\dagger_\gamma$ are not independent given  $X^\star _\gamma$. In the interest of simplicity, we have outlined algorithm \ref{alg:Groupknockoff}  for one knockoff copy, but it easily extends to creating $m$ knockoff copies, by generating  $\widetilde{X}^{\star (1)},\ldots \widetilde{X}^{\star (m)}$ in step 2, and $m$ independent copies in step 4. Finally, note that Algorithm \ref{alg:Groupknockoff} is stated for exact knockoff construction (requiring knowledge on the joint distribution of $X$), even if in the context of this paper, we will use it for second-order knockoffs.

\begin{thm}\label{thm:$M$-groupknockoffs}
	Knockoffs {\rm $\widetilde{X}$} generated under Algorithm {\rm\ref{alg:Groupknockoff}}, are valid 
 knockoffs  for groups $\{\mathcal{A}_\gamma\}_{\gamma=1}^g$ when the distribution of $X$ has group key conditional independence with respect to $\{\mathcal{A}_\gamma\}_{\gamma=1}^g$.
\end{thm} 

 Theorem \ref{thm:$M$-groupknockoffs}---proven in the supplemental section \ref{pr:$M$-groupknockoffs}---assures the validity of Algorithm \ref{alg:Groupknockoff}. Assuming that sampling from $F_\gamma$ in step 3 is easy, the computational advantages of algorithm {\rm\ref{alg:Groupknockoff}} lie in the reduction of  the number of the optimization variables. As group knockoffs have to be constructed only for $\{\mathcal{A}^\star_\gamma:\gamma\in [g]\}$,  depending on the criteria, we have to minimize either
$$L_{\text{SDP}}^\star(\textbf{S}^\star) = \sum_{\gamma=1}^g \frac{1}{|\mathcal{A}^\star_\gamma|^2} \sum_{i, j \in \mathcal{A}^\star_\gamma} |S_{ij} - \Sigma_{ij}|, \quad L_{\text{MVR}}^\star(\textbf{S}^\star) = \text{Tr}((\textbf{G}_{\textbf{S}^\star}^\star)^{-1}), \text{ or}\quad L_{\text{ME}}^\star(\textbf{S}^\star) = \log \det((\textbf{G}_{\textbf{S}^\star}^\star)^{-1}).$$
This reduces the number of optimization variables from $\sum_{\gamma=1}^g{|\mathcal{A}_\gamma|^2}$ to $\sum_{\gamma=1}^g{|\mathcal{A}^\star_\gamma|^2}$, which can be significant.

Algorithm \ref{alg:Groupknockoff}, leads to valid knockoffs with potential computational savings. But how ``good'' are these knockoffs? How do they compare to those constructed to satisfy any of the optimality criteria we considered in the previous section?
 Theorem \ref{thm:optimality} (proof  in the supplemental section \ref{pr:optimality}) offers some light on this topic: when ME is the criteria of choice, the group knockoffs constructed using the variance-covariance matrix $\bG_{\bS}$ from  Algorithm  \ref{alg:CCD_concise} have the same distribution as those constructed starting from $\textbf{G}_{\textbf{S}^\star}^\star$ and  following Algorithm \ref{alg:Groupknockoff}.
  \begin{thm}\label{thm:optimality}
 Let $\{\mathcal{A}_{\gamma}\}_{\gamma=1}^g$ be a collection of groups for the variables in  $X$; and let  $X$ have the group key conditional independence property with respect to $\{\mathcal{A}_\gamma\}_{\gamma=1}^g$. Let $\tilde{X}$ be  knockoffs generated according to Algorithm \ref{alg:Groupknockoff}, and such that the distribution of $(X^{\star},\tilde{X}^{\star})$  minimizes {\rm $L^\star_{\text{ME}}(S^\star)$}. Then, the distribution of $(X,\tilde{X})$ minimizes {\rm $L_{\text{ME}}(S)$}.
	\end{thm}


Theorems \ref{thm:$M$-groupknockoffs} and \ref{thm:optimality} taken together suggest that, if one is interested in ME group-knockoffs, a particularly convenient computational approach is available when the distribution of $X$ follows a precise form of conditional independence.

Unfortunately, however,  it is unrealistic to assume that the property described by Definition \ref{gkci} holds exactly, and that we have knowledge of  the variables in the $\star$ set. This being said, especially in the context of second-order (approximate) knockoffs, it is meaningful to try to identify a partition of the $p$ variables and a subset of key variables in each of the groups, so as to approximate the true distribution of $X$ with one that has the desired properties. 
The supplemental section \ref{heuristic} describes a heuristic approach to solve this problem. While we can make no general statement about its validity, we have studied its performance for genetic data, as we will detail in the following section.

\section{Group knockoffs for GWAS summary statistics }\label{sec:summarystat_gwas}
%
%
%
%

\color{black} 
%

Before applying our group knockoff algorithms, let us first discuss an application of second order group knockoffs which is a major motivation for this paper. 
If we have access to individual-level data $\bX \in \mathbb{R}^{n \times p}$ where each row is an i.i.d sample drawn from $X \sim \mathcal{N}(\bmu, \bSigma)$, then it is straightforward to estimate $\hat{\bmu}$ and $\hat{\bSigma}$ (see Supplemental \ref{sec:estimate_mu_and_sigma}). Unfortunately, sharing individual-level genotypes and phenotypes encounters numerous privacy and security issues, and introduces additional costs associated with storing and transmitting large datasets. Thus, despite considerable and successful efforts to encourage this practice, distribution of summary statistics continues to serve as an attractive option for geneticists. 

\cite{he2022ghostknockoff} described how conditional independence hypotheses can be tested using a knockoff framework even without access to $\bX$ and $\by$.  In a nutshell, if the feature importance statistic used in the test of $H_j$ (\ref{ci}) is the marginal Z-scores $Z_j = \frac{1}{\sqrt{n}}\bx_j^t\by \in \mathbb{R}$ ($\bx_j$ is the $j$th column of $\bX$), and the distribution of $(Z_1,\ldots, Z_p)$ is Gaussian, we can sample a value for  the corresponding $\tilde{Z}_j$  without having to generate knockoffs for all the distinct observations, but rather a {\em ghost knockoff} for $Z_j$ directly. In a companion work \citep{chen2024controlled} we show how a similar approach can be extended to feature importance statistics derived from a  lasso-like procedure, which takes as input  the  marginal Z-scores and covariance matrix $\bSigma$, where it is assumed that $(X_1,...,X_p) \sim \mathcal{N}(\bzero, \bSigma)$ and $\bSigma$ can be estimated. Although these methods enable a knockoff-based analysis of summary statistics data, the presence of tightly linked variants leads to highly correlated Z-scores, which can diminish power. As such, it could be useful to incorporate group knockoffs to this line of research, as detailed below. 

Even in this case, we rely on the fact that the distribution of  Z-scores is well-approximated by a Gaussian and we sample ghosts knockoffs $\tilde{Z}$ with a second order knockoff procedure. 
It is then useful to clarify how information on the relevant variance-covariance matrices can be gathered, how groups can be identified and how well the genotype distribution can be approximated with one satisfying  group-key conditional independence.

\subsection{Obtaining suitable estimates for variance-covariance matrix}\label{sec:obtaing_sigma_ukb}

Large consortiums such as Pan-UKB \citep{panukbb} and gnomAD \citep{chen2022genome}  make available sample variance-covariance  matrices calculated on genotypes of individuals from different human populations.  We  rely on  the European Pan-UKB panel containing $p\approx 24$ million variants across the human genome derived from $\approx 5 \times 10^5$ British samples. We restricted our attention to variants present in the UK Biobank GWAS genotype array \citep{sudlow2015uk} with minor allele frequency exceeding 0.01 in the Pan-UKB panel. 

Given $\hat{\bSigma}_{\text{PanUK}}$, we identified approximately independent blocks of SNPs by directly adapting the output of \texttt{ldetect} \citep{berisa2016approximately}. A total of 1703 blocks $\hat{\bSigma}_{1},...,\hat{\bSigma}_{1703}$ of size varying between   $\sim10^2$ and  $\sim 10^3$ were identified, see \ref{sec:LDmatrix_summarystats} for summary statistics. To ensure we obtained a reliable estimate of the true population SNP variance-covariance matrix, we regularized the resulting $\bSigma_i$s in multiple ways, as detailed in \ref{sec:LDregularization}. These matrices are then used for knockoff generation and computing feature importance statistics, as described below.

\subsection{Defining groups and key-variables}

Within each of the 1703 blocks, we partition the SNPs  into groups  using average linkage hierarchical clustering with correlation cutoff 0.5 (see supplemental section \ref{hcgroups}). To identify a set of key variables in each group such that the conditional independence described in definition \ref{gkci} might hold, we use Algorithm \ref{alg:rep} (unless otherwise specified, we use $c = 0.5$), which is motivated by the recent best subset selection algorithms \citep{sood2023statistical} .
Supplemental figures \ref{fig:LDsummary} report information on the size of the groups, and group key-variables identified. 

\subsection{Sampling knockoffs and inference}\label{sec:lasso_for_gwas_summarystat}

Given the  block structure of $\hat{\bSigma} = \operatorname{diag}(\hat{\bSigma}_{1},...,\hat{\bSigma}_{1703})$, we solve 1703 separate problems by applying Algorithm \eqref{alg:Groupknockoff} to each $\hat{\bSigma}_i$ in parallel to identify the $\bS_i$ that minimizes each of the three criteria  (SDP, ME, MVR).
Once these are obtained, we sampled $m=5$ knockoffs $\tilde{\bz}_i = (\tilde{\bz}_{i1},...,\tilde{\bz}_{im})\in \mathbb{R}^{mp_i}$ by
\begin{align}\label{eq:ghostknockoff}
	\tilde{\bz}_i &= \bP_i\bz_i + \mathcal{N}(\bzero, \bV_i), \\
	\bP_i &= \begin{pmatrix}
		\bI - \bS_i\hat{\bSigma}^{-1}_i\\
		\vdots\\
		\bI - \bS_i\hat{\bSigma}^{-1}_i
	\end{pmatrix}_{mp_i \times p_i}, \quad
	\bV_i = \begin{pmatrix}
		\bC_i & \bC_i - \bS_i & \cdots & \bC_i - \bS_i\\
		\bC_i - \bS_i & \bC_i & \cdots & \bC_i - \bS_i\\
		\vdots & & & \vdots \\
		\bC_i - \bS_i & \bC_i - \bS_i & \cdots & \bC_i
	\end{pmatrix}_{mp_i \times mp_i}\nonumber
\end{align}
where $\bC_i = 2\bS_i - \bS_i\bSigma^{-1}_i\bS_i$ as described in \cite{he2022ghostknockoff}. Here, the $\bS_i$s are block-diagonals in accord with group structure. The final z-score knockoffs are formed by $\tilde{\bz} = (\tilde{\bz}_1,...,\tilde{\bz}_{1703}) \in \mathbb{R}^{mp}$. 

Finally, we solve the pseudo-Lasso problem \citep{chen2024controlled} jointly over all blocks to define feature importance scores. This provides enhanced power compared to directly using the difference between $\bz$ and $\Tilde{\bz}$ (see Supplemental section \ref{sec:marginal_importance_score}). The unknown hyperparameter $\lambda$ is tuned via the pseudo-validation approach \citep{mak2017polygenic,zhang2021improved}, see \ref{sec:pseudovalidation} for details. In practice, we use the recently proposed BASIL algorithm \citep{qian2020fast} implemented in the \texttt{R} package \texttt{ghostbasil} \citep{yang2023ghostbasil} to carry out the Lasso regression step. Although we split knockoff construction over multiple blocks, the Lasso regression step includes $\approx 3.6$ million variables, including $0.6$ million Z scores and their knockoffs. 



\vspace*{.75cm}

\noindent {\huge \sc Results}

\section{Simulation studies}\label{sec:simulations}
We conduct two sets of simulations. In one, we use artificially generated covariance matrices $\bSigma$, so as to explore the performance of knockoff constructions under different and controlled forms of dependence. In the other, we use correlation matrices for SNP data derived from the Pan-UKB panel to investigate how well the  proposed methods lend themselves to the analysis of genetic data.


The knockoff score is defined as the  absolute value of the  group-wise Lasso coefficient difference: $Z_{\gamma} = \sum_{i \in \mathcal{A}_{\gamma}} |\beta_i|$ and $\tilde{Z}_{\gamma}^{(\ell)} = \sum_{i \in \mathcal{A}_{\gamma}} |\tilde{\beta}^{(\ell)}_{i}| $ for $\ell= 1,...,m$, where $\bbeta = (\bbeta, \tilde{\bbeta}^1,...,\tilde{\bbeta}^m)$ is the estimated effect sizes from performing Lasso regression on $(\bX, \tilde{\bX}^1,...,\tilde{\bX}^m)$. The feature importance score for group $\gamma$ is then defined as 
\begin{align*}
    W_\gamma = (Z_\gamma - \operatorname{median}(\tilde{Z}_\gamma^{(1)},...,\tilde{Z}_\gamma^{(m)}))I_{Z_\gamma \ge \operatorname{max}(\tilde{Z}^{(1)}_\gamma,...,\tilde{Z}^{(m)}_\gamma)}
\end{align*}
Here $I$ is the indicator function, and $W_\gamma$ is the feature-importance statistic first introduced in \cite{he2021identification}. Groups with $W_\gamma > \tau$ are selected, where $\tau$ is calculated from the multiple knockoff filter \citep{gimenez2019improving}.

\subsection{Power/FDR of different knockoffs constructions under special covariances}
\color{blue}

%

\color{black}

We consider five  types of covariance $\bSigma_{p \times p}$ with $p=1000$, always scaling $\bSigma$ back to a matrix with ones on the diagonal. Here we generate $m=5$ knockoffs for each experiment.  

\begin{description}
	\item \textbf{Block cov}. This simulation roughly follows \cite{dai2016knockoff}. In this basic setting, we define 200 contiguous blocks each of size 5, and the covariance $\bSigma$ is set as
	\begin{align*}
		\Sigma_{ij} = 
		\begin{cases}
			1 & i = j\\
			\rho & (i, j) \text{ in the same block}\\
			\gamma\rho & (i,j) \text{ in different blocks}
		\end{cases}
	\end{align*}
	where $\rho = 0.75$ and $\gamma = 0.25$. Note that the blocks corresponds to ``true" group structure, but  we do not leverage this, defining groups membership empirically as described above.
	\item \textbf{ER} (Erdos-Renyi). This simulation roughly follows the clustered ER simulation of \cite{li2021ggm}. Here we define 100 contiguous blocks $\bOmega^{1},...,\bOmega^{100}$ each of size $10 \times 10$. For each block $k$, $\bOmega^k$ is simulated from an Erdos-Renyi graph by letting $\Omega^k_{ii} = 1$ and $\Omega^k_{ij} = \Omega^k_{ji} = \omega_{ij}\phi_{ij}$ where $\omega_{ij}\sim\pm\text{Uniform}(0.3, 0.9)$ and $\phi_{ij}\sim\text{Bernoulli}(0.1)$. Intuitively, if features $i$ and $j$ are in the same block, then with with probability 0.1 they will have correlation $\omega_{ij}$. 
	We let $\bV = \operatorname{diag}(\bOmega^1,...,\bOmega^{100})$ and define
	\begin{align*}
		\bSigma = \bV + (|\lambda_{\min}(\bV)| + 0.1)\mathbf{I}.
	\end{align*}
	\item \textbf{ER(prec)}. Similar to the ER setting, we define
	\begin{align*}
		\bSigma = \left(\bV + (|\lambda_{\min}(\bV)| + 0.1)\mathbf{I}\right)^{-1}.
	\end{align*}
	\item \textbf{AR1}. This simulation follows \cite{spector2022powerful}. In the AR(1) setting, we simulate 
\begin{align*}
    \Sigma_{ij} = 
    \begin{cases}
        1 & i = j\\
        \exp\left\{-|\sum_{k=2}^i \log(\rho_k) - \sum_{k=2}^j \log(\rho_k)|\right\} & i \ne j.\\
    \end{cases}
\end{align*}
 We sample $\rho_j \sim \text{Beta}(3, 1)$ to generate a setting where neighboring features are highly correlated. To ensure positive definiteness, if $\lambda_{min}(\bSigma) < 0.001$, we add $(0.001-\lambda_{min}(\bSigma))\bI_{p}$ to $\bSigma$, where $\lambda_{min}(\bSigma)$ computes the minimum eigenvalue of $\bSigma$, and rescale back to a correlation matrix.
    \item \textbf{AR1(corr)}. Here, $\bSigma$ is the same as in AR(1) above, but the true coefficient vectors are simulated such that all $k$ non-zero $\beta_j$s are placed contiguously. This simulation aims to capture the genetic reality that many disease variants are clustered tigthly together in the same LD block (e.g. residing in the same gene) but each exerts an independent effect on disease outcome. 
\end{description}

For each of the 5 special covariance matrices, we simulate 100 copies of $\bSigma$, and the corresponding data matrix $\bX$ is formed by drawing $n$ independent samples from $\mathcal{N}(\mathbf{0}, \bSigma)$ where $n$ is varied as a parameter. The response is simulated as $\by = \bX\bbeta + \mathcal{N}(\mathbf{0}, \mathbf{I}_{p \times p})$. The true regression vector $\bbeta \in \mathbb{R}^p$ has $k=50$ non-zero coefficients randomly chosen across the $p$ features with effect size $\beta_j \sim \mathcal{N}(0, 1)$. Groups are defined on the basis of the observed data $\bX$ using average linkage hierarchical clustering with correlation cutoff $0.5$ (see supplementary section \ref{hcgroups}). To compute group power and group FDR, a group is considered a true null when it contains no causal features, and a group is correctly rejected if it contains at least 1 causal feature. Power is reported as the fraction of groups correctly discovered among all causal groups, while FDR is the fraction of falsely discovered groups among all discoveries.

\begin{figure}
    \centering
    \includegraphics[width=\linewidth]{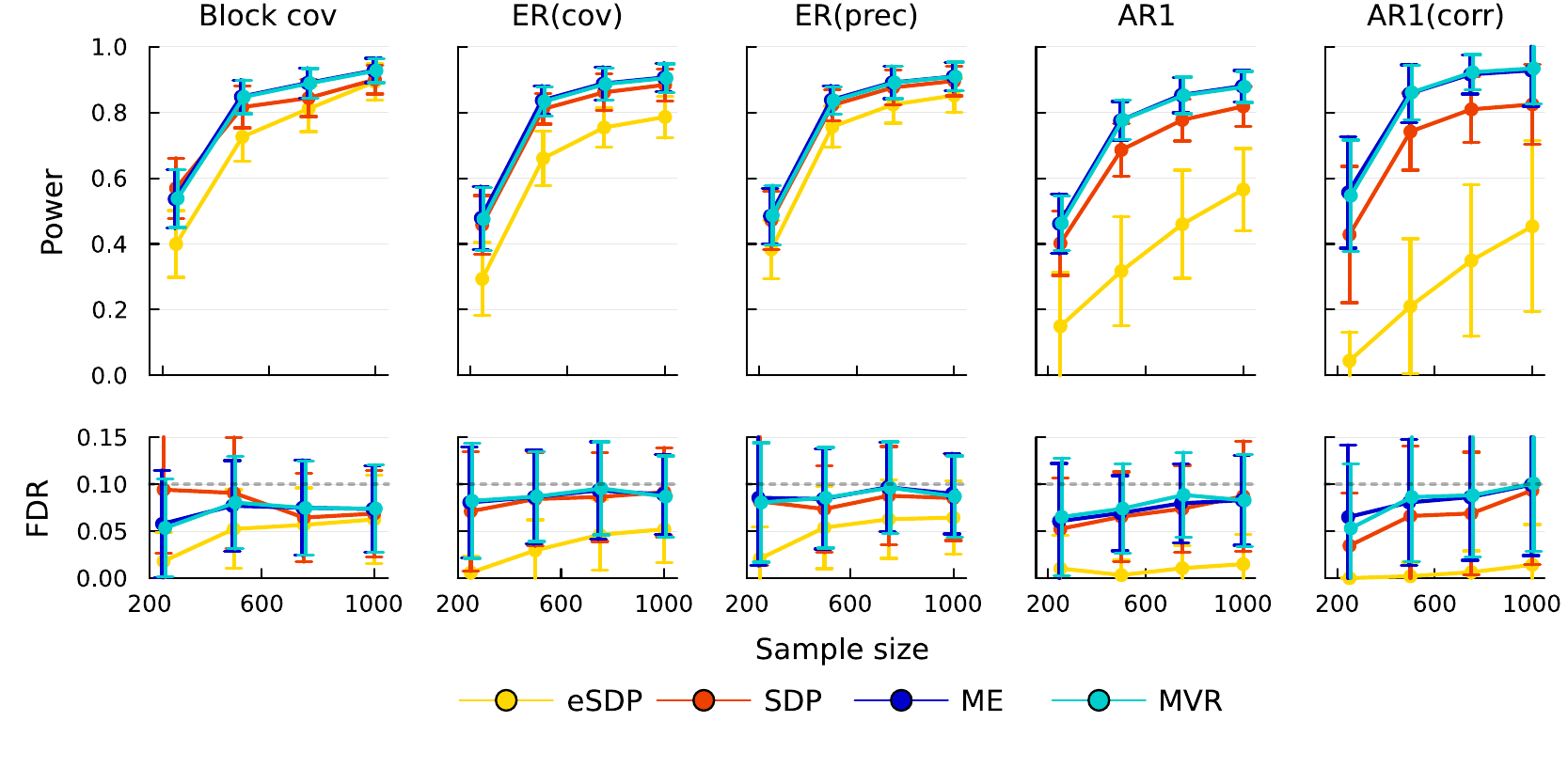}
    \caption{Power/FDR comparison of eSDP, SDP, MVR, and ME group knockoffs across different variance-covariance structure. Each point is average of 100 replicates. Timing results are available in Table \ref{tab:fig1_timings}.}
    \label{fig:sim1}
\end{figure}

Figure \ref{fig:sim1} compares the  performance of different knockoff constructions with reference to  power and FDR, averaging over 100 simulations. All methods control the FDR at target (10\%) level. Across different covariance matrices, ME and MVR solvers generally have the best power, followed by SDP, followed by eSDP knockoffs. This behavior is consistent with regular (non-grouped) knockoffs \citep{spector2022powerful}. For an indication of the computation time associated with the different methods, see Table \ref{tab:fig1_timings}. A more comprehensive analysis of timing is presented later. 

\begin{table}[]
    \centering
    \begin{tabular}{c|c|c|c|c}
        \hline
        Model & eSDP & SDP & ME & MVR \\
        \hline
Block Cov & $4.1 \pm 0.5$ & $244.2 \pm 141.2$ & $18.8 \pm 4.1$ & $33.6 \pm 5.9$\\
ER(cov) & $2.8 \pm 0.4$ & $3858.2 \pm 1035.7$ & $157.8 \pm 47.1$ & $124.3 \pm 38.9$\\
ER(prec) & $3.8 \pm 0.5$ & $6738.1 \pm 1755.9$ & $69.0 \pm 17.9$ & $92.0 \pm 26.4$\\
AR1 & $4.0 \pm 1.2$ & $668.8 \pm 418.3$ & $56.9 \pm 32.9$ & $92.6 \pm 51.4$\\
AR1(corr) & $4.3 \pm 1.3$ & $973.3 \pm 580.0$ & $172.4 \pm 106.0$ & $225.4 \pm 132.9$\\
        \hline
    \end{tabular}
    \caption{Group knockoff optimization times (in seconds) for Figure \ref{fig:sim1}, averaged over 100 simulations. There are $p=1000$ variables, and groups are constructed empirically based on the data matrix $\bX$. More comprehensive timing comparisons are presented in Figure \ref{fig:sim3}.}
    \label{tab:fig1_timings}
\end{table}

\subsection{Power/FDR of group-key conditional independence  and genetic data}\label{sec:ukb_data_sim}
%
%


Here, we use real UKB genotypes \citep{sudlow2015uk} in conjunction with covariance matrices extracted from Pan-UKB \citep{panukbb} to conduct simulations. The goal here is (a) to ensure that our approximately independent blocks of SNPs identified in section \ref{sec:obtaing_sigma_ukb} are really independent, (b) verify that the distribution of genotypes can be approximated by a Gaussian, and (c) explore to what extent the group-key conditional independent hypothesis is appropriate for these matrices. 

First, we randomly select $n=2000$ British samples from the UK Biobank and restricted our attention to $p=8844$ SNPs residing on chromosome 22. All SNPs are centered to mean 0 variance 1, and partitioned into 24 roughly independent blocks as described in section \ref{sec:obtaing_sigma_ukb}. Given this subset of individual-level genotypes, we simulate the response $\by = \bX\bbeta + \mathcal{N}(\bzero, \bI_{p \times p})$ where $\bbeta$ contain $k=100$ non-zero effects chosen randomly across the chromosome and effect sizes drawn from $\mathcal{N}(0, 0.15)$. 
After forming the phenotypes, we generate $m=1$ second order knockoffs where the covariance matrix $\bSigma$ is estimated as $\hat{\bSigma} = \operatorname{diag}(\hat{\bSigma}_{chr22, 1},...,\hat{\bSigma}_{chr22, 24})$ and each $\hat{\bSigma}_{chr22, i}$ is extracted from the corresponding region in the Pan-UKB. Importantly, the original genotypes $\bX$ have entries $\{0, 1, 2\}$ (prior to centering/scaling) but their knockoffs are generated from a Gaussian distribution with the covariance estimated without using the original $\bX$. We define groups with average linkage hierarchical clustering on each $\bSigma_{chr22, i}$ with correlation cutoff $0.5$, then identify key variables with Algorithm \ref{alg:rep} for $c \in \{0.25, 0.5, 0.75, 1.0\}$. Note that $c=1$ is equivalent to not using the conditional independence assumption, and when $c < 1.0$, it is possible for a causal variant to not be selected as a key variable. We ran 100 independent simulations with this setup and averaged the power/FDR. 

\begin{figure}
    \centering
    \includegraphics[width=\linewidth]{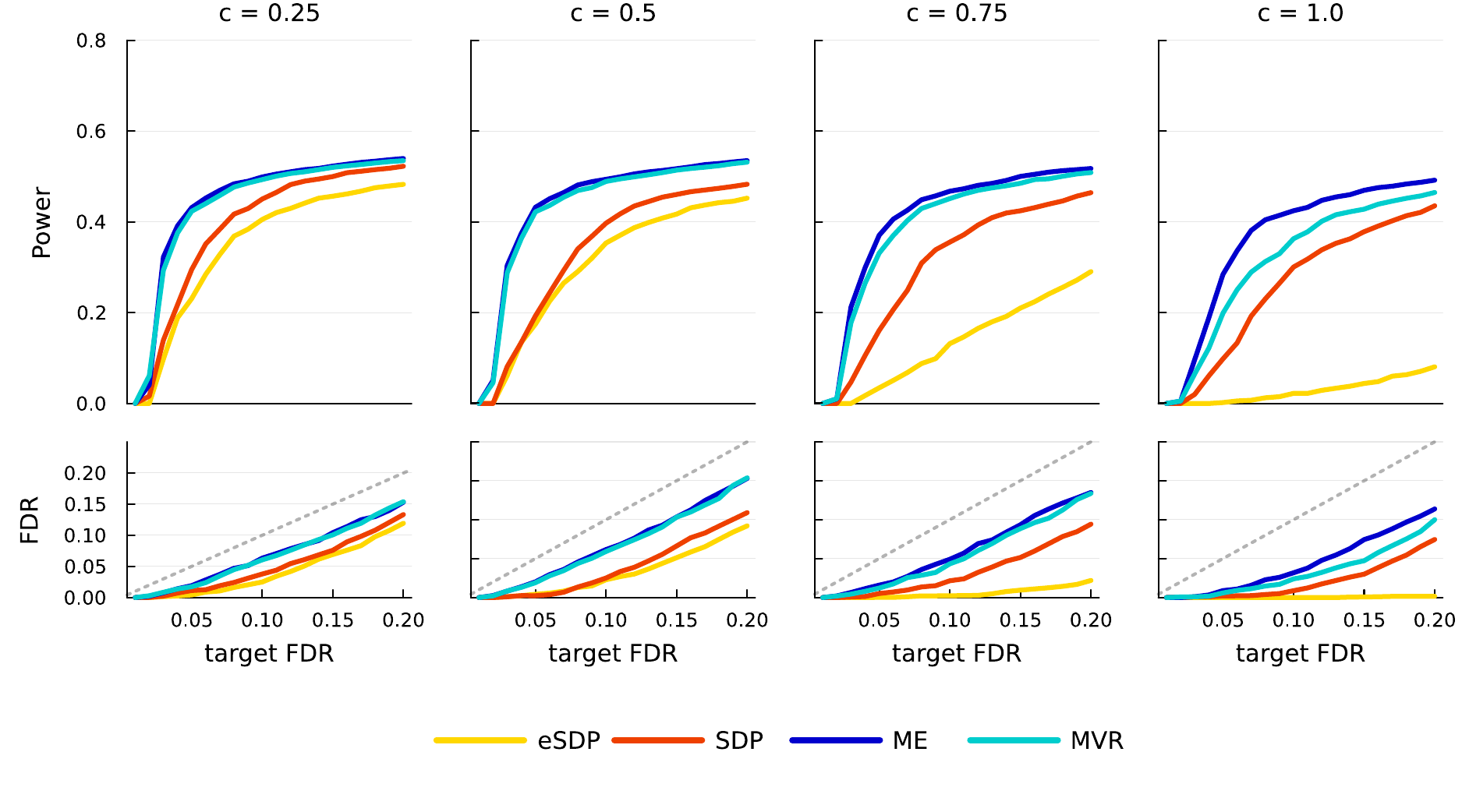}
    \caption{Power/FDR comparison of group knockoffs utilizing conditional independence assumption. Each curve is an average over 100 simulations. There are $n=2000$ samples and $p=8844$ genotypes residing on chromosome 22 of UKB. Dashed gray lines indicate the target FDR level. Phenotypes are simulated using real UKB genotypes, while correlation among SNPs is modeled using only LD matrices extracted from the Pan-UKB panel. Note that the threshold $c=1$ corresponds to regular group knockoffs.}
    \label{fig:sim2_new}
\end{figure}

Figure \ref{fig:sim2_new} summarizes power and FDR. In general, ME has the best power, followed by MVR, SDP, and finally eSDP. Utilizing conditional independence offers slightly better power than regular group knockoffs, without sacrificing empirical FDR. In this simulation, group-FDR is controlled for all threshold values $c$, although a separate simulation in Supplemental Figure \ref{fig:sim2} shows that $c=0.25$ could potentially lead to an inflated empirical FDR. The observed power boost is especially beneficial for eSDP constructions, likely because decreasing the number of variables within groups substantially relieves the eSDP constraint to a greater degree than for other methods. Overall, these results suggest the conditional independence assumption can approximate the genetic reality induced by linkage disequilibrium. Given its superior speed and promising empirical performance, we choose a threshold of $c = 0.5$ in our real data analysis of Albuminuria as shown later. 

\begin{figure}
    \centering
    \includegraphics[width=0.75\linewidth]{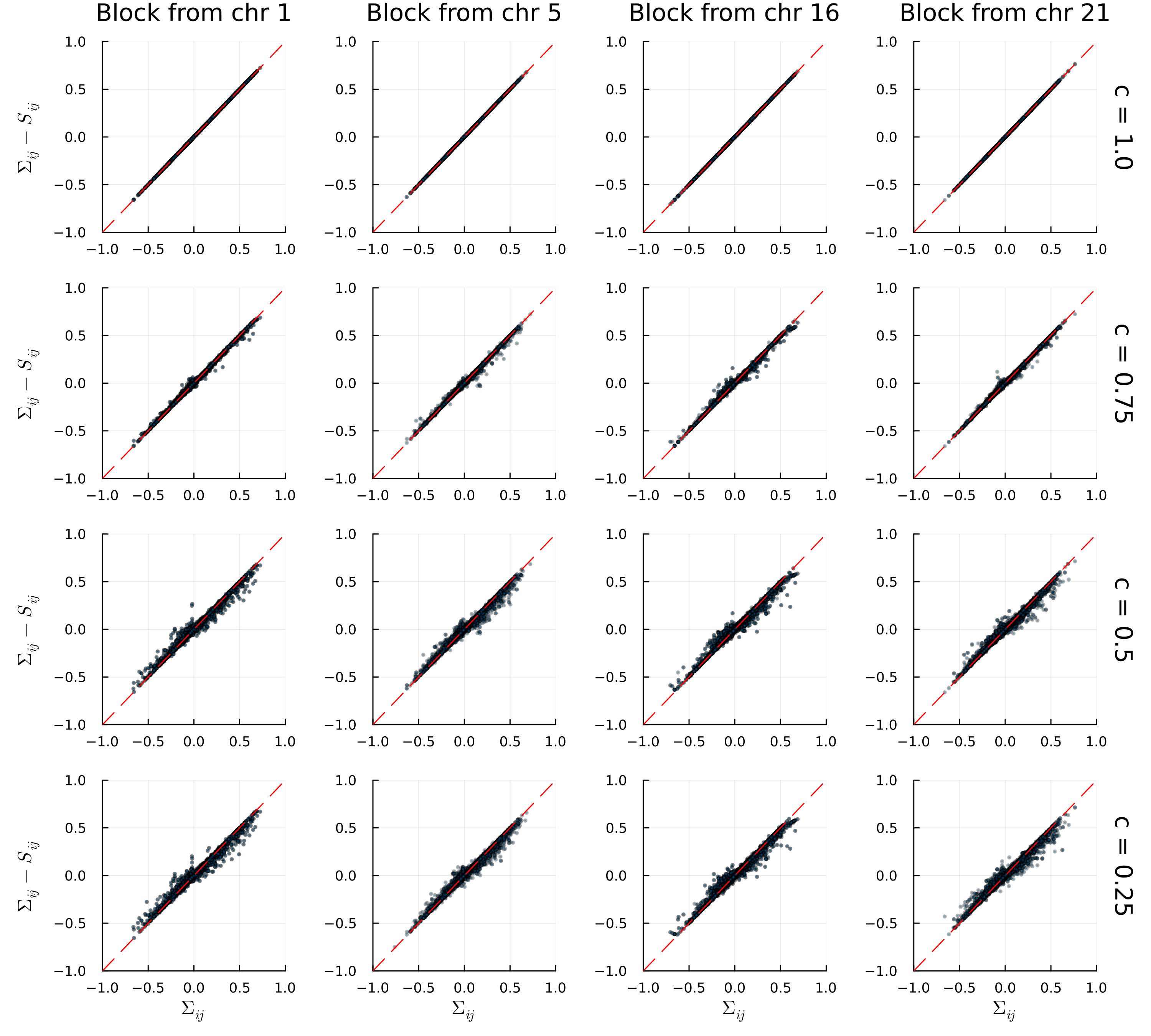}
    \caption{Visualization of group knockoff exchangeability by applying Algorithms \eqref{alg:Groupknockoff}-\eqref{alg:rep} on the Pan-UKB data. Each column corresponds to a randomly selected LD block, each row chooses a different threshold for selecting representatives, and each $i,j$ pair are indices from different groups. Here $\Sigma_{ij}$ represents $corr(X_i, X_j)$, while $\Sigma_{ij} - S_{ij}$ represents $corr(X_i, \tilde{X}_j)$. A threshold of 1 includes all variables within groups. The 4 selected regions contain 247, 511, 333, and 431 SNPs partitioned into 122, 279, 177, and 234 groups, respectively.}
    \label{fig:exchangeability}
\end{figure}

Figure \ref{fig:exchangeability} gives an illustration of exchangeability measures between knockoffs and original data for 4 randomly selected genomic regions. 
On the x-axis, we plot $\Sigma_{ij}$ which represents $corr(X_i, X_j)$, and on the y-axis we plot $\Sigma_{ij} - S_{ij}$ which represents $corr(X_i, \tilde{X}_j)$. When $(i,j)$ belong to different groups, $\tilde{X}_j$ should be exchangeable with $X_j$ in the sense that $corr(X_i, X_j) = corr(X_i, \tilde{X}_j)$, which will result in a point lying perfectly on the diagonal. Thus, decreasing the threshold value $c$ has the effect of producing less exchangeable knockoffs as a consequence of selecting fewer key variables per group. Since our block-diagonal approximation seems to produce slightly conservative FDR for regular group knockoffs, small deviations do not violate empirical FDR. 


\subsection{Compute times}

\begin{figure}
    \centering
    \includegraphics[width=\linewidth]{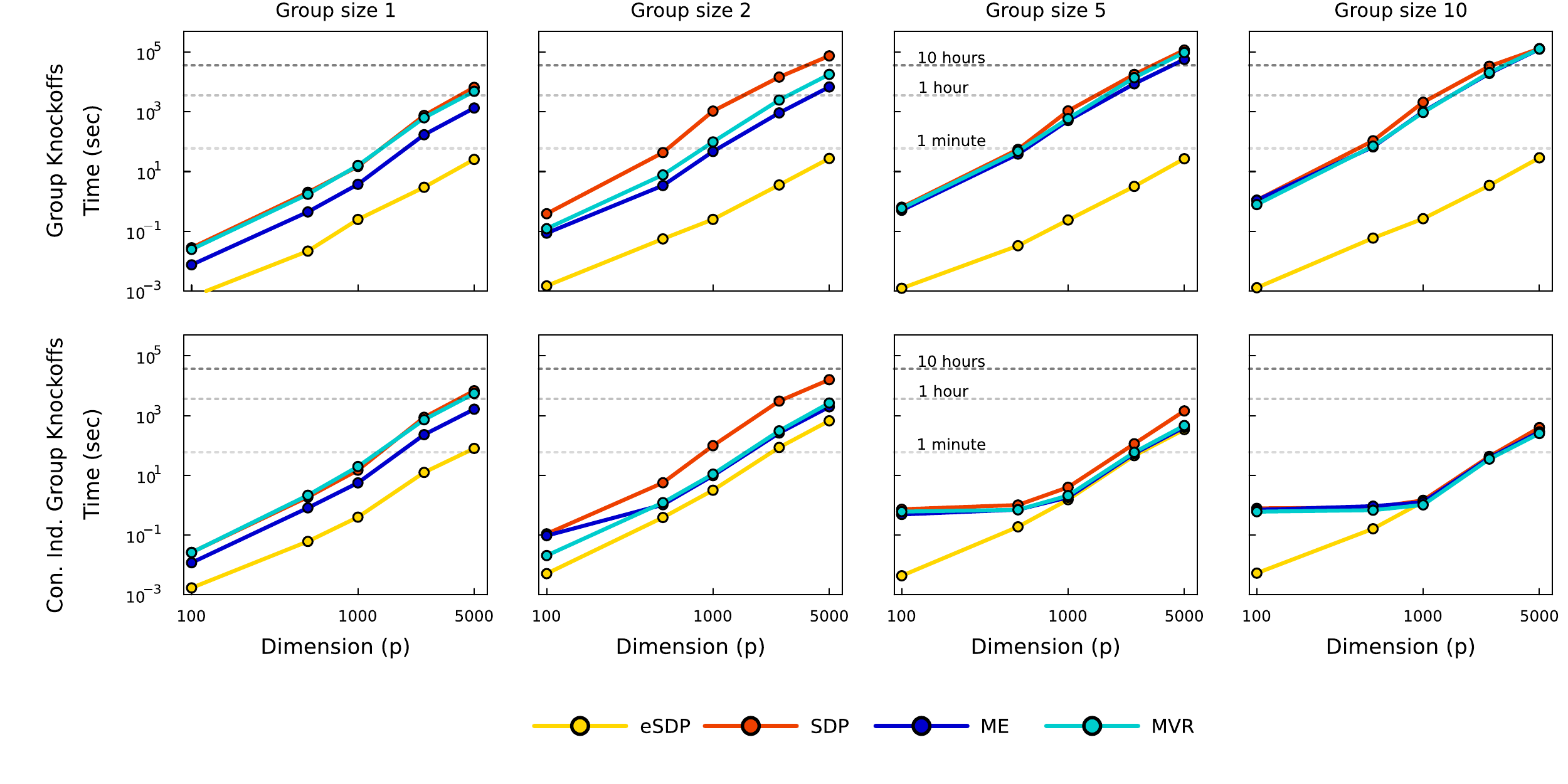}
    \caption{Computation time comparison of group knockoffs solvers (top) and group knockoffs solvers utilizing conditional independence assumption with threshold $c = 0.5$ (bottom). The gray dashed lines indicate 1-minute, 1-hour, and 10-hour marks. Each data point plotted is an average over 25 replicates. $\bSigma$ is drawn from the AR1 model. A group size of 1 corresponds to regular (non-grouped) knockoffs. The convergence tolerance was set to $0.0001$. }
    \label{fig:sim3}
\end{figure}

Timing results for Figure \ref{fig:sim1} are listed in Table \ref{tab:fig1_timings}. Figure \ref{fig:sim3} investigates the computational efficiency of our proposed algorithms in more detail. Two important parameters, namely, number of features $p$ and group sizes, are varied according to what seems appropriate for the GWAS analysis example. Data are generated according to the AR1 setting in Figure \ref{fig:sim1}, where the size of each group is fixed between 1 (no group structure) and 10 (each group contains 10 contiguous variables). When conditional independence assumption is utilized, we let $c = 0.5$. Equi-correlated constructions (eSDP) offer the best speed due to its convenient closed form formula \citep{dai2016knockoff}. Otherwise, ME solvers tend to run faster than MVR or SDP solvers. We find that SDP solver requires more iterations to converge, making it the slowest method in general. When comparing MVR and ME solvers, both require the same number of Cholesky updates, but each ME iteration requires only solving 1 forward-backward equation, while MVR requires 3. We wrote an efficient vectorized routine for performing Cholesky updates, and we use LAPACK \citep{lapack99} to perform required forward-backward solves. Careful benchmarks reveal the latter step constitute approximately $90$\% of compute time, which explains the timing difference between MVR and ME solvers.

\section{Albuminuria in the UK Biobank}\label{sec:gwas}

\color{black}

To demonstrate the potential of group knockoffs in analyzing real-world datasets, we our methodology to re-analyze an Albuminuria GWAS dataset \citep{haas2018genetic} with $n=382,500$ unrelated Europeans and $p=11,209,307$ SNPs. Elevated level of urine albumin concentration is a hallmark of diabetic kidney disease and is associated with multiple cardiovascular and metabolic diseases. Let us emphasize that the only required inputs are Z-scores and the Pan-UKB LD matrices, so knockoff-based conditional independence testing is easily accessible to everyone in the genetics community. 

\begin{figure}
	\centering
	\begin{subfigure}[b]{\textwidth}
		\includegraphics[width=0.9\linewidth]{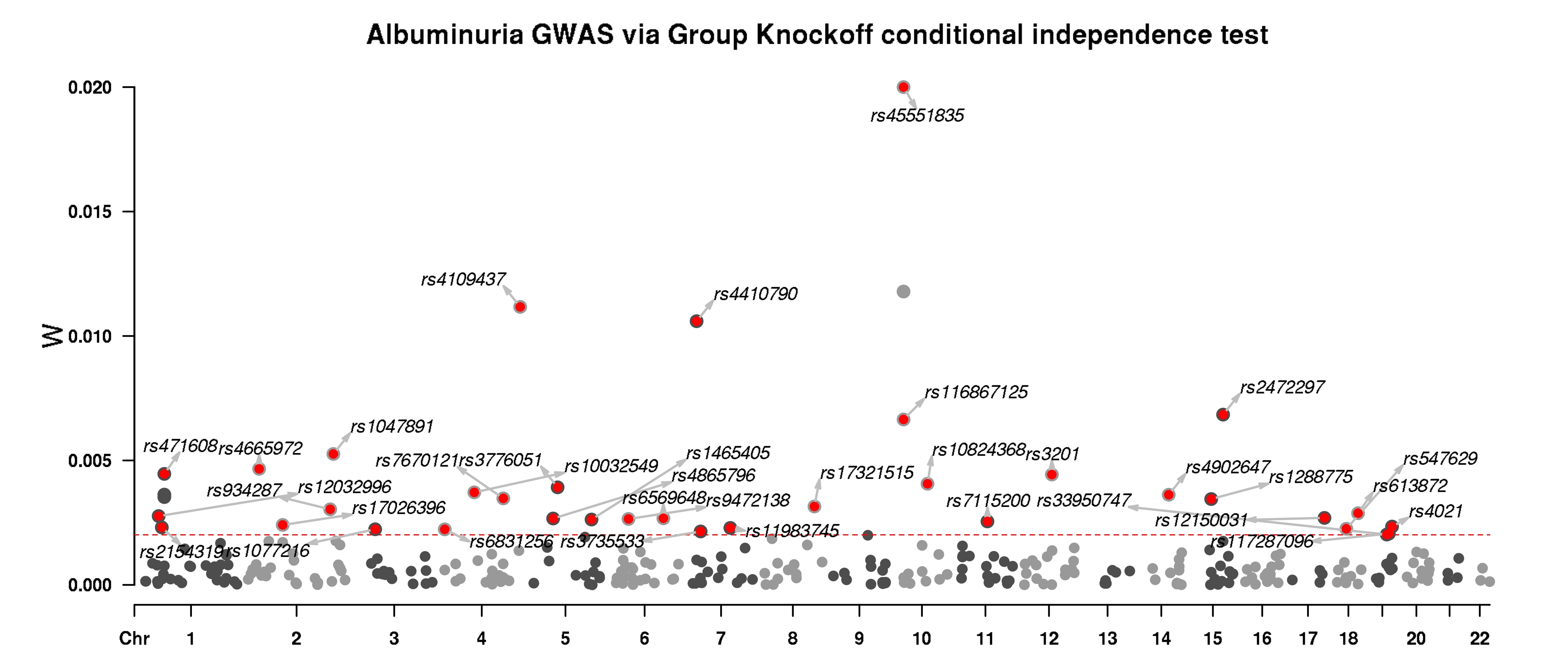}
	\end{subfigure}
	\hfill
	\begin{subfigure}[b]{\textwidth}
        \includegraphics[width=0.9\linewidth]{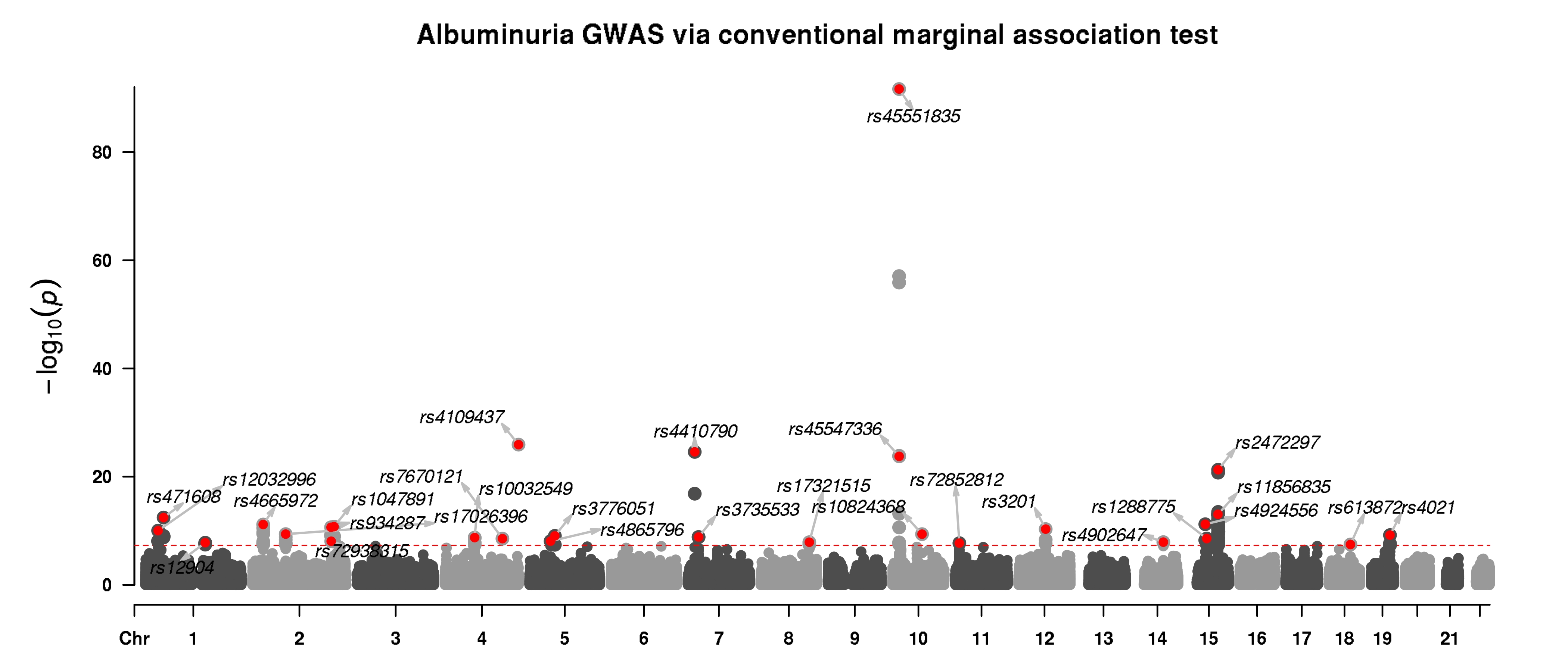}
	\end{subfigure}
	\caption{Summary statistics analysis of Albuminuria ($n = 382,500$ and $p=630,017$) using group-knockoff-ghost-Lasso (top) compared to a conventional marginal-association testing approach (bottom). In the top panel, each dot represents a group, where the SNP with the most significant marginal p-value within the group is plotted. Only an independent discovery is highlighted in red. A table formatted result is available as Supplemental Table \ref{table:ABtable1}.}
	\label{fig:gwas}
\end{figure}

 After matching the study Z-scores to the Pan-UKBB panel, and filtering for genotyped variants with minor allele frequency $\ge 0.01$, we retained $p=630,017$ Z-scores. We defined groups empirically using average linkage hierarchical clustering with correlation cutoff 0.5. For better speed and power, we use the maximum entropy (ME) solver and exploit the conditional independence assumption by selecting variants within groups such that $c = 0.5$, that is, the key variants explain 50\% of the variation within groups. This led to a maximum group size of $4$ (see supplemental for details and summary statistics). The result is visualized in Figure \ref{fig:gwas} via the \texttt{CMplot} package \citep{yin2021rmvp}.

Knockoff-based analysis discovers 7 additional independent signals (35 total while controlling FDR at $q=0.1$) compared to conventional approach which finds 28 independent signals passing genotype-wide significance threshold of $5 \times 10^{-8}$ (the full result can be accessed in Supplemental section \ref{sec:gwas_result_full}). The entire analysis completed in under $3$ hours, from knockoff construction to a genome-wide Lasso fit. Running genome-wide pseudo-Lasso with 3.6 million variables was the most memory intensive step, requiring roughly 40GB. An \textit{independent discovery} is defined as the most significant SNP within 1Mb region, and is highlighted in red. 
A number of our discoveries have marginal p-values close to the genome-wide threshold, such as rs1077216 in chromosome 3 ($p=6.8\times 10^{-8}$), rs6569648 and rs9472138 in chromosome 6 ($p=7.5\times 10^{-8}$ and $2.5\times 10^{-7}$), rs12150031 in chromosome 17 ($p=7.1\times 10^{-8}$), and rs117287096 in chromosome 19 ($p=4.6\times 10^{-7}$). Many of these discoveries have been mapped to functional genes (see Supplemental Table \ref{table:ABtable1} for a full list) that are directly or closely related to the disease phenotype. For instance, rs1077216 was previously reported by an independent study \citep{teumer2016genome} of the same trait also with p-value not passing genome-wide threshold, rs9472138 is significant for diastolic blood pressure \citep{surendran2020discovery}, and rs117287096 is highly significant with urinary sodium excretion \citep{pazoki2019gwas}. Other SNPs such as rs12150031 are not previously known, and therefore would be interesting candidates for follow-up studies. 


\section{Discussion}

We developed several algorithms for constructing second-order group knockoffs. Group knockoffs promote model selection at the group level, which leads to an improvement in power compared to standard knockoffs when there are highly correlated variables. Of the three group knockoff objectives considered, MVR/ME/SDP, we find that ME tends to exhibit the best power and computational efficiency in simulations with real or artificial data. These algorithms and pipelines are provided as individual open-sourced packages freely available to the scientific community. 

Compared to regular model-X knockoffs, group knockoffs require solving a more computationally intensive optimization problem because it requires optimizing significantly more variables. As such, we developed a number of algorithms that are computationally efficient and can flexibly fall back onto less general search spaces when it is not feasible to optimize every variable separately. Furthermore, we also propose a procedure that pre-selects key variables from each group such that the remaining variables are conditionally independent by groups. Group knockoff generation can be done separately for the key and non-key variables, which dramatically reduces the number of parameters for optimization. The validity of this procedure is confirmed theoretically, and we show empirically that it leads to no sacrifice in power or FDR. 

In our real-data example, we applied the group knockoff methodology to perform a summary statistics analysis on the albuminuria dataset, where the only required inputs are Z-scores and external LD matrices.
Running the pipeline on $p=630,017$ variables completed in under $3$ hours, from knockoff construction to a genome-wide Lasso fit. Thus, one can easily conduct knockoff-based conditional independence testing by (1) running standard marginal association testing, and (2) inputting the resulting Z-scores into our knockoff pipeline. 
Recent work \citep{qi2023robust} have shown that the first step can rely on commonly used linear mixed models, while the later step can be massively simplified if pre-computed knockoff statistics (e.g $\bSigma_i$s, $\bS_i$s,...,etc) are made widely accessible by storing them on the cloud. In a companion paper \citep{he2024in}, we try to provide such an ideal resource and conduct more in-depth analyses by applying it to over $60$ phenotypes. 

Finally, let us discuss a few limitations of the proposed methods and questions worthy of future research. Firstly, the group ME/MVR/SDP algorithms we propose only operate on Gaussian covariates, or in the context of second order knockoffs, covariates that can be approximated by a normal variable. 
Next, although selecting group-key variables partially overcomes many computational barriers, further speedup is possible if the covariance matrix can be factored into a low rank model $\bSigma = \bD + \bU\bU^t$ where $\bD$ is a diagonal and $\bU$ is low rank \citep{askari2021fanok}. 
One may also consider approaches to pinpoint the causal variant within a discovered group, for example by leveraging feature-level group knockoff filters \citep{gu2024powerful} or recent advances in signal localization \citep{gablenz2023catch}.
Finally, in our summary statistics analysis, it is unclear how knockoffs will respond when samples in a study have ancestral backgrounds that deviate too much from the ethnic backgrounds of the subject used to estimate the LD matrices. To overcome this uncertainty, we recommend analyzing homogeneous populations whose ethnic background matches the subject backgrounds from the LD matrices. As such, we will continue to explore improvements to group knockoffs. Given its promising empirical performance, we recommend it for general use with the understanding that analysts respect its limitations and complement its usage with standard feature selection tools.

\section{Acknowledgement}
B.C. and C.S. were supported by the grants R01MH113078, R56HG010812, NSF 2210392, and R01MH123157. J.G. and Z.H. were supported by NIH/NIA awards AG066206 and AG066515. Z.C. was supported by the Simons Foundation under award 814641.  T.M. was supported by a B.C. and E.J. Eaves Stanford Graduate Fellowship. E.C. was supported by the Office of Naval Research grant N00014-20-1-2157. 

The Pan-UKB data was accessed at \url{https://pan.ukbb.broadinstitute.org/docs/hail-format}. The UK Biobank data used in section \ref{sec:ukb_data_sim} was accessed under Material Transfer Agreement for UK Biobank Application 27837. 

\newpage
\printbibliography

\newpage
\setcounter{page}{0}
\setcounter{section}{0}
\setcounter{figure}{0}
\setcounter{equation}{0}
\setcounter{table}{0}

\noindent
{\Huge \sc Supplementary information}

\makeatletter
\renewcommand \thesection{S\@arabic\c@section}
\renewcommand\thetable{S\@arabic\c@table}
\renewcommand \thefigure{S\@arabic\c@figure}
\renewcommand{\theequation}{S.\arabic{equation}}

\makeatother


\section{The advantage of group-based inference}\label{Wgroups}

\begin{figure}[hb]
    \centering
    \includegraphics[width=0.8\linewidth]{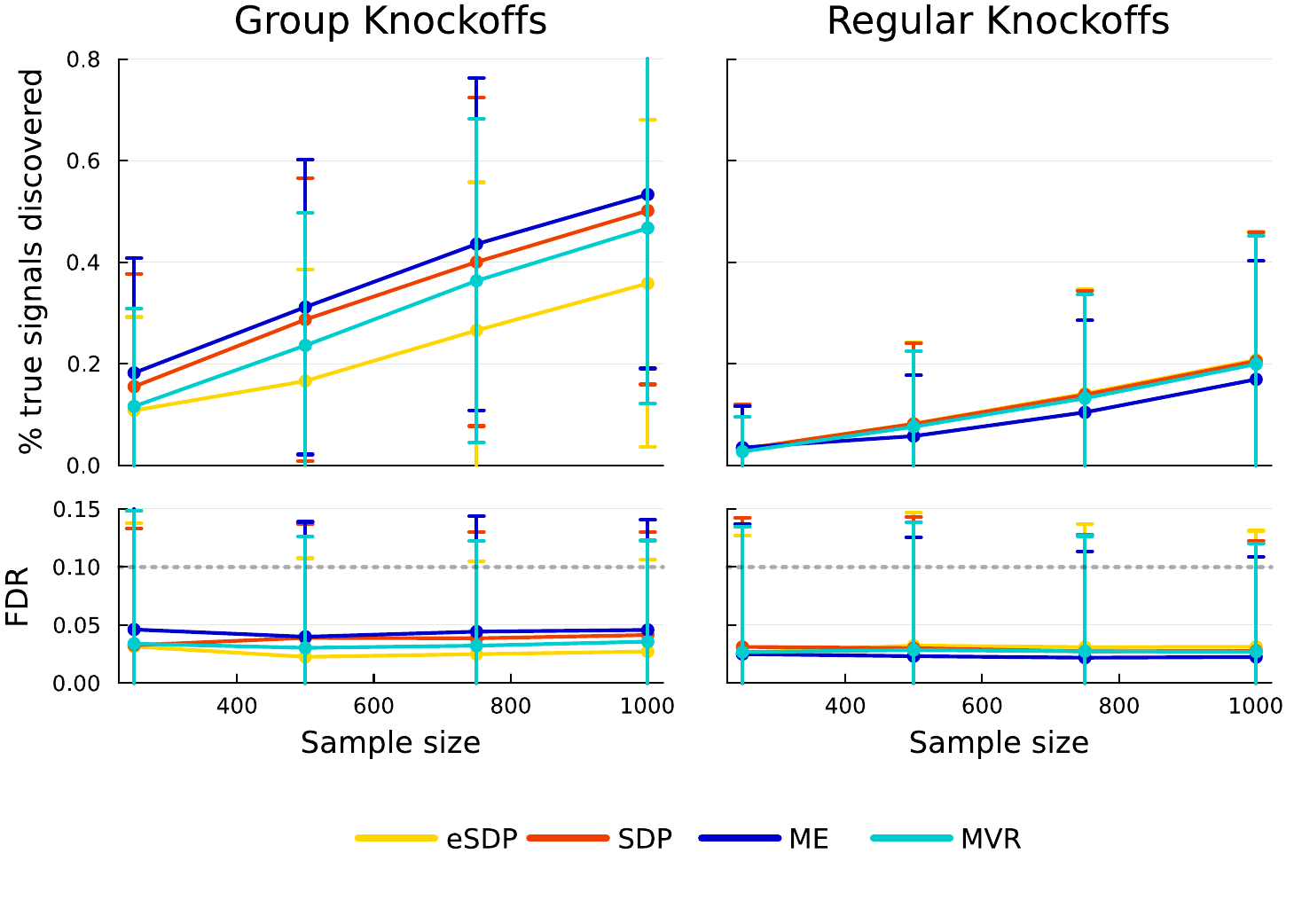}
    \caption{Group vs regular model-X knockoffs. When features are highly correlated, group knockoffs discover more causal features.}
    \label{fig:vs_ungroup}
\end{figure}

Although group knockoffs and regular model-X knockoffs test different hypotheses, it is still of practical interest to determine the porportion of true signals being discovered. In this section, we perform a basic simulation featuring a symmetric Toeplitz matrix
\begin{align}\label{eq:toeplitz_cov}
	\bSigma = 
	\begin{bmatrix}
		1 & \rho & \rho^2 & \cdots & \rho^{p-1} \\
		\rho & 1 & \rho & \cdots & \\
		\rho^2 & & & & \\
		\vdots & & & \ddots & \vdots\\
		\rho^{p-1} & & & \rho & 1
	\end{bmatrix}_{p \times p}
\end{align}
where correlation between features $\rho = 0.9$ and $p = 200$. The response $\by$ is simulated as $\by = \bX\bbeta + \mathcal{N}(\bzero, \bI)$ with $k=10$ causal effects randomly chosen across the $p$ features with effect size $\beta_j = \pm 0.25$. Then we generate model-X group and ungrouped knockoffs and compute the proportion of signals discovered and the grouped/ungrouped FDR based on the Lasso coefficient difference statistic. The result is visualized in Figure \ref{fig:vs_ungroup}. Due to the high-correlation between neighboring features, regular model-X knockoffs discovers much less causal features than group knockoffs. 

\section{Fully General Coordinate Descent}
\label{sec:supplementccd}

In general coordinate descent, we update $\bS$ with
\begin{eqnarray*}
    S_{ij}^{new} = S_{ij} + \delta.
\end{eqnarray*}
Here, we fill in algorithmic details for the general coordinate descent algorithm featured in Section \ref{sec:algs}. Details for PCA updates are provided in section \ref{sec:pca_ccd}. A summary is provided as Algorithm \eqref{alg:CCD}, which is a more detailed version of Algorithm \eqref{alg:CCD_concise} in the main text. 

\subsection{Fully general coordinate descent for ME (off-diagonal entries)} 

The objective for maximum entropy knockoffs can be simplified to maximizing the following \citep{gimenez2019improving}
\begin{align*}
    \max_{\bS} \log\det\left(\frac{m+1}{m}\bSigma - \bS\right) + m \log\det(\bS)     \quad \text{ subject to } 
    \begin{cases}
        \frac{m + 1}{m}\bSigma - \bS \succeq 0\\
        \bS \succeq 0
    \end{cases}.
\end{align*}
Consider updating $S_{ij}^{new} = S_{ij} + \delta$ where $i \neq j$. Our goal is to compute $\delta$. First note that by symmetry, we must also set $S_{ji}^{new} = S_{ij}^{new}$. Let $\be_i$ be the $i$th basis vector and $\bH = \be_i\be_j^t + \be_j\be_i^t$ be a $p \times p$ matrix that is zero everywhere except $H_{ij} = H_{ji} = 1$. Thus, we have $\bS^{new} = \bS + \delta \bH$. If we define $\bD = \frac{m+1}{m}\bSigma - \bS$, the objective becomes
\begin{eqnarray*}
    \log\det(\bD - \delta \bH) + m \log\det\left(\bS+ \delta \bH\right).
\end{eqnarray*}
Letting $\bH = \bU\bV$, where $\bU = \begin{bmatrix} \be_i & \be_j \end{bmatrix} \in \mathbb{R}^{p \times 2}$ and $\bV = \begin{bmatrix} \be_j^t \\ \be_i^t \end{bmatrix} \in \mathbb{R}^{2 \times p}$, the objective can be simplified by the matrix determinant lemma
\begin{eqnarray*}
    \log\det(\bD - \delta \bH)
    &=& \log\left(\det(\bI_2 - \delta \bV\bD^{-1}\bU)\det(\bD)\right) \amp = \amp \log\det(\bI_2 - \delta \bV\bD^{-1}\bU) + c,\\
    \log\det(\bS + \delta \bH)
    &=& \log\left(\det(\bI_2 + \delta \bV\bS^{-1}\bU)\det(\bS)\right) \amp = \amp \log\det(\bI_2 + \delta \bV\bS^{-1}\bU) + c.
\end{eqnarray*}
Due to the special structures of $\bU$ and $\bV$, we can explicitly compute
\begin{align*}
    \bV\bD^{-1}\bU 
    &= 
    \begin{bmatrix}
        a_{ji} & a_{jj}\\
        a_{ii} & a_{ij}
    \end{bmatrix}, \qquad
    \bV\bS^{-1}\bU = 
    \begin{bmatrix}
        b_{ji} & b_{jj}\\
        b_{ii} & b_{ij}
    \end{bmatrix},
\end{align*}
where
\begin{align*}
    a_{ij} &= \be_i^t\bD^{-1}\be_j, \quad 
    a_{ii} = \be_i^t\bD^{-1}\be_i, \quad
    a_{jj} = \be_j^t\bD^{-1}\be_j,\\
    b_{ij} &= \be_i^t\bS^{-1}\be_j, \quad 
    b_{ii} = \be_i^t\bS^{-1}\be_i, \quad
    b_{jj} = \be_j^t\bS^{-1}\be_j.
\end{align*}
Thus, evaluating the $2 \times 2$ determinants, the objective becomes
\begin{eqnarray}\label{eq:ccd_offdiag_ME}
    g(\delta) &=& \log\left((1 - \delta a_{ij})^2 - \delta^2 a_{ii}a_{jj}\right) + m\log\left((1+\delta b_{ij})^2 - \delta^2 b_{jj}b_{ii}\right).
\end{eqnarray}
This is a scalar-valued function with scalar inputs, so it is easy to optimize within an interval that defines the feasible region of $\delta$. In our software, we use Brent's method implemented in \texttt{Optim.jl} to solve this problem. Note that $\delta$ needs to reside within an interval to ensure the positive definite constraints are satisfied. This interval is derived in sections \ref{sec:feasible_region_diag} and \ref{sec:feasible_region_offdiag}, and we discuss how to efficiently obtain constants $a_{ij}, a_{ii}, a_{jj}, b_{ij}, b_{ii}, b_{jj}$ in section \ref{sec:efficiently_obtaining_const}. 

\subsection{Fully general coordinate descent for ME (diagonal entries)} 

Now we consider optimizing the diagonal entries, i.e. we want to find $\delta$ for the update $S_{jj}^{new} = S_{jj} + \delta \be_j\be_j^t$. Again letting $\bD = \frac{m+1}{m}\bSigma - \bS$, the objective becomes 
\begin{eqnarray*}
    g(\delta) 
    &=& \ln\det(\bD - \delta \be_j\be_j^t) + m\ln\det(\bS + \delta \be_j\be_j^t)\\
    &=& \ln\left[(1 - \delta \be_j\bD^{-1}\be_j)\det(\bD)\right] + m\ln\left[(1 + \delta \be_j^t\bS^{-1}\be_j)\det(\bS)\right]\\
    &\propto& \ln(1 - \delta \be_j\bD^{-1}\be_j) + m\ln(1 + \delta \be_j\bS^{-1}\be_j)
\end{eqnarray*}
where the second equality follows from the matrix determinant lemma. The first order optimality condition states
\begin{eqnarray*}
    \frac{d}{d\delta}g(\delta) 
    &=& \frac{-\be_j\bD^{-1}\be_j}{1 - \delta \be_j\bD^{-1}\be_j} + \frac{m \be_j\bS^{-1}\be_j}{1 + \delta \be_j\bS^{-1}\be_j} \amp=\amp 0.
\end{eqnarray*}
In the notation for computing off-diagonal entries, we have
\begin{eqnarray}\label{eq:ccd_diag_ME}
    \delta &=& \frac{m\be_j\bS^{-1}\be_j - \be_j\bD^{-1}\be_j}{(m+1)\be_j\bS^{-1}\be_j\be_j\bD^{-1}\be_j} \amp=\amp \frac{mb_{jj} - a_{jj}}{(m+1)b_{jj}a_{jj}}.
\end{eqnarray}

\subsection{Fully general coordinate descent for MVR (off-diagonal entries)}

In section \ref{sec:mvr_obj} we show that minimum variance-based reconstructability (MVR) knockoffs solve the problem
\begin{align*}
    \min_{\bS} \ 
    m^2\tr(\bS^{-1}) + \tr\left(\frac{m+1}{m}\bSigma - \bS\right)^{-1}       
    \quad \text{ subject to } 
    \begin{cases}
        \frac{m + 1}{m}\bSigma - \bS \succeq 0\\
        \bS \succeq 0
    \end{cases}.
\end{align*}
Again we consider updating $S^{new}_{ij} = S_{ij} + \delta$ where $i\ne j$. Let $\bD = \frac{m+1}{m}\bSigma - \bS$ and $\bH$ be the $p\times p$ matrix that is zero everywhere except $H_{ij} = H_{ji} = 1$. The objective becomes
\begin{align*}
    m^2\tr(\bS + \delta \bH)^{-1} + \tr\left(\bD - \delta \bH\right)^{-1}.
\end{align*}
Since $\bH = \bU\bV$, where $\bU = \begin{bmatrix} \be_i & \be_j \end{bmatrix} \in \mathbb{R}^{p \times 2}$ and $\bV = \begin{bmatrix} \be_j^t \\ \be_i^t\end{bmatrix} \in \mathbb{R}^{2 \times p}$, Woodbury's formula gives
\begin{eqnarray*}
    \tr(\bD - \delta \bH)^{-1} 
    &=& \tr(\bD - \delta \bU\bV)^{-1}\\
    &=& \tr\left(\bD^{-1} + \delta \bD^{-1}\bU(\bI_2 -\delta \bV\bD^{-1}\bU)^{-1}\bV\bD^{-1}\right)\\
    &=& \delta \tr\left(\bD^{-1}\bU(\bI_2 - \delta \bV\bD^{-1}\bU)^{-1}\bV\bD^{-1}\right) + c\\
    &=& \delta \tr\left(\bV\bD^{-2}\bU(\bI_2 - \delta \bV\bD^{-1}\bU)^{-1}\right) + c,\\
    \tr(\bS + \delta \bH)^{-1}
    &=& -\delta \tr\left(\bV\bS^{-2}\bU(\bI_2 + \delta \bV\bS^{-1}\bU)^{-1}\right)
\end{eqnarray*}
We already have explicit expressions for $\bV\bS^{-1}\bU$ and $\bV\bD^{-1}\bU$, thus
\begin{align*}
    (\bI_2 - \delta \bV\bD^{-1}\bU)^{-1} 
    &= 
    \begin{bmatrix}
        1 - \delta a_{ij} & -\delta a_{jj}\\
        -\delta a_{ii} & 1 - \delta a_{ij}
    \end{bmatrix}^{-1}
    = 
    \frac{1}{(1-\delta a_{ij})^2 - \delta^2 a_{ii}a_{jj}}
    \begin{bmatrix}
        1 - \delta a_{ij} & \delta a_{jj}\\
        \delta a_{ii} & 1 - \delta a_{ij}
    \end{bmatrix},\\
    (\bI_2 + \delta \bV\bS^{-1}\bU)^{-1} 
    &= 
    \begin{bmatrix}
        1 + \delta b_{ij} & \delta b_{jj}\\
        \delta b_{ii} & 1 + \delta b_{ij}
    \end{bmatrix}^{-1}
    = 
    \frac{1}{(1+\delta b_{ij})^2 - \delta^2 b_{ii}b_{jj}}
    \begin{bmatrix}
        1 + \delta b_{ij} & -\delta b_{jj}\\
        -\delta b_{ii} & 1 + \delta b_{ij}
    \end{bmatrix}.
\end{align*}
Similarly, $\bV\bS^{-2}\bU$ and $\bV\bD^{-2}\bU$ can be computed as
\begin{align*}
    \bV\bS^{-2}\bU 
    &= 
    \begin{bmatrix}
        c_{ji} & c_{jj}\\
        c_{ii} & c_{ij}
    \end{bmatrix}, \qquad
    \bV\bD^{-2}\bU = 
    \begin{bmatrix}
        d_{ji} & d_{jj}\\
        d_{ii} & d_{ij}
    \end{bmatrix},
\end{align*}
where
\begin{align*}
    c_{ij} &= \be_i^t\bS^{-2}\be_j, \quad 
    c_{ii} = \be_i^t\bS^{-2}\be_i, \quad
    c_{jj} = \be_j^t\bS^{-2}\be_j,\\
    d_{ij} &= \be_i^t\bD^{-2}\be_j, \quad 
    d_{ii} = \be_i^t\bD^{-2}\be_i, \quad
    d_{jj} = \be_j^t\bD^{-2}\be_j.
\end{align*}
The objective is therefore a scalar function of $\delta$
\begin{eqnarray}\label{eq:ccd_mvr_update}
    g(\delta) 
    &=& \frac{-m^2\delta}{(1 + \delta b_{ij})^2 - \delta^2b_{ii}b_{jj}}\tr\left(
    \begin{bmatrix}
        c_{ij} & c_{jj} \\ c_{ii} & c_{ij}
    \end{bmatrix}
    \begin{bmatrix}
        1 + \delta b_{ij} & -\delta b_{jj}\\ -\delta b_{ii} & 1 + \delta b_{ij}
    \end{bmatrix}
    \right) \amp + \amp \nonumber \\
    && \frac{\delta}{(1 - \delta a_{ij})^2 - \delta^2 a_{ii}a_{jj}} \tr\left(
    \begin{bmatrix}
        d_{ij} & d_{jj}\\d_{ii}& d_{ij}
    \end{bmatrix}
    \begin{bmatrix}
        1 - \delta a_{ij} & \delta a_{jj}\\
        \delta a_{ii} & 1 - \delta a_{jj}
    \end{bmatrix}
    \right) \nonumber \\
    &=& \frac{-m^2 \delta\left\{(c_{ij}b_{ij} - c_{jj}b_{ii} - c_{ii}b_{jj} + c_{ij}b_{ij})\delta + 2c_{ij}\right\}}{(1 + \delta b_{ij})^2 - \delta^2b_{ii}b_{jj}} \amp + \amp \nonumber \\
    && \frac{\delta\left\{\right(-d_{ij}a_{ij} + d_{jj}a_{ii} + d_{ii}a_{jj} - d_{ij}a_{ij})\delta + 2d_{ij}\}}{(1 - \delta a_{ij})^2 - \delta^2a_{ii}a_{jj}}.
\end{eqnarray}
If all the constants $a_{ii}, a_{ij}, a_{jj}, b_{ii}, b_{ij}, b_{jj}, c_{ii}, c_{ij}, c_{jj}, d_{ii}, d_{ij}, d_{jj}$ are known, the objective can be minimized using Brent's method similar to the maximum entropy case. If we hold the Cholesky factors $\bD = \bL\bL^t$ and $\bS = \bC\bC^t,$ then the first six of these constants can be evaluated in the same way as in \ref{sec:efficiently_obtaining_const}. To evaluate the other six constants, e.g. $d_{ij}$, note that
\begin{align*}
    d_{ij} = \be_i^t\bD^{-2}\be_j = \be_i^t(\bL\bL^t)^{-1}(\bL\bL^t)^{-1}\be_j \equiv \bu^t\bv.
\end{align*}
Here $\bu, \bv$ can be obtained by noting that $\bu = (\bL\bL^t)^{-1}\be_i \iff \bL\bL^t\bu = \be_i$ and using forward-backward substitution twice; that is, first solve for $\by$ in $\bL\by = \be_i$ and then for $\bu$ in $\bL^t\bu = \by$. 

Finally, we need to ensure that the proposed $\delta$ is feasible. Because the PSD constraint is the same as the maximum entropy case, the feasible region is already derived in sections \ref{sec:feasible_region_diag} and \ref{sec:feasible_region_offdiag}. Once we update $S^{new}_{ij} = S_{ij} + \delta$ and $S^{new}_{ji} = S_{ji} + \delta$, we perform a rank-2 update to maintain Cholesky equalities $\bL_{new}\bL_{new}^t = \frac{m+1}{m}\bSigma - \bS_{new}$ and $\bC_{new}\bC_{new}^t = \bS_{new}$. Again, this can be achieved with rank-1 updates in equation \eqref{eq:cholesky_updates}.

\subsection{Fully general coordinate descent for MVR (diagonal entries)}

Now we consider optimizing the diagonal entries, i.e. we want to find $\delta$ for the update $S_{jj}^{new} = S_{jj} + \delta \be_j\be_j^t$. Again letting $\bD = \frac{m+1}{m}\bSigma - \bS$, the objective becomes 
\begin{eqnarray*}
    g(\bS + \delta \be_j\be_j^t)
    &=& m^2 \tr[(\bS + \delta \be_j\be_j^t)^{-1}] + \tr[(\bD - \delta \be_j\be_j^t)^{-1}]\\
    &=& m^2 \tr\left[\bS^{-1} - \frac{\delta \bS^{-1}\be_j\be_j^t\bS^{-1}}{1 + \delta \be_j^t\bS^{-1}\be_j}\right] + \tr\left[\bD^{-1} + \frac{\delta \bD^{-1}\be_j\be_j^t\bD^{-1}}{1 - \delta \be_j^t\bD^{-1}\be_j}\right] \quad \text{(Sherman-Morrison)}\\
    &=& \frac{-m^2\delta}{1 + \delta \be_j^t\bS^{-1}\be_j}\tr(\bS^{-1}\be_j\be_j^t\bS^{-1}) + \frac{\delta}{1 - \delta \be_j^t\bD^{-1}\be_j}\tr(\bD^{-1}\be_j\be_j^t\bD^{-1}) + m^2\tr(\bS^{-1}) + \tr(\bD^{-1})\\
    &=& \frac{-m^2 \be_j^t\bS^{-2}\be_j \delta}{1 + \delta 
 \be_j^t\bS^{-1}\be_j} + \frac{\delta \be_j^t\bD^{-2}\be_j}{1 - \delta 
 \be_j^t\bD^{-1}\be_j} + m^2\tr(\bS^{-1}) + \tr(\bD^{-1}) \quad \text{(cyclic property of trace)}\\
    &\equiv& \frac{-m^2\delta c_{jj}}{1 + \delta b_{jj}} + \frac{\delta d_{jj}}{1 - \delta a_{jj}} + g(\bS),
\end{eqnarray*}
with the convention that $a_{jj} = \be_j^t\bD^{-1}\be_j, b_{jj}=\be_j^t\bS^{-1}\be_j, c_{jj} = \be_j^t\bS^{-2}\be_j,$ and $d_{jj} = \be_j^t\bD^{-2}\be_j$. The first-order optimality condition states
\begin{eqnarray*}
    0 &=& \frac{-(1+\delta b_{jj})m^2c_{jj} + \delta m^2c_{jj}b_{jj}}{(1 + \delta b_{jj})^2} + \frac{(1-\delta a_{jj})d_{jj} + \delta d_{jj} a_{jj}}{(1-\delta a_{jj})^2}\\
    &=& -m^2c_{jj}(1-\delta a_{jj})^2 + (1 + \delta b_{jj})^2 d_{jj}\\
    &=& \delta^2(-a_{jj}^2m^2c_{jj} + b_{jj}^2d_{jj}) + \delta(2a_{jj}m^2c_{jj} + 2b_{jj}d_{jj}) - m^2c_{jj} + d_{jj}.
\end{eqnarray*}
We apply the quadratic formula to find the roots of this objective, and enforce the boundary condition $-1/b_{jj} \le \delta \le 1/a_{jj}$ derived in section \ref{sec:feasible_region_diag}.

\subsection{Fully general coordinate descent for SDP}

Following the same definition of $\bH$ as MVR/ME case, the SDP objective is
\begin{align*}
    &\min_S \sum_{g \in G} \frac{1}{|\mathcal{A}_g|^2} \sum_{i, j \in \mathcal{A}_g}| \Sigma^{(g)}_{ij} - S_{ij}^{(g)} - \delta H_{ij}|.
\end{align*}
Only two entries of the objective depend on $\delta$:
\begin{align*}
    |\Sigma_{ij} - S_{ij} - \delta| + |\Sigma_{ji} - S_{ji} - \delta| = 2|\Sigma_{ij} - S_{ij} - \delta|.
\end{align*}
Computing the feasible region of $\delta \in [a, b]$ according to sections \ref{sec:feasible_region_diag} and \ref{sec:feasible_region_offdiag}, the solution is
\begin{eqnarray}\label{eq:ccd_sdp_update}
    \delta =
    \begin{cases}
        \Sigma_{ij} - S_{ij} & \Sigma_{ij} - S_{ij} \in [a, b]\\
        a & \Sigma_{ij} - S_{ij} \le a\\
        b & \Sigma_{ij} - S_{ij} \ge b
    \end{cases}.
\end{eqnarray}

\subsection{Feasible region of diagonal entries for general coordinate descent} \label{sec:feasible_region_diag}

What is the feasible region of $\delta$? To satisfy the PSD constraints, we must choose $\delta$ so that $\bD - \delta \be_j\be_j^t \succeq 0$ and $\bS + \delta \be_j\be_j^t \succeq 0$. Applying the matrix determinant lemma again, 
\begin{eqnarray*}
    \det(\bD - \delta \be_j\be_j^t) \ge 0 &\iff& (1 - \delta \be_j\bD^{-1}\be_j)\det(\bD) \ge 0,\\ 
    \det(\bS + \delta \be_j\be_j^t) \ge 0 &\iff& (1 + \delta \be_j\bS^{-1}\be_j)\det(\bS) \ge 0.
\end{eqnarray*}
Since $\bD$ and $\bS$ are positive definite, we must have
\begin{align}\label{eq:feasible_region_diag}
    \frac{1}{\be_j^t\bD^{-1}\be_j} \ge \delta \ge \frac{-1}{\be_j^t\bS^{-1}\be_j} \amp \iff \amp 
    \frac{1}{a_{jj}} \ge \delta \ge \frac{-1}{b_{jj}}.
\end{align}

\subsection{Feasible region of off-diagonal entries for general coordinate descent}\label{sec:feasible_region_offdiag}

$\delta$ must satisfy $\bD - \delta \bH \succeq 0$ and $\bS + \delta \bH \succeq 0.$ By convexity and continuity, these conditions define a closed interval of feasible values of $\delta$, with the endpoints satisfying either $\det(\bD - \delta \bH) = 0$ or $\det(\bS + \delta \bH) = 0$. In the first case, $\det(\bD - \delta \bH ) = (1 - \delta a_{ij})^2 - \delta^2 a_{ii}a_{jj},$ which is a quadratic function. Let $s_1, s_2$ be its roots and $d_1, d_2$ be the roots of $\det(\bS + \delta \bH)$. Then 
\begin{eqnarray*}
    s_1 &=& \frac{a_{ij} - \sqrt{a_{ii}a_{jj}}}{a_{ij}^2 - a_{ii}a_{jj}}, \quad 
    s_2 \amp=\amp \frac{a_{ij} + \sqrt{a_{ii}a_{jj}}}{a_{ij}^2 - a_{ii}a_{jj}},\\
    d_1 &=& \frac{-b_{ij} - \sqrt{b_{ii}b_{jj}}}{b_{ij}^2 - b_{ii}b_{jj}}, \quad
    d_2 \amp=\amp \frac{-b_{ij} + \sqrt{b_{ii}b_{jj}}}{b_{ij}^2 - b_{ii}b_{jj}},
\end{eqnarray*}
and the feasible region is defined by 
\begin{align}\label{eq:math_feasible_region}
    \max\{s_1, d_1\} \le \delta \le \min\{s_2, d_2\}.
\end{align}
In practice, we often need more stringent lower and upper bounds for computational reasons. Specifically, note that the updates $\bS + \delta \bH$ and $\bD - \delta \bH$ require us to maintain Cholesky equalities in equation \eqref{eq:cholesky_updates}, which is achieved via two separate rank-1 updates. Because we perform these actions sequentially, it is possible to violate positive definiteness in an intermediate step even though the overall update does not. Thus, we in fact need all of the following
\begin{align*}
    \begin{cases}
        \bD - \frac{\delta}{2}\left((\be_i + \be_j)(\be_i + \be_j)^t\right) \succeq 0 & \text{(first rank-1 update to }\bL)\\
        \bD - \delta(\be_i\be_j^t + \be_j\be_i^t) \succeq 0 & (\text{final rank 1 update to } \bL \text{, feasible region derived in \eqref{eq:math_feasible_region}})
    \end{cases}, 
\end{align*}
\begin{align*}
    \begin{cases}
        \bS + \frac{\delta}{2}\left((\be_i + \be_j)(\be_i + \be_j)^t\right) \succeq 0 & \text{(first rank-1 update to }\bC)\\
        \bS + \delta(\be_i\be_j^t + \be_j\be_i^t) \succeq 0 & (\text{final rank 1 update to } \bC, \text{ feasible region derived in \eqref{eq:math_feasible_region}})
    \end{cases}.
\end{align*}
Applying the matrix determinant lemma to the first terms of these equations,
\begin{eqnarray*}
    \det(\bD- \frac{\delta}{2} (\be_i + \be_j)(\be_i + \be_j)^t) \ge 0 &\iff& (1 - \frac{\delta}{2} (\be_i + \be_j)^t\bD^{-1}(\be_i + \be_j))\det(\bD) \ge 0,\\
    \det(\bS + \frac{\delta}{2} (\be_i + \be_j)(\be_i + \be_j)^t) \ge 0 &\iff& (1 + \frac{\delta}{2} (\be_i + \be_j)^t\bS^{-1}(\be_i + \be_j))\det(\bS) \ge 0.
\end{eqnarray*}
Solving for $\delta$, a different lower/upper bound emerges:
\begin{align*}
    \frac{2}{\be_i^t\bD^{-1}\be_i + \be_i\bD^{-1}\be_j + \be_j^t\bD^{-1}\be_i + \be_j\bD^{-1}\be_j} \ge \delta \ge \frac{-2}{\be_i^t\bS^{-1}\be_i + \be_i\bS^{-1}\be_j + \be_j^t\bS^{-1}\be_i + \be_j\bS^{-1}\be_j}.
\end{align*}
Thus, in the notation of the objective, $\delta$ must additionally satisfy the (computational) feasible region
\begin{align}\label{eq:feasible_region_offdiag}
    \frac{2}{a_{ii} + 2a_{ij} + a_{jj}} \ge \delta \ge \frac{-2}{b_{ii} + 2b_{ij} + b_{jj}}.
\end{align}

\subsection{Efficiently obtaining needed constants by maintaining Cholesky factors}\label{sec:efficiently_obtaining_const}

To efficiently evaluate constants such as $\be_i^t\bD^{-1}\be_j$ and $\be_i^t\bS^{-1}\be_j$, the natural way is to precompute and constantly update $\bD^{-1}, \bS^{-1}$ via Woodbury formulas. However, past work \citep{spector2022powerful, askari2021fanok} suggests that these low-rank updates are numerically unstable. The typical approach is to maintain two Cholesky decompositions $\bD = \bL\bL^t$ and $\bS = \bC\bC^t$, and proceed to extract necessary constants as described in section \ref{sec:algs}.

After updating $S_{ij}^{new} = S_{ij} + \delta$ and $S_{ji}^{new} = S_{ij}^{new}$, we need to update the Cholesky factors $\bC$ and $\bL$ to maintain equalities $\bL_{new}\bL_{new}^t = \frac{m+1}{m}\bSigma - \bS_{new}$ and $\bC_{new}\bC_{new}^t = \bS_{new}$. In general, if we have $\bL\bL^t = \bA$, we can obtain the Cholesky factor of $\bA + \bw\bw^t$ (i.e. a rank-1 update from $\bA$) in $\mathcal{O}(p^2)$ time. In light of this, lets write the equality we wish to maintain as:
\begin{equation}\label{eq:cholesky_updates}
\begin{split}
    \bL_{new}\bL_{new}^t 
    &= \frac{m+1}{m}\bSigma - \bS - \delta(\be_i\be_j^t + \be_j \be_i^t) \\
    &= \frac{m+1}{m}\bSigma - \bS - \frac{\delta}{2}\left((\be_i + \be_j)(\be_i + \be_j)^t - (\be_i - \be_j)(\be_i - \be_j)^t\right),\\
    \bC_{new}\bC_{new}^t &= \bS + \delta(\be_i\be_j^t + \be_j \be_i^t) = \bS + \frac{\delta}{2}\left((\be_i + \be_j)(\be_i + \be_j)^t - (\be_i - \be_j)(\be_i - \be_j)^t\right).
\end{split}
\end{equation}
Thus, the required rank-2 update can be achieved by first rank-1 updating using $\bw = \sqrt{\frac{\delta}{2}}(\be_i + \be_j)$ and then perform rank-1 downdate via $\bw = \sqrt{\frac{\delta}{2}}(\be_i-\be_j)$.

\subsection{Simplifying MVR objective}\label{sec:mvr_obj}

When generating $m$ multiple knockoffs, the MVR objective \citep{spector2022powerful} is
\begin{eqnarray*}
    \tr(\bG^{-1}_{\bS})
    &\equiv&
    \tr 
    \begin{pmatrix}
    \bSigma & \bSigma - \bS & \cdots & \bSigma - \bS\\
    \bSigma - \bS & \bSigma & \cdots & \bSigma - \bS\\
    \cdots & \cdots & \cdots & \cdots\\
    \bSigma - \bS & \bSigma - \bS & \cdots & \bSigma
    \end{pmatrix}^{-1}_{p(m+1) \times p(m+1)}.
\end{eqnarray*}
This can be simplified as follows
\begin{eqnarray*}
    \tr(\bG^{-1}_{\bS})
    \amp=\amp
    m\tr(\bS^{-1}) + \tr\left((m+1)\bSigma - m \bS)\right)^{-1}.
\end{eqnarray*}
Then using $(k\bA)^{-1} = k^{-1}\bA^{-1}$ for matrix $\bA$ and scalar $k$, we can scale the objective by $m$ to get the MVR objective
\begin{align*}
    m^2\tr(\bS^{-1}) + \tr\left(\frac{m+1}{m}\bSigma - \bS\right)^{-1}.
\end{align*}

\subsection{Initializating coordinate descent algorithms}

One plausible way to initialize the SDP/MVR/ME algorithms is to start at the equi-correlated solution $\bS_{\text{eSDP}}$ due to its convenient closed-form solution \citep{dai2016knockoff}. However, note that the equi-correlated solution solves for the largest $\bS$ matrix that satisfies $\frac{m+1}{m}\bSigma - \bS \succeq 0$, i.e. the smallest eigenvalue of $\bD \equiv \frac{m+1}{m}\bSigma - \bS$ is numerically 0. This can cause the initial Cholesky factorization for $\bD$ to fail. Thus, in \texttt{Knockoffs.jl}, we initialize the optimization problem with $\frac{1}{2}\bS_{\text{eSDP}}$, which circumvents numerical issues but also serves as a reasonable starting point. 

\subsection{Declaring convergence}

Each of the SDP/MVR/ME group knockoff problems have an objective function $g$. In \texttt{Knockoffs.jl}, we declare convergence when either one of the conditions below is met
\begin{enumerate}
	\item $\frac{|g^{new} - g|}{g} \le \epsilon$
	\item $\max_{i,j} |S^{new}_{ij} - S_{ij}| < 0.0001$
\end{enumerate}
where the default $\epsilon = 0.0001$. The first condition checks if the objective improves, and exits if improvement is small. The second condition is checking whether the optimization variables $\bS$ are changing sufficiently. This early exit criteria is motivated by the fact that model-X knockoffs control the FDR for any $\bS$, and thus the estimation to $\bS$ does not have to be very precise. In other words, if entries in $\bS$ are not really changing, then optimization halts even if the objective can still be improved. 

\subsection{Algorithm summary}

Algorithm \ref{alg:CCD} summarizes the group knockoff optimization procedure in the ME case. For MVR and SDP, the overall structure remains the same, but the computation of $\delta$ needs to be modified according to the relevant equations derived in the sections above. Also, note that the algorithm summary does not check for backtracking, which is done in practice for better stability.

\setcounter{supplementalalgorithm}{1} 
\renewcommand{\thealgorithm}{\thesupplementalalgorithm} 
{\begin{algorithm}
    \caption{Coordinate descent for ME group knockoffs (for MVR and SDP, only expression for $\delta$ changes but overall structure is the same)}\label{alg:CCD}
    \begin{algorithmic}[1]
	\STATE { \textbf{Input:} correlation matrix $\bSigma_{p \times p}$, group membership vector, and number of knockoff copies to generate $m$}
        \STATE{ \textbf{Initialize:} $\bS = \frac{1}{2}\bS_{\text{eSDP}}$ with $\bS_{\text{eSDP}}$ from  \citep{dai2016knockoff}} and $\bD = \frac{m+1}{m}\bSigma - \bS$
        \STATE{ \textbf{Compute:} ($\bv_1,...,\bv_p$) based on eigendecomposition of $\bSigma_{block}$ in eq \eqref{eq:sigma_block}}
        \STATE{ \textbf{Compute:} Cholesky factors $\bL\bL^t = \text{cholesky}(\bD)$ and $\bC\bC^t = \text{Cholesky}(\bS)$}
	\WHILE{Not converged}
            \STATE{\#\# PCA iterations}
            \FOR{$\bv_i \in (\bv_1,...,\bv_p)$} 
                \STATE Compute constants $\bv_i^t\bD^{-1}\bv_i, \bv_i^t\bS^{-1}\bv_i$ from $\bL, \bC$ via method derived in section \ref{sec:efficiently_obtaining_const}
                \STATE Compute $\delta$ in Eq \eqref{eq:pca_delta_ME}
		      \STATE Rank-1 update: $\bS_{new} = \bS + \delta \bv_i\bv_i^t$
                \STATE Update Cholesky factors $\bL_{new}\bL_{new}^t = \frac{m+1}{m}\bSigma - \bS_{new}$ and $\bC_{new}\bC_{new}^t = \bS_{new}$
		\ENDFOR
            \STATE{\#\# Full optimization}
            \FOR{$\gamma = \{1,...,g\}$}
                \FOR{$(i,j)$ in group $\gamma$}
                    \IF{$i = j$}
                        \STATE Compute constants $a_{jj}, b_{jj}$ from $\bL, \bC$ via method in section \ref{sec:efficiently_obtaining_const}
                        \STATE Compute $\delta$ by Eq \eqref{eq:ccd_diag_ME}
                        \STATE Clamp $\delta$ to be within feasible region derived in Eq \eqref{eq:feasible_region_diag}
                        \STATE Rank-1 update: $S^{new}_{ii} = S_{ii} + \delta$
                    \ELSE
                        \STATE Compute constants $a_{ii}, a_{jj}, a_{ij},b_{ii},b_{jj},b_{ij}$  from $\bL, \bC$ via method in section \ref{sec:efficiently_obtaining_const}
                        \STATE Compute $\delta$ by solving 1-D optimization problem in \eqref{eq:ccd_offdiag_ME}
                        \STATE Clamp $\delta$ to be within feasible region derived in Eq \eqref{eq:feasible_region_offdiag}
                        \STATE Rank-2 update: $S^{new}_{ij} = S^{new}_{ji} = S_{ij} + \delta$
                    \ENDIF
                    \STATE Update Cholesky factors $\bL_{new}\bL_{new}^t = \frac{m+1}{m}\bSigma - \bS_{new}$ and $\bC_{new}\bC_{new}^t = \bS_{new}$
                \ENDFOR
            \ENDFOR
        \ENDWHILE
        \STATE \textbf{Output:} Group-block-diagonal matrix $\bS$ satisfying $\frac{m+1}{m}\bSigma - \bS \succeq 0$ and $\bS \succeq 0$.
	\end{algorithmic}
\end{algorithm}}

\section{PCA-based coordinate descent}\label{sec:pca_ccd}




Recall that, in PCA optimization, we perturb $\bS$ via
\begin{align*}
    \bS^{new} = \bS + \delta \bv\bv^t,
\end{align*}
where $\bv$ is a precomputed vector such that $\bv^t\bv = 1$ and the outer product $\bv\bv^t$ respects the block diagonal structure of $\bS$. Here $\bv$ can be viewed as a direction and $\delta$ a step size. Thus, we naturally would like to have a set of different proposed directions. One option is to precompute $\bv_1,...,\bv_p$ where $\bv_i$ is the $i$th eigenvector (hence the name ``PCA-based") of the block diagonalized covariance matrix
\begin{eqnarray}
    \bSigma_{blocked} = \begin{bmatrix}
        \bSigma_1 & & \\
        & \ddots & \\
        & & \bSigma_g
    \end{bmatrix}_{p \times p}.
\end{eqnarray}
Eigendecomposition of $\bSigma_{blocked}$ is efficient due to its block structure, since we can just compute the eigenvectors for each block and pad them with zeros. We can obviously include more directions as long as the outer product $\bv\bv^t$ respects the block diagonal structure of $\bS$. For example, adding $p$ basis vectors (which allows updating just the diagonal entries) dramatically speeds up convergence. 

\subsection{PCA-based Coordinate descent for ME}

We will optimize
\begin{align*}
    \max_{\bS} \log\det\left(\frac{m+1}{m}\bSigma - \bS\right) + m \log\det(\bS) \quad \text{ subject to } 
    \begin{cases}
        \frac{m + 1}{m}\bSigma - \bS \succeq 0\\
        \bS \succeq 0\\
    \end{cases}.
\end{align*}
Consider updating $\bS^{new} = \bS + \delta\bv\bv^t$. If $\bD = \frac{m + 1}{m}\bSigma - \bS$, then the objective becomes
\begin{eqnarray*}
    g(\bS + \delta \bv\bv^t) 
    &=& \log\det(\bD - \delta \bv\bv^t) + m\log\det(\bS + \delta \bv\bv^t)\\
    &=& \log\left[(1 - \delta \bv^t\bD^{-1}\bv)\det(\bD)\right] + m\log\left[(1 + \delta \bv^t\bS^{-1}\bv)\det(\bS)\right]\\
    &\propto& \log(1 - \delta \bv^t\bD^{-1}\bv) + m\log(1 + \delta \bv^t\bS^{-1}\bv),
\end{eqnarray*}
where the second equality follows from the matrix determinant lemma. The first-order optimality condition is
\begin{eqnarray*}
    \frac{d}{d\delta}g 
    &=& \frac{-\bv^t\bD^{-1}\bv}{1 - \delta \bv^t\bD^{-1}\bv} + \frac{m \bv^t\bS^{-1}\bv}{1 + \delta \bv^t\bS^{-1}\bv} \amp=\amp 0.
\end{eqnarray*}
Thus,
\begin{eqnarray}\label{eq:pca_delta_ME}
    \delta &=& \frac{m\bv^t\bS^{-1}\bv - \bv^t\bD^{-1}\bv}{(m+1)\bv^t\bS^{-1}\bv\bv^t\bD^{-1}\bv}.
\end{eqnarray}
Simple algebraic manipulation reveals that $\delta$ satisfies
\begin{eqnarray*}
    \frac{-1}{\bv^t\bS^{-1}\bv} \le \delta \le \frac{1}{\bv^t\bD^{-1}\bv}.
\end{eqnarray*}
As derived in section \ref{sec:feasible_region_diag}, any $\delta$ within this range will satisfy the PSD constraints. Constants $\bv^t\bD^{-1}\bv$ and $\bv^t\bS^{-1}\bv$ can be extracted efficiently with the strategy outlined in section \ref{sec:efficiently_obtaining_const} as long as we have Cholesky factors of $\bS$ and $\bD$. Finally, we efficiently update the objective
\begin{eqnarray*}
    g(\bS^{new}) = \log(1 - \delta \bv^t\bD^{-1}\bv) + m \log(1 + \delta \bv^t \bS^{-1}\bv) + g(\bS).
\end{eqnarray*}

\subsection{PCA-based coordinate descent for MVR.} 

As shown above, the MVR objective can be written as
\begin{align*}
    \min_{\bS}  m^2 \tr(\bS)^{-1} + \tr\left(\frac{m+1}{m}\bSigma - \bS\right)^{-1} \quad \text{ subject to } 
    \begin{cases}
        \frac{M + 1}{M}\bSigma - \bS \succeq 0\\
        \bS \succeq 0
    \end{cases}.
\end{align*}
Consider updating $\bS^{new} = \bS + \delta\bv\bv^t$. If $\bD = \frac{m + 1}{m}\bSigma - \bS$, then the objective becomes
\begin{eqnarray*}
    g(\bS + \delta \bv\bv^t) 
    &=& m^2\tr\left((\bS + \delta\bv\bv^t)^{-1}\right) + \tr\left((\bD - \delta\bv\bv^t)^{-1}\right)\\
    &=& m^2\tr\left(\bS^{-1} - \delta \frac{\bS^{-1}\bv\bv^t\bS^{-1}}{1+\delta\bv^t\bS^{-1}\bv}\right) + \tr\left(\bD^{-1} + \delta \frac{\bD^{-1}\bv\bv^t\bD^{-1}}{1 - \delta\bv^t\bD^{-1}\bv}\right) \quad (\text{Sherman-Morrison})\\
   &=& \frac{-m^2\delta\bv^t\bS^{-2}\bv}{1 + \delta\bv^t\bS^{-1}\bv} + \frac{\delta\bv^t\bD^{-2}\bv}{1 - \delta\bv^t\bD^{-1}\bv} + g(\bS) \quad (\text{Cyclic property of trace}).
\end{eqnarray*}
The first-order optimality states
\begin{eqnarray*}
    0 &=& \delta^2(-m^2(\bv^t\bD^{-1}\bv)^2\bv^t\bS^{-2}\bv + (\bv^t\bS^{-1}\bv)^2\bv^t\bD^{-2}\bv) + \\
    && \delta(2m^2\bv^t\bD^{-1}\bv\bv^t\bS^{-2}\bv + 2\bv^t\bS^{-1}\bv\bv\bD^{-2}\bv) - m^2\bv^t\bS^{-2}\bv + \bv\bD^{-2}\bv.
\end{eqnarray*}
Thus, 
\begin{eqnarray}\label{eq:pca_mvr_update}
    \delta &=& \frac{-b \pm \sqrt{b^2-4ac}}{2a} \text{ where } \begin{cases}
        a &= -m^2(\bv^t\bD^{-1}\bv)^2\bv^t\bS^{-2}\bv + (\bv^t\bS^{-1}\bv)^2\bv^t\bD^{-2}\bv\\
        b &= 2m^2\bv^t\bD^{-1}\bv\bv^t\bS^{-2}\bv + 2\bv^t\bS^{-1}\bv\bv\bD^{-2}\bv\\
        c &= - m^2\bv^t\bS^{-2}\bv + \bv\bD^{-2}\bv
    \end{cases}.
\end{eqnarray}
A unique solution for $\delta$ exists given the boundary condition
\begin{eqnarray*}
    \frac{-1}{\bv^t\bS^{-1}\bv} \le \delta \le \frac{1}{\bv^t\bD^{-1}\bv}.
\end{eqnarray*}

\subsection{PCA-based coordinate descent for SDP} 

We will optimize
\begin{align*}
    \min_{\bS} \sum_{\gamma=1}^g \frac{1}{|\mathcal{A}_\gamma|^2} \sum_{i,j \in \mathcal{A}_\gamma} | \Sigma_{ij} - S_{ij} | 
    \quad \text{ subject to } 
    \begin{cases}
        \frac{m + 1}{m}\bSigma - \bS \succeq 0\\
        \bS \succeq 0
    \end{cases}.
\end{align*}
Consider updating $\bS^{new} = \bS + \delta\bv\bv^t$, the objective becomes
\begin{eqnarray}\label{eq:pca_sdp_update}
    g(\bS + \delta\bv\bv^t)
    &=& \sum_{\gamma=1}^g \frac{1}{|\mathcal{A}_\gamma|^2} \sum_{i,j \in \mathcal{A}_\gamma} | \Sigma_{ij} - S_{ij} - \delta v_iv_j|.
\end{eqnarray}
Although a closed form solution may exist for this problem, our software solves it numerically using Brent's method implemented in \texttt{Optim.jl}. This is a very fast operation because $\bv\bv^t$ respects the block diagonal structure of $\bS$, and thus only a single block in $\bS$ depends on $\delta$.


 	\section{Proof of Theorem \ref{thm:$M$-groupknockoffs}}\label{pr:$M$-groupknockoffs}
To prove Theorem \ref{thm:$M$-groupknockoffs}, we need to show that when knockoffs {\rm $\widetilde{X}$} are generated under Algorithm {\rm\ref{alg:Groupknockoff}}, the distribution of $(X,\tilde{X})$ satisfies both the conditional indepedendence and the group exchangeability. 

According to Algorithm {\rm\ref{alg:Groupknockoff}}, $\tilde{X}$ is generated using only information of $X$ without looking at the response $Y$. Thus, the conditional independence stands.

To prove group exchangeability of the distribution of $(X,\tilde{X})$, we rearrange elements of $X$ such that $X=({X}^{\star}_1,\ldots,{X}^{\star}_g,{X}^{\dagger}_1,\ldots,{X}^{\dagger}_g)$. According to step 2 of Algorithm \ref{alg:Groupknockoff}, it is clear that 
$$({X}^{\star}_1,\ldots,{X}^{\star}_g, \tilde{X}^{\star}_1,\ldots,\tilde{X}^{\star}_g)_{\text{swap}({\cal C})} \overset{d}{=}  ({X}^{\star}_1,\ldots,{X}^{\star}_g, \tilde{X}^{\star}_1,\ldots,\tilde{X}^{\star}_g) \;\;\;\; \forall\;\;  {\cal C}: \;\;{\cal C}=\cup_{\gamma\in {\cal S}} {\cal A}^{\star}_{\gamma}.$$
In other words, for any value ${x}^{\star}_1,\ldots,{x}^{\star}_g, \tilde{x}^{\star}_1,\ldots,\tilde{x}^{\star}_g$, it is equally possible for $$({X}^{\star},\tilde{X}^{\star})=({X}^{\star}_1,\ldots,{X}^{\star}_g, \tilde{X}^{\star}_1,\ldots,\tilde{X}^{\star}_g)\quad \text{and} \quad ({X}^{\star},\tilde{X}^{\star})_{\text{swap}({\cal C})}=({X}^{\star}_1,\ldots,{X}^{\star}_g, \tilde{X}^{\star}_1,\ldots,\tilde{X}^{\star}_g)_{\text{swap}({\cal C})}$$ to take values 
$({x}^{\star},\tilde{x}^{\star})=({x}^{\star}_1,\ldots,{x}^{\star}_g, \tilde{x}^{\star}_1,\ldots,\tilde{x}^{\star}_g)$.

By Definition \ref{gkci} and steps 3-4 of Algorithm \ref{alg:Groupknockoff}, the probability that 
$({X}^{\star}, {X}^{\dagger}, \tilde{X}^{\star},\tilde{X}^{\dagger})$ takes values 
$({x}^{\star}, {x}^{\dagger}, \tilde{x}^{\star},\tilde{x}^{\dagger})$ is equal to
\begin{equation}\label{prob1}
    \text{Pr}\Bigg\{({X}^{\star},\tilde{X}^{\star})=({x}^{\star},\tilde{x}^{\star})\Bigg\}\times\prod_{\gamma=1}^{g}\Bigg\{F_\gamma(\widetilde{x}^{\dagger}_{\gamma}|\widetilde{x}^{\star}_{\gamma})F_\gamma({x}^{\dagger}_{\gamma}|{x}^{\star}_{\gamma})\Bigg\}.
\end{equation}
Because 
$$
\text{Pr}\Bigg\{({X}^{\star},\tilde{X}^{\star})_{\text{swap}({\cal C})}=({x}^{\star},\tilde{x}^{\star})\Bigg\}\times\prod_{\gamma=1}^{g}\Bigg\{F_\gamma(\widetilde{x}^{\dagger}_{\gamma}|\widetilde{x}^{\star}_{\gamma})F_\gamma({x}^{\dagger}_{\gamma}|{x}^{\star}_{\gamma})\Bigg\}={\rm(\ref{prob1})},
$$
we have 
$$({X}^{\star}, {X}^{\dagger}, \tilde{X}^{\star},\tilde{X}^{\dagger})_{\text{swap}({\cal C})}\overset{d}{=}({X}^{\star}, {X}^{\dagger}, \tilde{X}^{\star},\tilde{X}^{\dagger})$$
and thus the group exchangeability stands.

\section{Proof of Theorem \ref{thm:optimality}}\label{pr:optimality}

To prove Theorem \ref{thm:optimality}, we need to 
\begin{itemize}
    \item first derive sufficient and necessary conditions of the minimizers of $L_{\text{ME}}(\textbf{S})=\log\det(\textbf{G}^{-1}_\textbf{S})$ in the case that $X$ has the group key conditional independence property with respect to $\{\mathcal{A}_\gamma\}_{\gamma=1}^g$;
    \item and then show the variance-covariance matrix of $(X,\tilde{X})$ generated according to Algorithm \ref{alg:Groupknockoff} and minimizing {\rm $L^\star_{\text{ME}}(S^\star)$} satisfies all sufficient and necessary conditions.
\end{itemize}

\subsection{Sufficient and Necessary Conditions of the minimizer of $L_{\text{ME}}(\textbf{S})$}

For simplicity, we rearrange elements of $X$ such that $X=({X}^{\star}_1,\ldots,{X}^{\star}_g,{X}^{\dagger}_1,\ldots,{X}^{\dagger}_g)$ whose variance-covariance matrix is $$\boldsymbol{\Sigma}=\left(
\begin{array}{c;{2pt/2pt}c}
			\boldsymbol{\Sigma}^\star&\boldsymbol{\Sigma}^{\star\dagger}\\
					\hdashline[2pt/2pt](\boldsymbol{\Sigma}^{\star\dagger})^T&\boldsymbol{\Sigma}^\dagger\\	\end{array}\right)
     =\left(
\begin{array}{cccc;{2pt/2pt}cccc}
    \boldsymbol{\Sigma}^{\star\star}_{11}&\boldsymbol{\Sigma}^{\star\star}_{12}&\cdots&\boldsymbol{\Sigma}^{\star\star}_{1g}&\boldsymbol{\Sigma}^{\star\dagger}_{11}&\boldsymbol{\Sigma}^{\star\dagger}_{12}&\cdots&\boldsymbol{\Sigma}^{\star\dagger}_{1g}\\
    \boldsymbol{\Sigma}^{\star\star}_{21}&\boldsymbol{\Sigma}^{\star\star}_{22}&\cdots&\boldsymbol{\Sigma}^{\star\star}_{2g}&\boldsymbol{\Sigma}^{\star\dagger}_{21}&\boldsymbol{\Sigma}^{\star\dagger}_{22}&\cdots&\boldsymbol{\Sigma}^{\star\dagger}_{2g}\\
    \vdots&\vdots&\ddots&\vdots&\vdots&\vdots&\ddots&\vdots\\
    \boldsymbol{\Sigma}^{\star\star}_{g1}&\boldsymbol{\Sigma}^{\star\star}_{g2}&\cdots&\boldsymbol{\Sigma}^{\star\star}_{gg}&\boldsymbol{\Sigma}^{\star\dagger}_{g1}&\boldsymbol{\Sigma}^{\star\dagger}_{g2}&\cdots&\boldsymbol{\Sigma}^{\star\dagger}_{gg}\\
    \hdashline[2pt/2pt]
    \boldsymbol{\Sigma}^{\dagger\star}_{11}&\boldsymbol{\Sigma}^{\dagger\star}_{12}&\cdots&\boldsymbol{\Sigma}^{\dagger\star}_{1g}&\boldsymbol{\Sigma}^{\dagger\dagger}_{11}&\boldsymbol{\Sigma}^{\dagger\dagger}_{12}&\cdots&\boldsymbol{\Sigma}^{\dagger\dagger}_{1g}\\
    \boldsymbol{\Sigma}^{\dagger\star}_{21}&\boldsymbol{\Sigma}^{\dagger\star}_{22}&\cdots&\boldsymbol{\Sigma}^{\dagger\star}_{2g}&\boldsymbol{\Sigma}^{\dagger\dagger}_{21}&\boldsymbol{\Sigma}^{\dagger\dagger}_{22}&\cdots&\boldsymbol{\Sigma}^{\dagger\dagger}_{2g}\\
    \vdots&\vdots&\ddots&\vdots&\vdots&\vdots&\ddots&\vdots\\
    \boldsymbol{\Sigma}^{\dagger\star}_{g1}&\boldsymbol{\Sigma}^{\dagger\star}_{g2}&\cdots&\boldsymbol{\Sigma}^{\dagger\star}_{gg}&\boldsymbol{\Sigma}^{\dagger\dagger}_{g1}&\boldsymbol{\Sigma}^{\dagger\dagger}_{g2}&\cdots&\boldsymbol{\Sigma}^{\dagger\dagger}_{gg}\\
\end{array}\right).
$$
By (\ref{G_S}), we have correspondingly $$\textbf{G}_\textbf{S}=\left(
\begin{array}{c;{2pt/2pt}c}
					\textbf{G}_{\textbf{S}^\star}^\star&\textbf{G}_{\textbf{S}}^{\star\dagger}\\
					\hdashline[2pt/2pt](\textbf{G}_{\textbf{S}}^{\star\dagger})^T&\textbf{G}_{\textbf{S}}^\dagger\\
				\end{array}\right)$$
    where 
    \begin{align*}
    \textbf{G}_{\textbf{S}^\star}^\star&=    \left(\begin{array}{cc}\boldsymbol{\Sigma}^\star&\boldsymbol{\Sigma}^\star-\textbf{S}^\star\\ \boldsymbol{\Sigma}^\star-\textbf{S}^\star&\boldsymbol{\Sigma}^\star\end{array}\right),\quad \textbf{S}^\star=\text{diag}(\textbf{S}^{\star}_1,\textbf{S}^{\star}_2,\ldots,\textbf{S}^{\star}_g),\\
    \textbf{G}_{\textbf{S}^\star}^{\star\dagger}&=    \left(\begin{array}{cc}\boldsymbol{\Sigma}^{\star\dagger}&\boldsymbol{\Sigma}^{\star\dagger}-\textbf{S}^{\star\dagger}\\ \boldsymbol{\Sigma}^{\star\dagger}-\textbf{S}^{\star\dagger}&\boldsymbol{\Sigma}^{\star\dagger}\end{array}\right),\quad \textbf{S}^{\star\dagger}=\text{diag}(\textbf{S}^{{\star\dagger}}_1,\textbf{S}^{{\star\dagger}}_2,\ldots,\textbf{S}^{{\star\dagger}}_g),\\
    \textbf{G}_{\textbf{S}}^{\dagger}&=    \left(\begin{array}{cc}\boldsymbol{\Sigma}^{\dagger}&\boldsymbol{\Sigma}^{\dagger}-\textbf{S}^{\dagger}\\ \boldsymbol{\Sigma}^{\dagger}-\textbf{S}^{\dagger}&\boldsymbol{\Sigma}^{\dagger}\end{array}\right),\quad \textbf{S}^{\dagger}=\text{diag}(\textbf{S}^{{\dagger}}_1,\textbf{S}^{{\dagger}}_2,\ldots,\textbf{S}^{{\dagger}}_g).
    \end{align*}
Consider a transformation matrix $$\textbf{Q}=\left(
\begin{array}{cc;{2pt/2pt}cc}
					\textbf{I}&\textbf{0}&\textbf{Q}^{\star\dagger}&\textbf{0}\\
\textbf{0}&\textbf{I}&\textbf{0}&\textbf{Q}^{\star\dagger}\\
\hdashline[2pt/2pt]
\textbf{0}&\textbf{0}&\textbf{I}&\textbf{0}\\
\textbf{0}&\textbf{0}&\textbf{0}&\textbf{I}\\
				\end{array}\right),\quad \text{where } \textbf{Q}^{\star\dagger}=-\text{diag}((\boldsymbol{\Sigma}^{\star\star}_{11})^{-1}\boldsymbol{\Sigma}^{\star\dagger}_{11},(\boldsymbol{\Sigma}^{\star\star}_{22})^{-1}\boldsymbol{\Sigma}^{\star\dagger}_{22},\ldots,(\boldsymbol{\Sigma}^{\star\star}_{gg})^{-1}\boldsymbol{\Sigma}^{\star\dagger}_{gg}).$$
    Because $X$ has group key conditional independence property with respect to $\{\mathcal{A}_\gamma\}_{\gamma=1}^g$, we have for any $\gamma_1\neq \gamma_2$,
    $$\begin{cases}
    \boldsymbol{\Sigma}^{\star\dagger}_{\gamma_1\gamma_2}=\boldsymbol{\Sigma}^{\star\star}_{\gamma_1\gamma_2}(\boldsymbol{\Sigma}^{\star\star}_{\gamma_2\gamma_2})^{-1}\boldsymbol{\Sigma}^{\star\dagger}_{\gamma_2\gamma_2}
    \\\boldsymbol{\Sigma}^{\dagger\dagger}_{\gamma_1\gamma_2}=\boldsymbol{\Sigma}^{\dagger\star}_{\gamma_1\gamma_2}(\boldsymbol{\Sigma}^{\star\star}_{\gamma_2\gamma_2})^{-1}\boldsymbol{\Sigma}^{\star\dagger}_{\gamma_2\gamma_2}.
    \end{cases}$$
Therefore, 
\begin{equation}\label{help}
    \begin{cases}
    \boldsymbol{\Sigma}^\star\textbf{Q}^{\star\dagger}+\boldsymbol{\Sigma}^{\star\dagger}=\textbf{0}\\
    (\boldsymbol{\Sigma}^{\star\dagger})^T\textbf{Q}^{\star\dagger}+\boldsymbol{\Sigma}^{\dagger}=\text{diag}(\boldsymbol{\Sigma}^{\dagger\dagger|\star}_{11|1},\boldsymbol{\Sigma}^{\dagger\dagger|\star}_{22|2},\ldots,\boldsymbol{\Sigma}^{\dagger\dagger|\star}_{gg|g}),
\end{cases}
\end{equation}
leading to a transformed variance-covariance matrix $$\check{\textbf{G}}_\textbf{S}=\textbf{Q}^T\textbf{G}_\textbf{S}\textbf{Q}=\left(
\begin{array}{c;{2pt/2pt}c}
					\textbf{G}_{\textbf{S}^\star}^\star&\check{\textbf{G}}_{\textbf{S}}^{\star\dagger}\\
					\hdashline[2pt/2pt](\check{\textbf{G}}_{\textbf{S}}^{\star\dagger})^T&\check{\textbf{G}}_{\textbf{S}}^\dagger\\
				\end{array}\right)$$
where 
    \begin{align*}    \check{\textbf{G}}_{\textbf{S}^\star}^{\star\dagger}&=    \left(\begin{array}{cc}\textbf{0}&-\check{\textbf{S}}^{\star\dagger}\\ -\check{\textbf{S}}^{\star\dagger}&\textbf{0}\end{array}\right),\quad\check{\textbf{S}}^{\star\dagger}=\textbf{S}^\star\textbf{Q}^{\star\dagger}+\textbf{S}^{\star\dagger},\\
    \check{\textbf{G}}_{\textbf{S}}^{\dagger}&=    \left(\begin{array}{cc}\text{diag}(\boldsymbol{\Sigma}^{\dagger\dagger|\star}_{11|1},\boldsymbol{\Sigma}^{\dagger\dagger|\star}_{22|2},\ldots,\boldsymbol{\Sigma}^{\dagger\dagger|\star}_{gg|g})&\text{diag}(\boldsymbol{\Sigma}^{\dagger\dagger|\star}_{11|1},\boldsymbol{\Sigma}^{\dagger\dagger|\star}_{22|2},\ldots,\boldsymbol{\Sigma}^{\dagger\dagger|\star}_{gg|g})-\check{\textbf{S}}^{\dagger}\\ \text{diag}(\boldsymbol{\Sigma}^{\dagger\dagger|\star}_{11|1},\boldsymbol{\Sigma}^{\dagger\dagger|\star}_{22|2},\ldots,\boldsymbol{\Sigma}^{\dagger\dagger|\star}_{gg|g})-\check{\textbf{S}}^{\dagger}&\text{diag}(\boldsymbol{\Sigma}^{\dagger\dagger|\star}_{11|1},\boldsymbol{\Sigma}^{\dagger\dagger|\star}_{22|2},\ldots,\boldsymbol{\Sigma}^{\dagger\dagger|\star}_{gg|g})\end{array}\right),\\ \check{\textbf{S}}^{\dagger}&=(\textbf{Q}^{\star\dagger})^T\textbf{S}^\star\textbf{Q}^{\star\dagger}+(\textbf{Q}^{\star\dagger})^T\textbf{S}^{\star\dagger}+(\textbf{S}^{\star\dagger})^T\textbf{Q}^{\star\dagger}+\textbf{S}^{\dagger},
    \end{align*}

As a result, we observe that
\begin{enumerate}
    \item Minimizing $L_{\text{ME}}(\textbf{S})=\log\det(\textbf{G}^{-1}_\textbf{S})$ is equivalent to maximizing $\det(\textbf{G}_\textbf{S})$.
    \item Since $\det(\textbf{Q})=1$, we have $\det(\check{\textbf{G}}_\textbf{S})=\det(\textbf{G}_\textbf{S})$ and thus minimizing $L_{\text{ME}}$ is equivalent to maximizing $\det(\check{\textbf{G}}_\textbf{S})$.
    \item There exists a one-to-one correspondence between $(\textbf{S}^{\star},\textbf{S}^{\star\dagger},\textbf{S}^{\dagger})$ and $(\textbf{S}^{\star},\check{\textbf{S}}^{\star\dagger},\check{\textbf{S}}^{\dagger})$.
    \item For any $(\textbf{S}^{\star},\check{\textbf{S}}^{\dagger})$, $\det(\check{\textbf{G}}_\textbf{S})$ is maximized if and only if $\check{\textbf{S}}^{\star\dagger}=\textbf{0}$ or equivalently,
    \begin{equation}
        \label{solution1}
        \textbf{S}^{{\star\dagger}}=-\textbf{S}^{\star}\textbf{Q}^{\star\dagger}.
    \end{equation}
    \item Given (\ref{solution1}), for any $\textbf{S}^{\star}$, $\det(\check{\textbf{G}}_\textbf{S})$ is maximized if and only if $\check{\textbf{S}}^{\dagger}=\text{diag}(\boldsymbol{\Sigma}^{\dagger\dagger|\star}_{11|1},\boldsymbol{\Sigma}^{\dagger\dagger|\star}_{22|2},\ldots,\boldsymbol{\Sigma}^{\dagger\dagger|\star}_{gg|g})$ or equivalently,
    \begin{equation}
        \label{solution2}
        \begin{aligned}[b]
            \textbf{S}^{{\dagger}}&=\text{diag}(\boldsymbol{\Sigma}^{\dagger\dagger|\star}_{11|1},\boldsymbol{\Sigma}^{\dagger\dagger|\star}_{22|2},\ldots,\boldsymbol{\Sigma}^{\dagger\dagger|\star}_{gg|g})-(\textbf{Q}^{\star\dagger})^T\textbf{S}^{\star}-\textbf{S}^{\star}\textbf{Q}^{\star\dagger}-(\textbf{Q}^{\star\dagger})^T\textbf{S}^{\star}\textbf{Q}^{\star\dagger}\\
            &=\text{diag}(\boldsymbol{\Sigma}^{\dagger\dagger|\star}_{11|1},\boldsymbol{\Sigma}^{\dagger\dagger|\star}_{22|2},\ldots,\boldsymbol{\Sigma}^{\dagger\dagger|\star}_{gg|g})+(\textbf{Q}^{\star\dagger})^T\textbf{S}^{\star}\textbf{Q}^{\star\dagger}
        \end{aligned}
    \end{equation}
    for $\gamma=1,\ldots,g$.
    \item Given (\ref{solution1})-(\ref{solution2}), $\det(\check{\textbf{G}}_\textbf{S})$ is maximized if and only if $\det(\textbf{G}_{\textbf{S}^\star}^\star)$ is maximized or equivalently, 
    \begin{equation}
        \label{solution3}
        L^\star_{\text{ME}}(S^\star)\text{ is minimized.}
    \end{equation}
\end{enumerate}
In summary, (\ref{solution1})-(\ref{solution3}) are
sufficient and necessary conditions of the minimizer of $L_{\text{ME}}(\textbf{S})=\log\det(\textbf{G}^{-1}_\textbf{S})$ if $X$ has the group key conditional independence property with respect to $\{\mathcal{A}_\gamma\}_{\gamma=1}^g$.

In the following, we need to show that the variance-covariance matrix of $(X,\tilde{X})$ generated according to Algorithm \ref{alg:Groupknockoff} and minimizing {\rm $L^\star_{\text{ME}}(S^\star)$} satisfies (\ref{solution1})-(\ref{solution3}).
\begin{itemize}
\item It is trivial that (\ref{solution3}) is satisfied.
    \item Because $(X,\tilde{X})$ generated according to Algorithm \ref{alg:Groupknockoff}
    has the group key conditional independence property with respect to $\{\mathcal{A}_\gamma,\tilde{\mathcal{A}}_\gamma\}_{\gamma=1}^g$, we have by (\ref{help}),
    \begin{align*}
        \text{Cov}\left\{(X_\star,\tilde{X}_\star),(X_\dagger,\tilde{X}_\dagger)\right\}&=\text{Cov}\left\{(X_\star,\tilde{X}_\star),(X_\star,\tilde{X}_\star)\right\}\begin{pmatrix}
        -\textbf{Q}^{\star\dagger}&\textbf{0}\\
        \textbf{0}&-\textbf{Q}^{\star\dagger}
    \end{pmatrix}\\
    &=\left(\begin{array}{cc}\boldsymbol{\Sigma}^\star&\boldsymbol{\Sigma}^\star-\textbf{S}^\star\\ \boldsymbol{\Sigma}^\star-\textbf{S}^\star&\boldsymbol{\Sigma}^\star\end{array}\right)\begin{pmatrix}
        -\textbf{Q}^{\star\dagger}&\textbf{0}\\
        \textbf{0}&-\textbf{Q}^{\star\dagger}
    \end{pmatrix}\\
    &=    \left(\begin{array}{cc}\boldsymbol{\Sigma}^{\star\dagger}&\boldsymbol{\Sigma}^{\star\dagger}-(-\textbf{S}^{\star\dagger}\textbf{Q}^{\star\dagger})\\ \boldsymbol{\Sigma}^{\star\dagger}-(-\textbf{S}^{\star\dagger}\textbf{Q}^{\star\dagger})&\boldsymbol{\Sigma}^{\star\dagger}\end{array}\right),
    \end{align*}
    and thus (\ref{solution1}) is satisfied.
    In addition, by (\ref{help}),
    \begin{align*}
        &\text{Cov}\left\{(X_\dagger,\tilde{X}_\dagger),(X_\dagger,\tilde{X}_\dagger)\right\}\\=&\begin{pmatrix}
        -(\textbf{Q}^{\star\dagger})^T&\textbf{0}\\
        \textbf{0}&-(\textbf{Q}^{\star\dagger})^T
    \end{pmatrix}\text{Cov}\left\{(X_\star,\tilde{X}_\star),(X_\dagger,\tilde{X}_\dagger)\right\}\\
    &+\left(\begin{array}{cc}\text{diag}(\boldsymbol{\Sigma}^{\dagger\dagger|\star}_{11|1},\boldsymbol{\Sigma}^{\dagger\dagger|\star}_{22|2},\ldots,\boldsymbol{\Sigma}^{\dagger\dagger|\star}_{gg|g})&\textbf{0}\\ \textbf{0}&\text{diag}(\boldsymbol{\Sigma}^{\dagger\dagger|\star}_{11|1},\boldsymbol{\Sigma}^{\dagger\dagger|\star}_{22|2},\ldots,\boldsymbol{\Sigma}^{\dagger\dagger|\star}_{gg|g})\end{array}\right)\\
    =&    \begin{pmatrix}
        -(\textbf{Q}^{\star\dagger})^T&\textbf{0}\\
        \textbf{0}&-(\textbf{Q}^{\star\dagger})^T
    \end{pmatrix}\left(\begin{array}{cc}\boldsymbol{\Sigma}^{\star\dagger}&\boldsymbol{\Sigma}^{\star\dagger}-(-\textbf{S}^{\star\dagger}\textbf{Q}^{\star\dagger})\\ \boldsymbol{\Sigma}^{\star\dagger}-(-\textbf{S}^{\star\dagger}\textbf{Q}^{\star\dagger})&\boldsymbol{\Sigma}^{\star\dagger}\end{array}\right)\\
    &+\left(\begin{array}{cc}\text{diag}(\boldsymbol{\Sigma}^{\dagger\dagger|\star}_{11|1},\boldsymbol{\Sigma}^{\dagger\dagger|\star}_{22|2},\ldots,\boldsymbol{\Sigma}^{\dagger\dagger|\star}_{gg|g})&\textbf{0}\\ \textbf{0}&\text{diag}(\boldsymbol{\Sigma}^{\dagger\dagger|\star}_{11|1},\boldsymbol{\Sigma}^{\dagger\dagger|\star}_{22|2},\ldots,\boldsymbol{\Sigma}^{\dagger\dagger|\star}_{gg|g})\end{array}\right)\\
    = &   {\footnotesize\left(\begin{array}{cc}\boldsymbol{\Sigma}^{\dagger}&\boldsymbol{\Sigma}^{\dagger}-((\textbf{Q}^{\star\dagger})^T\textbf{S}^{\star}\textbf{Q}^{\star\dagger}+\text{diag}(\boldsymbol{\Sigma}^{\dagger\dagger|\star}_{11|1},\boldsymbol{\Sigma}^{\dagger\dagger|\star}_{22|2},\ldots,\boldsymbol{\Sigma}^{\dagger\dagger|\star}_{gg|g}))\\ \boldsymbol{\Sigma}^{\dagger}-((\textbf{Q}^{\star\dagger})^T\textbf{S}^{\star}\textbf{Q}^{\star\dagger}+\text{diag}(\boldsymbol{\Sigma}^{\dagger\dagger|\star}_{11|1},\boldsymbol{\Sigma}^{\dagger\dagger|\star}_{22|2},\ldots,\boldsymbol{\Sigma}^{\dagger\dagger|\star}_{gg|g}))&\boldsymbol{\Sigma}^{\dagger}\end{array}\right)}
    \end{align*}
    and thus (\ref{solution2}) is satisfied.
\end{itemize}

\section{Heuristic strategy to identify groups and group-key variables to approximately achieve conditional independence}\label{heuristic}

There are many ways for defining groups and selecting their representatives. In this section, we present 2 plausible strategies for constructing groups using interpolative decomposition or hierarchical clustering. Then we discuss the intuition of Algorithm \ref{alg:rep} for selecting representatives from each group. 

\subsection{Defining groups by Hierarchical clustering}\label{hcgroups}

The standard approach for defining groups is to use hierarchical clustering, e.g. as featured in \citep{sesia2021false}. In this approach, we use the absolute value of the empirical correlation matrix as the distance matrix
\begin{align*}
    d_{ij} = 1 - |corr(X_i, X_j)| = 1 - |\Sigma_{ij}|,
\end{align*}
where $d_{ij} \in [0, 1]$ represents the distance between features $i$ and $j$. Thus, $d_{ii} = 0$, and more correlated variables are considered ``closer". Groups are defined by applying (single, average, or complete linkage) hierarchical clustering to the resulting distance matrix, with some correlation cutoff. When contiguous groups are desired, we can use adjacency-constrained hierarchical clustering rather than standard hierarchical clustering. 

\subsection{Defining groups by Interpolative Decomposition (ID)}

Given design matrix $\bX \in \mathbb{R}^{n\times p}$ with highly correlated columns, the ID approach selects a subset $J \subset \{1,...,p\}$ of $k$ columns from $\bX$ such that
\begin{eqnarray*}
    \underset{n\times (p-k)}{\bX_{J^c}} &\approx& \underset{n \times k}{\bX_J} \amp \times \amp \underset{k \times (p-k)}{\bT}.
\end{eqnarray*}
The selected columns in $J$ are sometimes called the \textit{skeleton} columns, the non selected columns $J^c = \{1,...,p\} \backslash J$ are called the \textit{redundant} columns, and $\bT$ is the interpolation matrix. Thus, ID is trying to represent the redundant columns via the skeleton columns. 

To define groups, we will let the $J$ skeleton columns be ``group centers". For a variable $i$ that is not a cluster center, we put it in the same group as cluster center $j$ that is most correlated with $i$, i.e. $\max_j |corr(x_i, x_j)|$. If we would like to enforce contiguous groups, we simply assign $i$ to its left or right center, whichever is more correlated with $i$, without breaking the adjacency constraint. 

\subsubsection{Interpolative decomposition for covariance matrices}

In some applications we only have the covariance matrix $\bSigma$. To define groups, we can compute the Cholesky $\bSigma = \bA^t\bA$ where $\bA$ is an upper triangular matrix, and apply the ID procedure to $\bA$. This produces the following algorithm
\begin{enumerate}
    \item Let $\bSigma$ be the correlation matrix and $S$ be the set of selected representatives
    \item Compute Cholesky $\bSigma = \bA^t\bA$ where $\bA$ is upper triangular
    \item Apply interpolative decomposition to $\bA = \bC\bZ$ where $\bC$ is a column permutation of $\bA$ and $\bZ$ is some interpolation matrix
    \item Choose $S$ to be first $k$ columns of $\bC$
    \item Increase $k$ until: for all $j \notin S, max(c_j) < 0.25$ where $c_j = \Sigma_{jj} - \bSigma_{j,S}\bSigma_{SS}^{-1}\bSigma_{S,j}$
\end{enumerate}

Here the constant $c_j$ is the sum of squares residuals of regression the $j$ variant on those in $S$. To see why, let $\ba_j$ denote the $j$th column of $\bA$, $\bA_S$ denote the columns of $\bA$ that are in $S$, then
\begin{eqnarray*}
    c_j 
    &=& (\ba_j - \bA_S\hat{\bbeta})^t(\ba_j - \bA_S\hat{\bbeta})\\
    &=& (\ba_j - \bA_S(\bA_S^t\bA_S)^{-1}\bA_S\ba_j)^t(\ba_j - \bA_S(\bA_S^t\bA_S)^{-1}\bA_S\ba_j)\\
    &=& \ba_j^t\ba_j - 2\ba_j^t\bA_S^t(\bA_S^t\bA_S)^{-1}\bA_S\ba_j + \ba_j^t\bA_S^t(\bA_S^t\bA_S)^{-1}\bA_S^t\bA_S(\bA_S^t\bA_S)^{-1}\bA_S\ba_j\\
    &=& \ba_j^t\ba_j - \ba_j^t\bA_S^t(\bA_S^t\bA_S)^{-1}\bA_S\ba_j\\
    &=& \Sigma_{jj} - \bSigma_{j,S}\bSigma_{S,S}^{-1}\bSigma_{S,j}.
\end{eqnarray*}
At first glance, step (5) requires computing $\bSigma_{S,S}^{-1}$ afresh. However, assuming we have its expression at the previous $k$, we can compute $\bSigma_{S,S}^{-1}$ by taking advantage of the block-matrix inversion formula which circumvents inverting a $k \times k$ matrix.

\subsection{Selecting group-key variables to exploit conditional independence}

Algorithm \ref{alg:rep} provides a heuristic algorithm for identifying group-key conditional independence described in Definition \eqref{gkci}, somewhat motivated by the algorithms presented in \cite{sood2023statistical}. It proceeds as follows.

 Firstly, we can think of $\mathcal{A}_\gamma^\star$ as \textit{key}  variables selected for group $\gamma$, and $\mathcal{A}_\gamma^\dagger$ are the non-selected variables (i.e. redundant variables) in group $\gamma$. Consider variant $j$ belonging in group $\gamma$ but not yet selected as a key. The quantity
\begin{eqnarray*}
    \eta_j &=& \bSigma_{j,\mathcal{A}_\gamma^\star}
    \bSigma_{\mathcal{A}_\gamma^\star,\mathcal{A}_\gamma^\star}^{-1}
    \bSigma_{\mathcal{A}_\gamma^\star,j}
\end{eqnarray*}
is the variation of the $j$th variant explained by key variants in group $\gamma$. To see why, consider regressing the $j$th variant on those in $\mathcal{A}_\gamma^\star$. Letting $\bH$ denote the hat matrix and $\hat{\bx}_j$ the predicted values of regressing $\bx_j$ onto those in $\mathcal{A}_\gamma^\star$, we have
\begin{align*}
    \bSigma_{j,\mathcal{A}_\gamma^\star}\bSigma_{\mathcal{A}_\gamma^\star,\mathcal{A}_\gamma^\star}^{-1}\bSigma_{\mathcal{A}_\gamma^\star,j}
    &= \bx_j^t\bX_{\mathcal{A}_\gamma^\star}(\bX_{\mathcal{A}_\gamma^\star}^t\bX_{\mathcal{A}_\gamma^\star})^{-1}\bX_{\mathcal{A}_\gamma^\star}^t\bx_j = \bx_j^t\bH\bx_j = \bx_j^t\bH^2\bx_j
    = \|\bH\bx_j\|^2 = \|\hat{\bx}_j\|^2.
\end{align*}
Analogously, the quantity $\zeta_j = \bSigma_{j,-\mathcal{A}_\gamma^\dagger}\bSigma_{-\mathcal{A}_\gamma^\dagger,-\mathcal{A}_\gamma^\dagger}^{-1}\bSigma_{-\mathcal{A}_\gamma^\dagger,j}$ is the variation of the $j$th variant explained by $\mathcal{A}_\gamma^\star \cup -\mathcal{A}_\gamma$ (i.e. key variables plus all variants outside group $\gamma$). Thus, $\frac{\eta_j}{\zeta_j}$ is the proportion of variation explained by $\mathcal{A}_\gamma^\star$ vs variation explained by $\mathcal{A}_\gamma^\star \cup -\mathcal{A}_\gamma$, which is supposed to be 1 under the conditional independence assumption in Definition \ref{gkci}. The algorithm proceeds by increasing the number of key variables until
\begin{eqnarray*}
    \underset{j \notin \mathcal{A}_\gamma^\star, j \in \mathcal{A}_\gamma}{\text{mean }} \  \frac{\bSigma_{j,\mathcal{A}_\gamma^\star}\bSigma_{\mathcal{A}_\gamma^\star,\mathcal{A}_\gamma^\star}^{-1}\bSigma_{\mathcal{A}_\gamma^\star,j}}{\bSigma_{j,-\mathcal{A}_\gamma^\dagger}\bSigma^{-1}_{-\mathcal{A}_\gamma^\dagger,-\mathcal{A}_\gamma^\dagger}\bSigma_{-\mathcal{A}_\gamma^\dagger,j}} &\ge& c \in [0, 1].
\end{eqnarray*}
When this condition in not met, we search through all variables $j$ in $\mathcal{A}_g^\dagger$ and find the one that can explain the most amount of the remaining variation, elect that variant as a member of $\mathcal{A}_g^\star$, and repeat the process.

{\begin{algorithm}
		\caption{Searching for covariates to explain most of between-group dependencies with respect to the group structure $\{\mathcal{A}_\gamma:\gamma\in [g]\}$.}\label{alg:rep}
		\begin{algorithmic}[1]
			\STATE { \textbf{Input:} Random variables $X$, covariance matrix $\bSigma$ and the threshold $c\in [0,1]$ (the target proportion of between-group dependencies to be explained).}
			\FOR{$\gamma\in [g]$}
			\STATE Initialize $\mathcal{A}^\star_\gamma=\emptyset$ and $\mathcal{A}^\dagger_\gamma=\mathcal{A}_\gamma$.
			\FOR{$j\in \mathcal{A}^\dagger_\gamma$}
			\STATE Compute $\eta_j=0$ and  $\zeta_j={\bSigma}_{j,-\mathcal{A}^\dagger_\gamma}{\bSigma}_{-\mathcal{A}^\dagger_\gamma,-\mathcal{A}^\dagger_\gamma}^{-1}{\bSigma}_{-\mathcal{A}^\dagger_\gamma,j}$.
			\ENDFOR
			\WHILE{$\sum_{j\in \mathcal{A}^\dagger_\gamma}\eta_j/\zeta_j<c|\mathcal{A}^\dagger_\gamma|$} 
		\STATE Find $j^*$ via
		$$j^*=\arg\max_{j\in \mathcal{A}^\dagger_\gamma}\sum_{j^\dagger\in \mathcal{A}^\dagger_\gamma\setminus\{j\}}{{\bSigma}_{j^\dagger,\mathcal{A}^\star_\gamma\cup\{j\}}{\bSigma}_{\mathcal{A}^\star_\gamma\cup\{j\},\mathcal{A}^\star_\gamma\cup\{j\}}^{-1}{\bSigma}_{\mathcal{A}^\star_\gamma\cup\{j\},j^\dagger}}.$$
		\STATE Update $\mathcal{A}^\star_\gamma\leftarrow \mathcal{A}^\star_\gamma\cup\{j^*\}$ and $\mathcal{A}^\dagger_\gamma\leftarrow \mathcal{A}^\dagger_\gamma\setminus\{j^*\}$.
		\FOR{$j\in  \mathcal{A}^\dagger_\gamma$}
		\STATE Update $\eta_j={\bSigma}_{j, \mathcal{A}^\star_\gamma}{\bSigma}_{ \mathcal{A}^\star_\gamma, \mathcal{A}^\star_\gamma}^{-1}{\bSigma}_{ \mathcal{A}^\star_\gamma,j}$ and $\zeta_j={\bSigma}_{j,-\mathcal{A}^\dagger_\gamma}{\bSigma}_{-\mathcal{A}^\dagger_\gamma,-\mathcal{A}^\dagger_\gamma}^{-1}{\bSigma}_{-\mathcal{A}^\dagger_\gamma,j}$
		\ENDFOR
		\ENDWHILE
		\ENDFOR
		\STATE \textbf{Output:} $\{\mathcal{A}^\star_\gamma:\gamma\in [g]\}$.
	\end{algorithmic}
\end{algorithm}}

\section{Practical strategy for estimating $\hat{\bmu}$ and $\hat{\bSigma}$}\label{sec:estimate_mu_and_sigma}

In practice, we are often given individual level data $\bX \in \mathbb{R}^{n \times p}$ and asked to generate second-order model-X knockoffs. This relies on obtaining suitable estimates for $\hat{\bmu} \in \mathbb{R}^p$ and $\hat{\bSigma}\in\mathbb{R}^{p\times p}$. We always use the sample mean $\mu_j = \sum_i X_{ij} / n$ to estimate $\hat{\bmu}$, but when $p \gg n$, the sample covariance or maximum-likelihood based estimators are ill-suited to estimate $\hat{\bSigma}$ \citep{schafer2005shrinkage}. 

In \texttt{Knockoffs.jl}, we use a linear shrinkage estimator of the form 
\begin{align*}
	\hat{\bSigma} &= (1 - \lambda)\bS + \lambda \bF
\end{align*}
where $\bF$ is a \textit{target} matrix of appropriate dimensions, $\lambda \in [0,1]$ is a shrinkage intensity, and $\bS$ is the sample covariance estimator. Several choises of $\bF$ are possible \citep{schafer2005shrinkage}. By default, we use the common choice $F_{ii} = S_{ii}$ and $F_{ij} = 0$ and compute $\lambda$ via Lediot-Wolf shrinkage \citep{ledoit2003honey}. These features, including more choices for $\bF$ and estimating $\lambda$, are implemented in the Julia package \texttt{CovarianceEstimation.jl}.

\section{Genotype variance covariances, groups and $\star$ variables identification from Pan-UKB}\label{sec:LDmatrix_summarystats}

\begin{figure}
    \centering
    \includegraphics[width=0.5\linewidth]{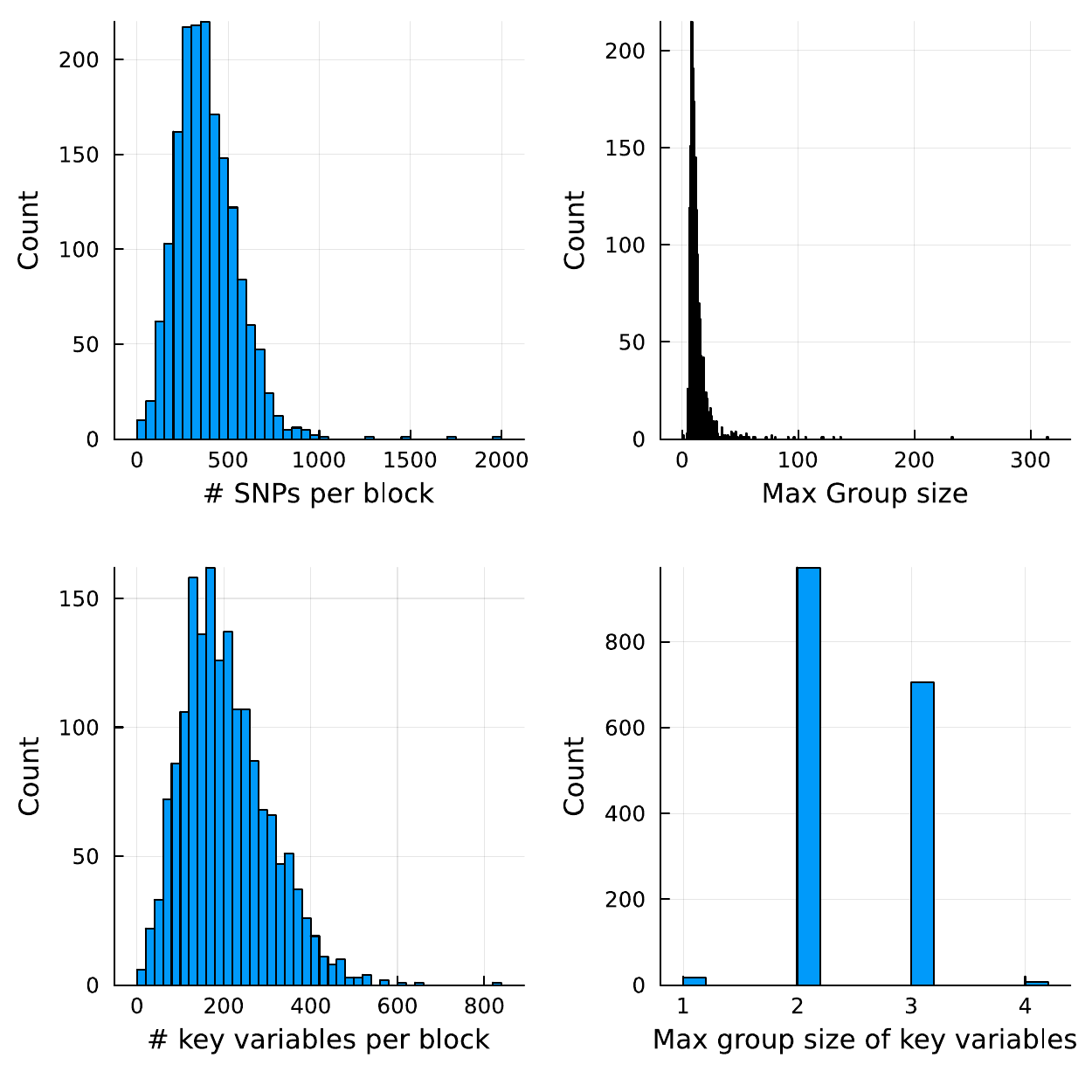}
    \caption{Summary statistics for the Pan-UKBB LD panels after partitioning into 1703 quasi-independent blocks. We restrict each region to typed SNPs present on the UKB-genotyping array. For each region, we compute the number of SNPs, number of representatives selected, maximum group size in each region (determined by average linkage hierarchical clustering with correlation cutoff 0.5), and the maximum group size in each region after selecting representatives.}
    \label{fig:LDsummary}
\end{figure}

Here we provide some summary statistics on the Pan-UKB matrices featured in our real data analysis of Albuminuria GWAS data.  Figures \ref{fig:LDsummary} provides some summary statistics on the 1703 independent blocks, restricting to the genotyped SNPs. Most blocks have around 500 SNPs per block, while possessing group sizes of up to 300 variables per group. After the identification of group-key variables by running Algorithm \eqref{alg:rep}, the number of (key) variables per block is reduced to around 200 SNPs per block, and the maximum group size becomes at most 4. Thus, the identification of group-key variables significantly reduced the number of parameters that need to be optimized ($>90,000$ to $16$ for the largest group) for GWAS summary statistics analysis. 



\section{Regularizations applied to Pan-UKB LD matrices}\label{sec:LDregularization}

Regularization to the LD matrices were done in multiple steps. 

First, given the downloaded $\hat{\bSigma}_{\text{PanUK}}$, we identified approximately independent  blocks of SNPs with the \texttt{ldetect} software \citep{berisa2016approximately}.  A total of 1703 blocks $\hat{\bSigma}_{1},...,\hat{\bSigma}_{1703}$ of size varying between   $\sim10^2$ and  $\sim 10^3$ were identified. Next, after reading the relevant portions of the data in $\hat{\bSigma}_{i}$, we force $\hat{\bSigma}_{i}$ to be positive definite by computing its eigen-decomposition and setting all eigenvalues to be $\ge 10^{-5}$. Next, we solve for $\bS_1,...,\bS_{1703}$ by applying Algorithm \eqref{alg:Groupknockoff} to each regularized $\hat{\bSigma}_i$ separately. This action delivers
\begin{eqnarray*}
    \bG_{\bS_i} &=& 
    \begin{bmatrix}
		\bSigma_i & \bSigma_i - \bS_i & \cdots & \bSigma_i - \bS_i\\
		\bSigma_i - \bS_i & \bSigma_i & \cdots & \bSigma_i - \bS_i\\
		\vdots & \vdots & \ddots & \vdots\\
		\bSigma_i - \bS_i & \cdots & \cdots & \bSigma_i
	\end{bmatrix} \in \bbR^{p_i(m + 1) \times p_i(m + 1)}.
\end{eqnarray*}
for each $i = \{1, ..., 1703\}$ which represents the covariance matrix for $(\bz_i,\tilde{\bz}_{i1},...,\tilde{\bz}_{im})$. This allows us to assemble the overall covariance matrix 
\begin{eqnarray*}
    \bA_{\text{unregularized}} &=& \begin{bmatrix}
        \bG_{\bS_1} & & \\
        & \ddots & \\
        & & \bG_{\bS_{1703}}
    \end{bmatrix} \in \mathbb{R}^{p(m+1) \times p(m+1)}.
\end{eqnarray*}
Finally, we set 
\begin{eqnarray}\label{eq:matrixA_for_lasso}
\bA &=& \bA_{\text{unregularized}} + 0.01\bI
\end{eqnarray}
and plug $\bA$ into the lasso solver \citep{yang2023ghostbasil}.

\section{Additional simulations using the Pan-UKB panel}

Here we use covariance matrices extracted from Pan-UKB \citep{panukbb} to conduct simulations. Given the pre-processed data $\bSigma_{1},...,\bSigma_{1703}$ described in section \ref{sec:obtaing_sigma_ukb}, we randomly select with replacement 500 covariances $\bSigma_{i}$ and generate corresponding design matrices $\bX_{i} \in \mathbb{R}^{n \times p_i}$ with $n=250$ independent samples. For each replicate, we assume $k=10$  non-zero effects, with  $\beta_j \sim N(0, 0.5)$ where the non-zero $\beta_j$s are randomly chosen across the $p_i$ features. Then the response is simulated as $\by_i = \bX_i\bbeta_i + N(\mathbf{0}, \mathbf{I}_{p \times p})$ as before. To explore to which extend the group-key conditional independence hypothesis is appropriate for these matrices, we first define groups with average linkage hierarchical clustering with correlation cutoff $0.5$, then identified key variables with Algorithm \ref{alg:rep} and considering four different levels for $c \in \{0.25, 0.5, 0.75, 1.0\}$. Note that $c=1$ is equivalent to not using the conditional independence assumption. Importantly, it is possible for a causal variant to not be selected as a key variable. We ran one simulation for each of the selected covariance matrices, and averaged the power/FDR across 500 simulations. 

\begin{figure}
    \centering
    \includegraphics[width=\linewidth]{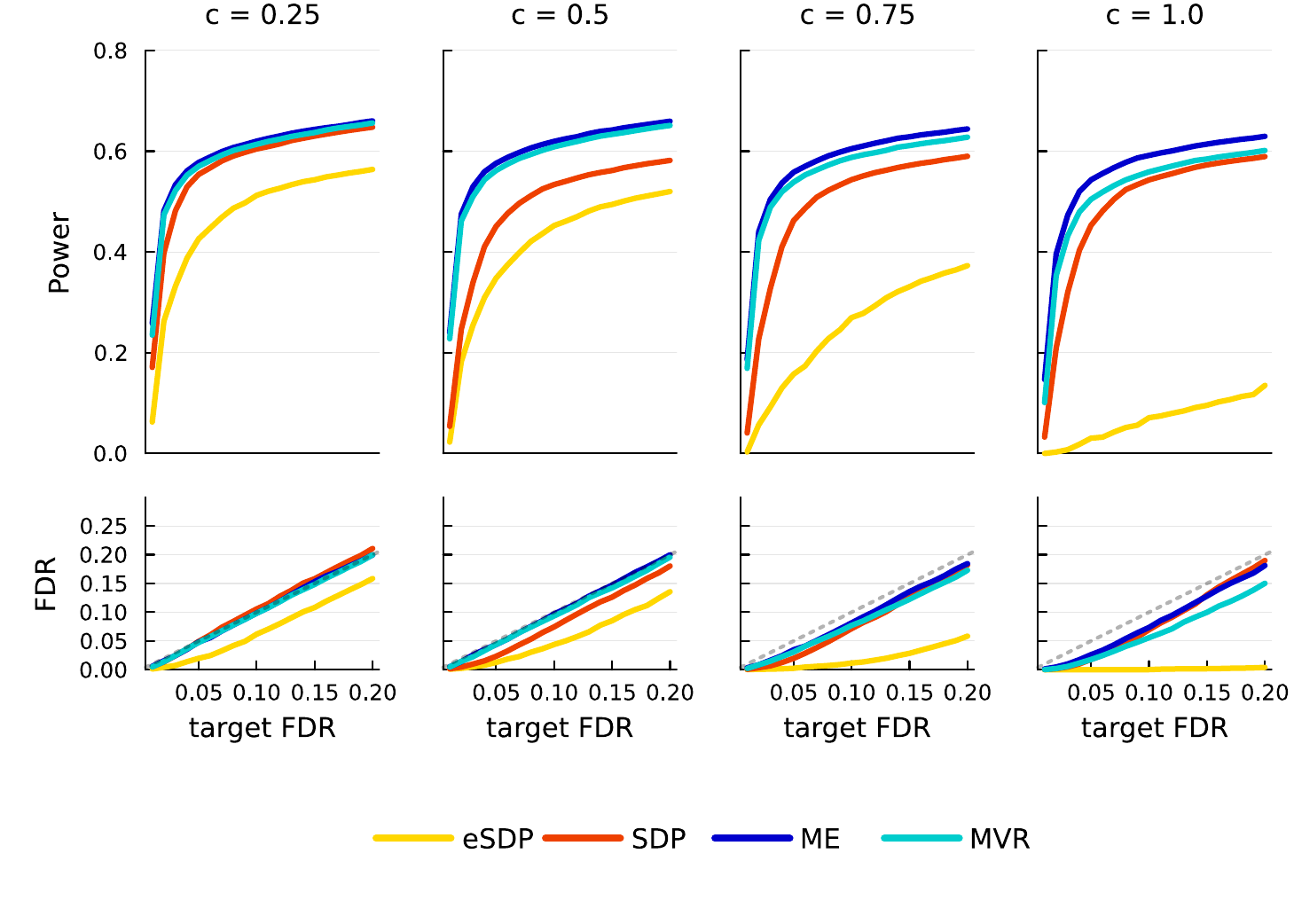}
    \caption{Power/FDR comparison of group knockoffs utilizing conditional independence assumption using 500 different covariance matrices extracted from the Pan-UKBB LD panel. FDR is controlled when the threshold $c$ for selecting group representatives exceeds 0.5. Note that a threshold of 1 corresponds to regular group knockoffs.}
    \label{fig:sim2}
\end{figure}

Figure \ref{fig:sim2} summarizes power and FDR level achieved in a simulation constructed starting from genetic variance-covariance matrices. In general, ME have the best power, followed by MVR, SDP, and finally eSDP. Utilizing conditional independence offers slightly better power than regular group knockoffs. Group-FDR can be controlled on GWAS data when the selected representatives explain at least 50\% of variation within groups, while a threshold of 25\% leads to slightly inflated empirical FDR. 

\section{Marginal correlation as feature importance statistics}\label{sec:marginal_importance_score}

\begin{figure}
    \centering
    \includegraphics[width=\linewidth]{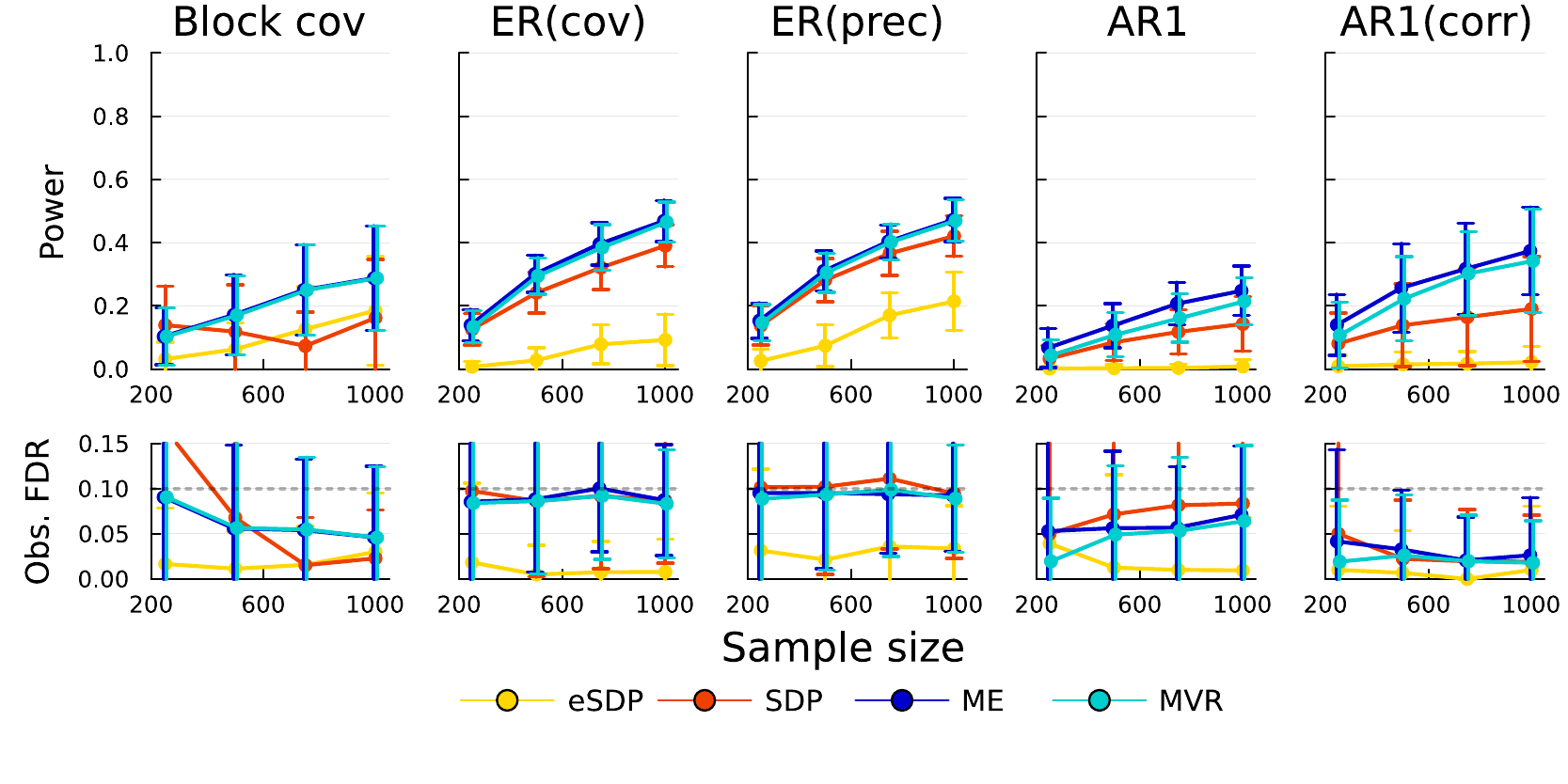}
    \caption{Power/FDR comparison of Equi, SDP, MVR, and ME group knockoffs on simulated covariances. This is the same simulation as Figure \ref{fig:sim1} except we use the squared marginal Z-statistics as feature importance scores. }
    \label{fig:marginal}
\end{figure}

Figure \ref{fig:sim1} explored the performance of our proposed algorithms using Lasso coefficient difference statistic as feature importance scores. In practice, marginal association are used more often in genome-wide association studies (GWAS). Therefore, here we use squared marginal correlation as importance measure. For each feature $i$, we compute its knockoff scores as
\begin{eqnarray*}
	T_i &=& \frac{1}{n}(\bx^t\by)^2, \quad \tilde{T}_i \amp=\amp \frac{1}{n}(\tilde{\bx}^t\by)^2,
\end{eqnarray*}
where both $\by$ and $\bx_i$ have been standardized to mean 0 variance 1. Then the feature importance score for each group $ \gamma$ can be computed as $Z_\gamma = \sum_{i \in \mathcal{A}_\gamma}T_i$. The result is presented in Figure \ref{fig:marginal}.

\section{Tuning Lasso hyperparameter without individual level data}\label{sec:pseudovalidation}

Optimizing the Lasso objective typically involves cross-validating for the hyper parameter $\lambda$. In the absence of individual level data, we adopt the pseudo-summary statistics approach \citep{mak2017polygenic,zhang2021improved}. Given $(\br, \tilde{\br})$ where $\br = \frac{1}{\sqrt{n}}\bz$ and $\tilde{\br} = \frac{1}{\sqrt{n}}\tilde{\bz}$ and the matrix $\bA$ in equation \eqref{eq:matrixA_for_lasso}, we create training and validation summary statistics
\begin{eqnarray*}
	\begin{pmatrix}
		\br \\ \tilde{\br}
	\end{pmatrix}_t
	&=&
	\begin{pmatrix}
		\br \\ \tilde{\br}
	\end{pmatrix}
	+\sqrt{\frac{n_v}{n \times n_t}}N(0, \bA),\\
	\begin{pmatrix}
		\br \\ \tilde{\br}
	\end{pmatrix}_v
	&=&
	\frac{1}{n_v}\left[n
	\begin{pmatrix}
		\br \\ \tilde{\br}
	\end{pmatrix}
	- 
	n_t
	\begin{pmatrix}
		\br \\ \tilde{\br}
	\end{pmatrix}_t
	\right].
\end{eqnarray*}
If the sample size is $n$, we let $n_t = 0.8n, n_v = 0.2n$, and choose $\lambda$ that maximizes
\begin{align*}
	f(\lambda) &= 
	\frac{\bbeta_{\lambda}^t \begin{pmatrix}
	\br\\\tilde{\br}
	\end{pmatrix}_v}{\sqrt{\bbeta_{\lambda}^t\bA\bbeta_\lambda}},
\end{align*} where $\bbeta_{\lambda}$ is trained on the training summary statistics.

\section{Full list of discoveries from Albuminuria GWAS}\label{sec:gwas_result_full}

Table \ref{table:ABtable1} list the SNPs discovered by our 
second-order Ghost knockoff analysis of Albuminuria GWAS. Note that for each discovered group, we only list the group-key variant with the most significant marginal Z score. The full result, including all SNPs within groups and non-significant SNPs, can be accessed from our online GitHub page. To compare against previous studies, we manually searched each discovered variant against the NHGRI-EBI GWAS catalog \citep{macarthur2017new} and list genes that have been mapped to the SNP. 

\begin{table}
\centering
\footnotesize
\begin{tabular}{c|c|c|c|c|c|c}
    \hline
    \hline
    rsID & Variant & AF & Z & p-value & W & Mapped gene \\
    \hline
rs12032996&1:33454985:G:A&0.162&-6.477&9.33E-11&0.0028&PHC2,TLR12P\\
rs2154319&1:41280098:T:C&0.218&-5.121&3.04E-07&0.0023&FOXO6,SCMH1\\
rs12727019&1:47495360:T:C&0.087&-6.131&8.75E-10&0.0036&\\
rs7540974&1:47503824:G:A&0.881&6.097&1.08E-09&0.0035&\\
rs471608&1:47519020:T:G&0.185&7.271&3.57E-13&0.0045&\\
rs934287&2:202843584:A:G&0.813&6.688&2.26E-11&0.003&ICA1L\\
rs1047891&2:210675783:C:A&0.316&-6.729&1.71E-11&0.0053&CPS1\\
rs4665972&2:27375230:T:C&0.607&-6.858&6.96E-12&0.0047&SNX17\\
rs17026396&2:85532004:T:C&0.43&-6.248&4.15E-10&0.0024&RPSAP22,PARTICL\\
rs1077216&3:46850671:C:T&0.069&5.396&6.81E-08&0.0022&MYL3\\
rs7670121&4:148207444:A:G&0.24&5.941&2.84E-09&0.0035&NR3C2\\
rs4109437&4:189848068:G:A&0.038&10.685&1.20E-26&0.0112&FRG1-DT\\
rs6831256&4:3471412:A:G&0.421&5.251&1.51E-07&0.0022&DOK7\\
rs10032549&4:76476862:A:G&0.537&-6.013&1.82E-09&0.0037&SHROOM3\\
rs1465405&5:148736639:T:G&0.248&-5.348&8.90E-08&0.0026&\\
rs4865796&5:53976834:G:A&0.692&5.764&8.20E-09&0.0027&ARL15\\
rs3776051&5:64993329:A:G&0.229&6.132&8.70E-10&0.0039&CWC27\\
rs6569648&6:130027974:C:T&0.759&5.379&7.50E-08&0.0027&L3MBTL3\\
rs9472138&6:43844025:C:T&0.291&5.156&2.52E-07&0.0026&VEGFA,LINC02537\\
rs11983745&7:100626055:T:C&0.2&-5.137&2.79E-07&0.0023&\\
rs4410790&7:17244953:T:C&0.634&10.394&2.63E-25&0.0106&AHR\\
rs3735533&7:27206274:T:C&0.926&6.053&1.42E-09&0.0021&HOTTIP\\
rs17321515&8:125474167:A:G&0.476&-5.687&1.29E-08&0.0032&TRIB1,LINC00861\\
rs1801239&10:16877053:T:C&0.104&16.025&8.51E-58&0.0118&CUBN\\
rs45551835&10:16890385:G:A&0.014&20.385&2.28E-92&0.0226&CUBN\\
rs116867125&10:16991505:G:A&0.019&7.487&7.06E-14&0.0066&CUBN\\
rs10824368&10:76115357:G:A&0.226&6.246&4.22E-10&0.0041&\\
rs7115200&11:72041114:T:G&0.44&5.309&1.10E-07&0.0026&NUMA1\\
rs3201&12:69579612:T:C&0.345&-6.573&4.92E-11&0.0044&\\
rs4902647&14:68787474:C:T&0.464&-5.691&1.26E-08&0.0036&ZFP36L1,RNU6-921P\\
rs1288775&15:45369480:T:A&0.256&-5.95&2.68E-09&0.0034&GATM\\
rs2472297&15:74735539:C:T&0.266&9.642&5.31E-22&0.0068&CYP1A1,CYP1A2\\
rs12150031&17:81453198:C:G&0.391&-5.389&7.09E-08&0.0027&\\
rs547629&18:26627838:A:G&0.607&5.143&2.71E-07&0.0023&\\
rs613872&18:55543071:G:T&0.826&-5.517&3.44E-08&0.0029&TCF4\\
rs33950747&19:35848345:C:T&0.075&4.956&7.21E-07&0.002&\\
rs117287096&19:40835508:G:A&0.022&-5.042&4.61E-07&0.0021&CYP2A6,CYP2F2P\\
rs4021&19:48750004:A:G&0.278&-6.184&6.25E-10&0.0023&FUT1\\
    \hline
    \hline
\end{tabular}
\caption{List of discovered conditionally-independent groups for Albuminuria GWAS. Variant = chromosome:position:ref-allele:alt-allele where position uses HG38 coordinates, AF = alternate allele frequency, Z = marginal Z score, W = Lasso coefficient difference statistic. }
\label{table:ABtable1}
\end{table}

\section{Distribution of $S_{ij}$ and minimum eigenvalues of $\bG_{\bS}$}

Other than comparing power and FDR, one often employs heuristics to assess the quality of knockoff solvers. One commonly used metric is to look at the distribution of the diagonal elements of $\bS$, where larger values are considered ``better". As discussed in \cite{spector2022powerful}, this heuristic could fail and very large $\bS$ values can imply \textit{lower} power. 

\begin{figure}
    \centering
    \includegraphics[width=0.6\linewidth]{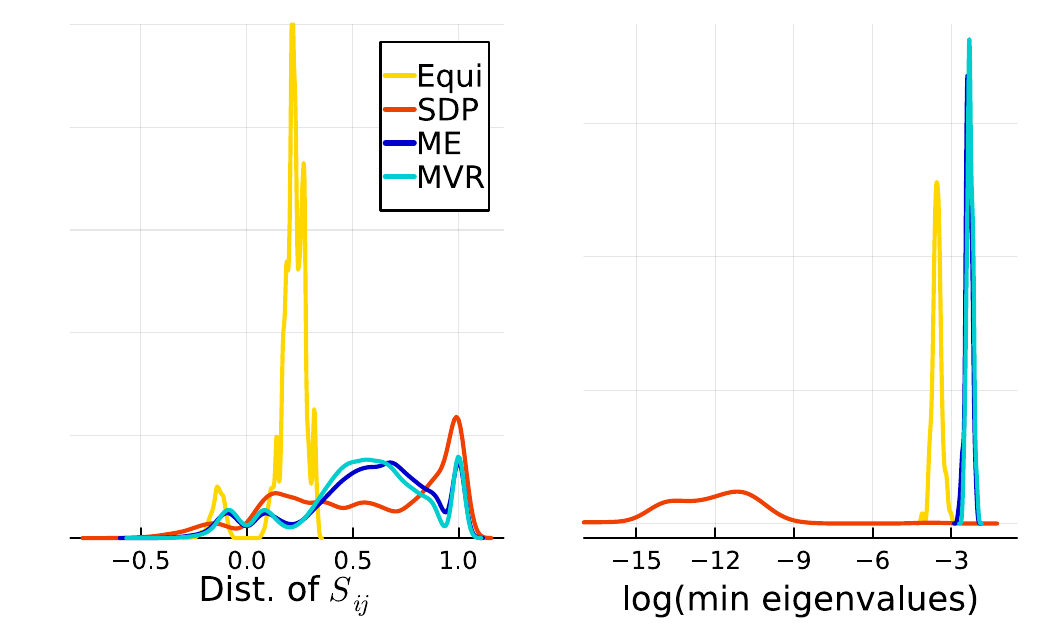}
    \caption{Comparison between methods for generating Gaussian knockoffs. On the left, we plot the distribution of non-zero values in $\bS$ matrix. On the right, we plot the log of the minimum eigenvalue of the covariance matrix $\bG_{\bS}$}
    \label{fig:S_and_eval_dist}
\end{figure}

To verify this phenomenon, here we compare the four group knockoff solvers and plot the distribution of non-zero entries of $\bS$ and the minimum eigenvalue of $\bG_{\bS}$ in Figure \ref{fig:S_and_eval_dist}. The simulation here replicates the ER(prec) simulation in Figure \ref{fig:sim1} except we generate $m=1$ knockoff for simplicity. One observes that SDP solvers do produce rather large $\bS$ values, with a heavy tail for values closer to 1. However, we know from Figure \ref{fig:sim1} that SDP solvers exhibit worse power than ME/MVR in this example. Furthermore, the minimum eigenvalue of $\bG_{\bS_{\text{SDP}}}$ appears much closer to 0 than that of MVR/ME solvers, causing $\bG_{\bS_{\text{SDP}}}$ to become numerically singular. Thus, the columns of the concatenated design matrix $[\bX \tilde{\bX}]$ are linearly dependent, exemplifying the reconstruction effect discussed in \cite{spector2022powerful}. In both plots, MVR and ME results are quite similar, as they ought to be since $L_{\text{MVR}}$ and $L_{\text{ME}}$ are very similar in the Gaussian case. This explains their comparable power and FDR. 

%

\end{document}